\documentclass[fleqn,usenatbib]{mnras}

\usepackage{newtxtext,newtxmath}

\usepackage[T1]{fontenc}

\DeclareRobustCommand{\VAN}[3]{#2}
\let\VANthebibliography\thebibliography
\def\thebibliography{\DeclareRobustCommand{\VAN}[3]{##3}\VANthebibliography}

\usepackage{graphicx}	% Including figure files
\usepackage{amsmath}	% Advanced maths commands
\usepackage{pgffor}
\usepackage{xtab,booktabs}
\usepackage{longtable}

\newcommand{\mi}{\relax \ifmmode {\mu{\mbox m}}\else $\mu$m\fi}
\newcommand{\sii}{\relax \ifmmode {\mbox S\,{\scshape ii}}\else S\,{\scshape ii}\fi}
\newcommand{\siii}{\relax \ifmmode {\mbox S\,{\textsc {iii}}}\else S\,{\scshape iii}\fi}
\newcommand{\siv}{\relax \ifmmode {\mbox S\,{\textsc {iv}}}\else S\,{\scshape iv}\fi}
\newcommand{\nii}{\relax \ifmmode {\mbox N\,{\scshape ii}}\else N\,{\scshape ii}\fi}
\newcommand{\neii}{\relax \ifmmode {\mbox Ne\,{\textsc {ii}}}\else Ne\,{\scshape ii}\fi}
\newcommand{\neiii}{\relax \ifmmode {\mbox Ne\,{\textsc {iii}}}\else Ne\,{\scshape iii}\fi}
\newcommand{\oiii}{\relax \ifmmode {\mbox O\,{\scshape iii}}\else O\,{\scshape iii}\fi}
\newcommand{\oii}{\relax \ifmmode {\mbox O\,{\scshape ii}}\else O\,{\scshape ii}\fi}
\newcommand{\oi}{\relax \ifmmode {\mbox O\,{\scshape i}}\else O\,{\scshape i}\fi}
\newcommand{\ha}{\relax \ifmmode {\mbox H}\alpha\else H$\alpha$\fi}
\newcommand{\hep}{\relax \ifmmode {\mbox H}\epsilon\else H$\epsilon$\fi}
\newcommand{\hdel}{\relax \ifmmode {\mbox H}\delta\else H$\delta$\fi}
\newcommand{\hgam}{\relax \ifmmode {\mbox H}\gamma\else H$\gamma$\fi}
\newcommand{\doclogO}{\relax \ifmmode {\mbox 12 + log(O/H)}\else  12 + log(O/H)}

\newcommand{\pa}{\relax \ifmmode {\mbox Pa}\alpha\else Pa$\alpha$\fi}
\newcommand{\hb}{\relax \ifmmode {\mbox H}\beta\else H$\beta$\fi}
\newcommand{\rdostres}{\relax \ifmmode {\,\mbox{R}}_{\rm 23}\else \,\mbox{R}$_{\rm 23}$\fi}
\newcommand{\ergs}{\relax \ifmmode {\,\mbox{erg\,s}}^{-1}\else \,\mbox{erg\,s}$^{-1}$\fi}
\newcommand{\me}{\relax \ifmmode {\,}^{-1}\else \,$^{-1}$\fi}

\newcommand{\msun}{\relax \ifmmode {\,\mbox{M}}_{\odot}\else \,\mbox{M}$_{\odot}$\fi}

\newcommand{\cmtres}{\relax \ifmmode {\,\mbox{cm}}^{-3}\else \,\mbox{cm}$^{-3}$\fi}
\newcommand{\cmdos}{\relax \ifmmode {\,\mbox{cm}}^{-2}\else \,\mbox{cm}$^{-2}$\fi}
\newcommand{\cmseis}{\relax \ifmmode {\,\mbox{cm}}^{-6}\else \,\mbox{cm}$^{-6}$\fi}
\newcommand{\arcminut}{$^{\prime}$}

\newcommand{\chidos}{$\chi^{2}_{\rm r}$}
\newcommand{\stol}{small-to-large}
\newcommand{\DstoDl}{$D_{S}/D_{L}$}
\newcommand{\hii}{\relax \ifmmode {\mbox H\,{\scshape ii}}\else H\,{\scshape ii}\fi}
\title[Dust grain size evolution in local galaxies: a comparison between observations and simulations]{Dust grain size evolution in local galaxies: a comparison between observations and simulations}

\author[M. Relaño et al.]{
M. Relaño$^{1,2}$\thanks{E-mail: mrelano@ugr.es},
I. De Looze$^{3,4}$,
A. Saintonge $^{4}$,
K.-C., Hou$^{5}$,
L.\,E.\,C.Romano$^{6,7}$,
K. Nagamine$^{6,8,9}$,
H. Hirashita$^{10}$, \newauthor 
S. Aoyama$^{10}$, 
I. Lamperti$^{11}$,
U. Lisenfeld$^{1,2}$,
M.~W.~L. Smith$^{12}$,
J. Chastenet$^{3}$,
T. Xiao$^{13}$,
Y. Gao$^{14}$,\newauthor
M. Sargent$^{15}$,
and S.\,A. van der Giessen$^{3}$
\\
$^{1}$Dept. F\'{i}sica Te\'orica y del Cosmos, Universidad de Granada, Spain\\
$^{2}$Instituto Universitario Carlos I de F\'isica Te\'orica y Computacional, Universidad de Granada, 18071, Granada, Spain\\
$^{3}$Sterrenkundig Observatorium, Universiteit Gent, Krijgslaan 281 S9, B-9000 Gent, Belgium\\
$^{4}$Department of Physics and Astronomy, University College London, Gower Street, London WC1E 6BT, UK\\
$^{5}$Physics Department, Ben-Gurion University of the Negev, Be'er-Sheva 84105, Israel \\
$^{6}$Theoretical Astrophysics, Department of Earth and Space Science, Osaka University, 1-1 Machikaneyama, Toyonaka, Osaka 560-0043, Japan\\  
$^{7}$Physik-Department, Technische Universität München, James-Franck-Straße, 85748 Garching, Germany \\
$^{8}$Kavli IPMU (WPI), The University of Tokyo, 5-1-5 Kashiwanoha, Kashiwa, Chiba 277-8583, Japan \\
$^{9}$Department of Physics and Astronomy, University of Nevada, Las Vegas, 4505 S. Maryland Pkwy, Las Vegas, NV 89154-4002, USA\\
$^{10}$Institute of Astronomy and Astrophysics, Academia Sinica, Astronomy-Mathematics Building, AS/NTU, No. 1, Section 4, Roosevelt Road, Taipei 10617, Taiwan \\
$^{11}$Centro de Astrobiología (CSIC-INTA), Ctra. de Ajalvir, km 4, 28850 Torrejón de Ardoz, Madrid, Spain\\ 
$^{12}$Cardiff Hub for Astrophysics Research \&\ Technology, School of Physics \&\ Astronomy, Cardiff University, Queens Buildings, Cardiff, CF24 3AA, UK \\
$^{13}$Department of Physics, Zhejiang University, Hangzhou, Zhejiang 310027, China \\
$^{14}$Purple Mountain Observatory and Key Lab of Radio Astronomy, Chinese Academy of Sciences, Nanjing 210034, China \\  
$^{15}$Astronomy Centre, Department of Physics and Astronomy, University of Sussex, Brighton BN1 9QH, England 
}

\date{Accepted XXX. Received YYY; in original form ZZZ}

\pubyear{2022}

\begin{document}
\label{firstpage}
\pagerange{\pageref{firstpage}--\pageref{lastpage}}
\maketitle

% Abstract of the paper
\begin{abstract}
The evolution of the dust grain size distribution has been studied in recent years with great detail in cosmological hydrodynamical simulations taking into account all the channels under which dust evolves in the interstellar medium. We present a systematic analysis of the observed spectral energy distribution of a large sample of galaxies in the local universe in order to derive not only the total dust masses but also the relative mass fraction between small and large dust grains (\DstoDl). Simulations reproduce fairly well the observations except for the high stellar mass regime where dust masses tend to be overestimated. We find that $\sim45$\,\% of galaxies exhibit \DstoDl\ consistent with the expectations of simulations, while there is a sub-sample of massive galaxies presenting high \DstoDl\ ($\log(D_{S}/D_{L})\sim-0.5$), and deviating from the prediction in simulations. For these galaxies, which also have high molecular gas mass fractions and metallicities, coagulation is not an important mechanism affecting the dust evolution. Including diffusion, transporting large grains from dense regions to a more diffuse medium where they can be easily shattered,  would explain the observed high  \DstoDl\ values in these galaxies. With this study we reinforce the use of the  \stol\ grain mass ratio to study the relative importance of the different mechanisms in the dust life cycle. Multi-phase hydrodynamical simulations with detailed feedback prescriptions and more realistic subgrid models for the dense phase could help to reproduce the evolution of the dust grain size distribution traced by observations.
\end{abstract}


%%%%%%%%%%%%%%%%%%%%%%%%%%%%%%%%%%%%%%%%%%%%%%%%%%

%%%%%%%%%%%%%%%%% BODY OF PAPER %%%%%%%%%%%%%%%%%%

\section{Introduction}\label{sec:intro}

Dust is a key component of galaxies and it is directly linked to their evolution across time in the Universe. It absorbs the ultraviolet light and reemits the radiation in the infrared hampering the observations of the light coming from stars. Dust grains act as catalysts for star formation as it is on the surface of the dust grains where the molecular hydrogen forms \citep{1971ApJ...163..155H}. Since star formation occurs in molecular clouds dust is a key component for star formation and hence plays an active role in galaxy evolution. Interstellar dust can be ejected from galaxies, providing an additional cooling channel and playing an extra role in the evolution of galaxies \citep[e.g.][]{2019MNRAS.487.4870V}.

Dust is created in low-intermediate mass stars 
\citep[e.g.,][]{2013MNRAS.434.2390N,2014MNRAS.438.2328N,2017MNRAS.467.4431D,2019A&A...626A.100B} and supernovae (SNe) 
\citep[e.g.,][]{2015ApJ...800...50M,2019MNRAS.488..164D,2017ApJ...836..129T,2020MNRAS.493.2706C,2001MNRAS.325..726T} 
and during its lifetime it is affected by processes that destroy it, modify its physical properties and enhance its total amount in the interstellar medium (ISM). Dust can grow in the ISM by accretion of gas phase metals onto its surface. This mechanism, enhanced in high metallicity and dense environments, has been claimed to be an important source of dust production in local and high-redshift galaxies \citep[e.g.,][]{2017MNRAS.471.1743D}. However, the debate is still open as the physics under which the gas phase metals are finally incorporated into dust grains is not fully understood  \citep[e.g.,][]{2021MNRAS.502.2438P}. Dust temperatures in dense environments might be higher than previously thought \citep[e.g.,][]{2020MNRAS.498.4192F,2021MNRAS.508L..58B}, which would make accretion to be inefficient. Besides, dispersion of molecular clouds could potentially make the metals return to the gas phase on short time scales \citep[][]{2016MNRAS.463L.112F}. All this shows that to understand the evolution of the interstellar dust it is important not only to study the total amount, but also how the dust properties of the dust grains change with time and environment. 

The study of the interstellar dust has significantly improved in the last decades in two main directions. Firstly, it has been possible to cover observationally the full spectral energy distribution (SED) of galaxies at low and high redshifts \citep[e.g.][Chastenet et al. in prep.]{2014MNRAS.441.1017R,2018A&A...609A..37C,2019ApJS..244...24L, 2019ApJS..244...40L,2021MNRAS.507..129S}; observed SEDs of statistically significant samples of nearby galaxies allow to explore the main trends of the total dust mass with other physical properties of galaxies such as stellar mass, star formation rate or gas mass content \citep[e.g.,][]{2010MNRAS.403.1894D,2015MNRAS.452..397C,2017MNRAS.464.4680D}, and detailed spatially resolved SEDs have been analysed to pinpoint the dust properties at small linear scales \citep[e.g.][]{2014ApJ...780..172D,2014ApJ...797...85G,2017A&A...601A..55C}. Secondly, an impressive amount of work regarding dust evolution has been theoretically done in the following ways: linking the physics of dust in models of chemical evolution of galaxies \citep[e.g.,][]{2021MNRAS.505.3228D, 2021A&A...649A..18G, 2020MNRAS.496.3668D, 2017MNRAS.471.1743D, 2017MNRAS.471.4615G,2015MNRAS.449.3274F,2014MNRAS.441.1040R,2013EP&S...65..213A}, in hydrodynamical simulations including dust evolution in individual galaxies \citep[e.g.,][]{2020MNRAS.491.3844A,2017MNRAS.466..105A,2016ApJ...831..147Z,2015MNRAS.449.1625B,2013MNRAS.432.2298B}, in simulations of cosmological volumes \citep[e.g.,][]{2021MNRAS.503..511G,2020MNRAS.494.1071G,2019MNRAS.490.1425L,2018MNRAS.478.4905A,2017MNRAS.468.1505M}, and in galaxy formation semi-analytical models (SAM) \citep[e.g.,][]{2020MNRAS.493.2490T,2019MNRAS.489.4072V,2017MNRAS.471.3152P}. Post processing simulations with a dust radiative transfer approach \citep[e.g.][]{2021ApJS..252...12N,2016MNRAS.462.1057C,2020MNRAS.494.2823T,2021MNRAS.506.5703K}  have allowed to perform a deeper study on the physical properties of dust and its role in galaxy evolution. This significant progress in theory and observations enables us to explore how the relation between metals, stars, dust and gas has evolved across time in the universe \citep{2020ARA&A..58..363P,2021MNRAS.503.4537F}. 

In particular, hydrodynamical simulations include the main ingredients needed to explain the evolution of interstellar dust: stellar dust production, dust growth in the ISM, dust destruction and coagulation. Astration, (i.e. removal of dust in the star formation process) has also been added in most of the simulations\footnote{\citet{2018MNRAS.478.2851M} neglected astration in their simulations but recent studies \citep{2021MNRAS.503..511G,2017MNRAS.466..105A} highlight the importance of including this mechanism in the evolution of dust in the ISM.}. There are, however, limitations in how these mechanisms are incorporated into dust evolution models, which brings different outcomes for the simulated results: i) Dust growth has been added using different prescriptions \citep[e.g.][]{2016ApJ...831..147Z,1998ApJ...501..643D}. The total amount of dust grown in the ISM can vary depending on whether a limitation of the minimum amount of element species to form the dust grain \citep[{\it key element approximation}, see][]{2021MNRAS.503..511G,2016ApJ...831..147Z} is taken into account in the accretion process or not, and whether the grain size distribution is allowed to evolve when grain growth is taking place \citep{2021MNRAS.502.2438P}. ii) Stellar dust production has been incorporated with different dust condensation efficiencies and metal yields, e.g. \citet{2021MNRAS.503..511G} use condensation efficiencies for different chemical elements, while \citet{2017MNRAS.466..105A} apply a single factor to form dust for all the metals produced by the stars. 

Of particular interest is how the molecular gas mass fraction have been treated in the simulations. Accretion and coagulation occur in  dense environments, most probably related to molecular clouds. The dense gas phase is difficult to be added in the models due to the limited resolution of the simulations. In some studies \citep{2017MNRAS.466..105A,2019MNRAS.485.1727H}, the dense gas mass has been included in the simulations assuming that a fixed mass fraction of cold and dense gas particles are in the form of dense clouds where accretion and coagulation take place. But this parametrisation produces a global dense gas fraction that is significantly lower than the molecular gas mass fraction estimated for galaxies using CO observations \citep[e.g., ][]{2017ApJS..233...22S}. To overcome this difficulty, \citet{2021MNRAS.503..511G} introduced multiphase particles that allow a more accurate treatment of the dense gas mass fraction in accretion and coagulation processes. In \citet{2018MNRAS.474.1545C} a subgrid post-processing model was applied to explore the effect of dust evolution on the molecular gas content. Recently, \citet{2022MNRAS.tmp.1338R} study the evolution of the dust and molecular gas using numerical simulations of an isolated Milky-Way like galaxy and allowing the mass fraction of cold and dense gas particles in the form of dense clouds to vary with the density of the particle. This more realistic approach gives as a result a global dense gas fraction in better agreement with the observed molecular gas mass fractions in local galaxies. 

A step forward in studying the evolution of dust is to incorporate how the different dust grains are modified along the dust life in the ISM. In particular, the evolution of the grain size distribution is shaped by the most relevant mechanisms affecting the dust at each time step in the galaxy evolution \citep{2013MNRAS.432..637A}. In the simulations, dust is produced by stellar sources mainly in the form of large dust grains. Asymptotic Giant Branch (AGB) stars are thought to produce large (radius, $a$\,$\gtrsim$\,0.1\,\mi) dust grains \citep[e.g.,][]{1997A&A...326..305W,2012ApJ...745..159Y,2012MNRAS.424.2345V}, and dust produced by SN would have a higher contribution of large grains \citep[e.g.,][]{2014Natur.511..326G,2015MNRAS.446.2089W,2016MNRAS.456.1269B,2020MNRAS.491.6020P} as the reverse shock seems to be more effective in destroying small rather than large grains \citep{2007ApJ...666..955N,2007MNRAS.378..973B}. Grain growth via accretion of metals in the gas phase is favoured when the number of small grains is large \citep{2012MNRAS.422.1263H}, while fragmentation of dust grains associated with shattering creates a large number of small grains \citep[]{2009MNRAS.394.1061H,2004ApJ...616..895Y,1996ApJ...469..740J}. Finally, grain-grain collisions can lead to coagulation of dust grains, moving the grain size distribution towards the large radius regime \citep{2014MNRAS.437.1636H,2009A&A...502..845O}. 

Including the evolution of the grain size distribution in hydrodynamical simulations is a very expensive computing task (see \citealt{2018MNRAS.478.2851M} for a first attempt to implement the evolution of the full dust grain size distribution in cosmological simulations) that has been alleviated by the two-grain size approximation proposed by \citet{2015MNRAS.447.2937H}. This approximation is a robust representation of the evolution of the full grain size distribution, as was demonstrated in \citet{2020MNRAS.491.3844A}. The approximation has been successfully applied in numerous situations. \citet{2017MNRAS.466..105A} and \citet{2017MNRAS.469..870H} have applied it in SPH simulations of individual galaxies, \citet{2021MNRAS.503..511G} used it to study the dust evolution in a galaxy with zoom-in cosmological simulations of galaxy formation, and \citet{2018MNRAS.479.2588G} used the approximation to simulate dust evolution in galaxy cluster formation. \citet{2018MNRAS.478.4905A} incorporated the two-grain size formalism to study the relative contribution of small and large grains in the circum-galactic (CGM) and intergalactic (IGM) medium. The two-grain size approximation has also been used to analyse extinction curves as a function of redshift in \citet{2019MNRAS.485.1727H}. Moreover, \citet{2019MNRAS.482.2555H} post-processed the simulation of an isolated spiral galaxy performed in \citet{2017MNRAS.466..105A} and studied how the full grain size distribution evolves in dense and diffuse medium. They found that grain growth and coagulation occurring in the dense ISM are important to recover the grain size distribution that reproduces the Milky Way extinction curve.

Physical dust properties such as dust mass, temperature and dust mass function predicted from the simulations have been compared in general with observations provided by the literature (see, for example, \citealt{2021arXiv210800830V, 2019MNRAS.484.1852A}). However, in the case of the most sophisticated simulations including evolution of the grain size distribution \citep[e.g.,][]{2021MNRAS.503..511G,2019MNRAS.485.1727H} it is necessary to perform a comparison on how the small-to-large grain mass ratio obtained from observations varies as a function of the different properties of the galaxy. This requires a systematic methodology to extract the relative amount of small and large grains from the observed SEDs. In \citet{2020A&A...636A..18R} we performed a comparison of the \stol\ grain mass ratio derived from fitting the observed SEDs of a sample of galaxies with the simulations performed by \citet{2019MNRAS.485.1727H}. We found good agreement between observations and simulations but the characteristics of the galaxy sample used in  \citet{2020A&A...636A..18R} lacked of a wide range of parameters to test the validity of the theoretical assumptions of the simulations, and therefore we were not able to explore the full range of parameters that the simulations covered. We furthermore obtained the radial variation of the \stol\ grain mass ratio in a sample of three nearby galaxies and found good correlation with the predictions of SPH simulations of individual galaxies. In a recent paper, \citet{2021MNRAS.503..511G} compare the radial trend of the \stol\ grain mass ratio predicted for a spiral galaxy by zoom-in cosmological simulations with the radial trend derived from the observations in  \citet{2020A&A...636A..18R}, finding very good agreement between simulations and observations. 

This study attempts to overcome the lack of a systematic comparison between observations and simulations that include evolution of the grain size distribution. We apply a rigorous methodology to extract the relative mass fraction of small and large dust grains from the observed SEDs in a large sample of galaxies extending the range of galaxy properties to cover the high mass and high metallicity end. The observational analysis is treated in a consistent way for all our galaxy samples. Dust masses and \stol\ grain mass ratios are obtained with the same methodology for all the objects and comparisons with previously derived dust masses are carefully performed. We analyse the relation of the dust masses and \stol\ grain mass ratios with other galaxy properties to infer under which physical conditions a dust formation/destruction mechanism might dominate the evolution of the interstellar dust. Dust masses and \stol\ grain mass ratios are given in this paper to provide the future simulations and semi-analytical models with an observational data set that can be used to set up constrains on the prescriptions and initial conditions generally used in these type of studies. 

The paper is organised as follows: in Section\,\ref{sec:samp} we present the galaxy sample we have studied in this paper, as well as the selection procedure to obtain a final subsample with well determined observational SEDs. In Section\,\ref{sec:fit} we present our fitting methodology. Section\,\ref{sec:results} shows the dust mass and \stol\ grain mass ratios derived from the SED fitting. In Section\,\ref{sec:compsim} we compare our results with the predictions of hydrodynamical simulations. We discuss the results in Section\,\ref{sec:discussion} and present our conclusions in Section\,\ref{sec:conc}.   
  
\section{Galaxy sample}
\label{sec:samp}
The galaxy sample used in this study is selected from a combination of surveys that provide the integrated IR fluxes from $\sim$\,3\mi, to 500\,\mi\ and span a wide range of physical properties. We refer the reader to the main paper regarding each survey and we will highlight here the most relevant survey characteristics for this study. 

JINGLE \citep{2018MNRAS.481.3497S} is a James Clerk Maxwell Telescope (JCMT) legacy survey assembling galaxies in the local Universe with the aim to study systematically the cold interstellar medium. The sample consists of 193 ($z=0.01-0.05$) SDSS-selected galaxies covering homogenously the star formation rate$-$stellar mass (SFR-M$_{\rm star}$) plane between 10$^9$ and 10$^{11}$\,M$_{\odot}$. The sample is required to have detections in  \textit{Herschel} SPIRE 250\,\mi\ and 350\,\mi\ bands. Most of the JINGLE galaxies are classified as late-type spirals or irregular galaxies. Observations with JCMT SCUBA-2 850\,\mi\ \citep{2019MNRAS.486.4166S} were performed to map the FIR range of the dust emission spectrum and RxA CO J=2-1 observations have been done so far for 63 JINGLE galaxies (Xiao et al. in prep). We rely on the aperture photometry and estimated errors of the  \textit{Herschel},  \textit{WISE} and SCUBA-2 maps presented in \citet{2019MNRAS.486.4166S}, which cover from 3.4\,\mi\ from \textit{WISE} to 500\,\mi\ from \textit{Herschel}\footnote{\citet{2019MNRAS.486.4166S} include also photometry from SCUBA-2 maps at 850\,\mi. For homogeneity with the rest of the galaxy sample we decided to exclude the SCUBA-2 fluxes.}. PACS photometry does not cover the 70\,\mi\ band, therefore we obtain 
IRAS\,60\,\mi\ fluxes using SCANPI\footnote{\url {https://irsa.ipac.caltech.edu/applications/Scanpi/}}
following the same methodology as in \citet{2003AJ....126.1607S}. We find that 31 galaxies had no-detections in IRAS\,60\,\mi. We furthermore do not take into account IRAS\,60\,\mi\ fluxes below 3\,$\sigma$. Finally, after inspecting the observed SEDs, we found that some galaxies present higher fluxes in the IRAS\,60\,\mi\ band than in PACS\,100\,\mi. These fluxes might suffer from contamination from other sources, as the angular size of the JINGLE galaxies is lower than the typical angular resolution at IRAS\,60\,\mi\ \citep[1.5\arcminut,][]{2018A&A...609A..37C}. We decided to eliminate the IRAS\,60\,\mi\ fluxes in this situation. Out of 193 JINGLE galaxies, 115 galaxies have IRAS\,60\,\mi\ fluxes in their SEDs. The H{\sc{i}} masses and uncertainties were obtained from the ALFALFA catalog \citep{2018ApJ...861...49H} and JINGLE H{\sc{i}} observations at Arecibo, as described in \citet{2020MNRAS.496.3668D}. H$_{2}$ gas masses were derived for 63 galaxies from RxA CO J=2-1 observations (Xiao et al. in preparation). Stellar masses and star formation rates with the corresponding uncertainties were inferred from \texttt{MagPhys} \citep{2008MNRAS.388.1595D} as presented in \citet{2018MNRAS.481.3497S}. 

The \textit{Herschel} Reference Survey \citep[HRS,][]{2010PASP..122..261B,2012A&A...540A..52C,2012A&A...543A.161C,2014A&A...565A.128C,2014A&A...564A..66B} consists on 322 $K$-band selected galaxies in a volume-limited sample with distances between 15 and 25\,Mpc. The sample contains a wide range of morphological types and environments (more than half of the HRS sample consists of cluster galaxies). \citet{2018A&A...609A..37C} obtained the photometry using a Comprehensive \& Adaptable Aperture Photometry Routine (CAAPR) in all IR available bands (\textit{WISE},  \textit{Spitzer} and \textit{Herschel}) for 288 HRS galaxies in common with the Dustpedia sample \citep{2017PASP..129d4102D}. We use these set of fluxes to perform our SED fitting. Stellar masses and star formation rates with the corresponding uncertainties were inferred from \texttt{MagPhys} in \citet{2017MNRAS.464.4680D}. The H{\sc{i}} and H$_{2}$ masses (using a Galactic standard value $\rm X_{CO}$ factor) and corresponding uncertainties were taken from \citet{2014A&A...564A..65B}. We assume a 15\% uncertainty for H{\sc{i}} masses as given in \citet{2014A&A...564A..66B}. We assume the same oxygen abundances as in \citet{2020MNRAS.496.3668D} which were extracted from \citet{2013A&A...550A.115H} using the  O3N2 calibrator. 

HiGH galaxies \citep{2017MNRAS.464.4680D} were selected from the \textit{Herschel} Astrophysical Terahertz Large Area Survey (H-ATLAS, \citealt{2010PASP..122..499E}) based on H{\sc{i}} detections. The sample is formed by 40 galaxies with distances between 11.3 and 159\,Mpc. HiGH galaxies are blue, low surface brightness gas-rich galaxies which are actively forming stars. They have low stellar masses (M$_{\rm star}$$\leq$10$^{9}$\,M$_{\odot}$) which indicates an early evolution stage. We rely on the IR fluxes (\textit{WISE},  \textit{Spitzer}, \textit{Herschel}, and IRAS\,60\,\mi), as well as the H{\sc{i}} gas masses and the stellar and star formation rates derived using \texttt{MagPhys} reported in \citet{2017MNRAS.464.4680D}. There are no H$_{2}$ masses reported in the literature for these galaxies.  

The Key Insights on Nearby Galaxies: A Far-Infrared Survey with \textit{Herschel} \citep[KINGFISH,][]{2011PASP..123.1347K} is a sample of 61 nearby (D < 30\,Mpc) galaxies covering a wide range of morphologies, galaxy properties and local ISM environments. Aperture photometry of the KINGFISH galaxies have been done in \citet{2017ApJ...837...90D} where the SEDs have been covered in \textit{WISE}, \textit{Spitzer} and \textit{Herschel} (including PACS\,70\,\mi\ and MIPS\,70\,\mi). We use the H{\sc{i}} gas masses and H$_{2}$ gas masses (assuming a Galactic standard $\rm X_{CO}$ factor) reported in \citet{2014A&A...563A..31R}. Stellar mass and star formation rates inferred from \texttt{MagPhys} have been taken from \citet{2019A&A...621A..51H}. 

The Dwarf Galaxy Survey (DGS, \citealt{2013PASP..125..600M}) is a sample of 48 dwarf galaxies especially designed to study the dust properties in low metallicity environments from photometric and spectroscopic observations with \textit{Herschel} \citep{2013A&A...557A..95R}. The galaxies span a wide range in metallicity from 12 + log(O/H) = 7.2\,$-$\,8.4 and have in general low dust and gas content. We rely on the updated version of the IR fluxes presented in \citet{2015A&A...582A.121R} for \textit{WISE}, \textit{Spitzer} and \textit{Herschel} (including PACS\,70\,\mi). The H{\sc{i}} and H$_{2}$ gas masses have been reported in \citet{2014A&A...563A..31R}. For H$_{2}$ gas masses we use those given in  \citet{2014A&A...563A..31R} assuming a Galactic standard $\rm X_{CO}$ factor. There is no fitting with  \texttt{MagPhys} for this galaxy sample in the literature, therefore we rely on the stellar mass and star formation rates derived in \citet{2015A&A...582A.121R}. In this paper, the stellar masses were derived following \citet{2012AJ....143..139E} and the star formation rates were obtained with the linear combination of \ha\ and TIR luminosities, following the prescription from \citet{2009ApJ...703.1672K}. We use the correlation found by  \citet{2019A&A...621A..51H} between stellar masses derived from the photometry in 3.6\,\mi\ and the predictions from \texttt{MagPhys} to corroborate that the stellar masses presented in  \citet{2015A&A...582A.121R} were in agreement within the expected values from \texttt{MagPhys}. Unfortunately we cannot do the same check with star formation rates as the correlations found in \citet{2019A&A...621A..51H} were obtained either using FUV+TIR or \ha+24\,\mi, and there are no FUV nor 24\,\mi\ observations for the whole DGS sample. 

The combined galaxy sample allows us to explore a wide range of galaxy properties when comparing with simulations. JINGLE galaxies cover relatively well the high stellar mass regime (M$_{\rm star}\sim$10$^{10}$-10$^{11}$\,M$_{\odot}$) and star formation rate range (SFR\,$\sim$\,0.1-10\,M$_{\odot}$\,yr$^{-1}$). HRS galaxies are stellar mass selected, therefore they are more evolved galaxies with typically lower gas mass fractions. HiGH and DGS galaxies are galaxies with active star formation ($\rm\log\,(sSFR )\gtrsim$\,-10$^{-1}$\,Gyr$^{-1}$, with sSFR\,=\,$\rm SFR/M_{star}$ the specific star formation rate), cover the lower range in stellar mass (M$_{\rm star}\,\sim\,$10$^{7}$-10$^{11}$\,M$_{\odot}$), and are typically at early evolution stages. KINGFISH galaxies cover a wide range of morphology and stellar mass being typically disk galaxies with active star formation. The average values of the main properties of JINGLE, HRS, HiGH and KINGFISH are presented in table\,1 of \citet{2020MNRAS.496.3668D} and for DGS in table\,1 of \citet{2015A&A...582A.121R}, we refer the reader to those papers to obtain more detailed information on the galaxy parameters covered by each survey. 

\subsection{Final selected sample}

In order to make accurate estimates of the \stol\ grain mass ratio we would ideally need the full 3-500\,\mi\ wavelength range of the SED covered with well detected band fluxes. In Table\,\ref{tab:bands} we present a summary of the observed bands for our galaxy samples. Most of the galaxy samples have observations covering the 3-500\,\mi\ wavelength range, however some galaxies do not have emission detected in all the bands. In order to create a final selection of galaxies with well covered SEDs we have applied the following criteria to the aforementioned samples: $i$) when observations from \textit{Spitzer} 3.6/4.5\,\mi\ and \textit{WISE} 3.5/4.6\,\mi\ are available we select those galaxies that have at least fluxes above 3\,$\sigma$ in any of these instruments. $ii$) The same criteria is applied to \textit{Spitzer} 24\,\mi\ and \textit{WISE} 22\,\mi\ when both observations are available, we reject galaxies not having fluxes above 3\,$\sigma$ in any of these bands. $iii$) IRAS\,60\,\mi\ fluxes are taken when \textit{Herschel} 70\,\mi\ is not available and we reject galaxies not having detected fluxes at least in one of these bands. $iv$) Finally, we also require to have detected fluxes in all \textit{Herschel} 100, 160, 250, 350 and 500\,\mi\ bands. Although we might introduce some bias towards galaxies with bright and well-detected IR emission, with these criteria we guarantee that the SEDs of the selected galaxies are well covered in the 3-500\,\mi\ wavelength range and a robust fit can be performed. Once we created the galaxy sample we inspected by eye each fit and we removed the following galaxies: M81DwB from the KINGFISH sample and NGC\,5253, NGC\,1705, NGC\,625 and VIIZw403 from the DGS sample, as these galaxies could not be fitted with the constrains assumed in the initial parameters of the fitting procedure (see Section\,\ref{sec:fit}). NGC\,2366,  NGC\,4861(and IIZw40) from DGS were also eliminated because our best fits defined their SEDs with only the emission from very small grains (VSGs) (and big grains, BGs), which we consider as unphysical fits. The same occurred for NGC\,5713 from HiGH. Our final sample consists of 247 galaxies: 10 from DGS, 98 JINGLE galaxies, 53 from KINGFISH, 13 HiGH galaxies and 72 HRS galaxies.  
\begin{table*}
\caption{Available observations for our galaxy sample}\label{tab:bands}
\begin{tabular}{ccccccccc}\hline
Galaxy sample & \textit{Spitzer}-IRAC & \textit{WISE} & \textit{Spitzer}-MIPS & IRAS\,60\,\mi\ & \textit{Herschel}\,70\,\mi\ &  \textit{Herschel}\,100\,\mi\ & \textit{Herschel}\,160\,\mi\ & \textit{Herschel}\,SPIRE \\
 \hline
JINGLE & & \checkmark & \checkmark & \checkmark & \checkmark & \checkmark & \checkmark & \checkmark \\
HRS &  \checkmark & \checkmark & \checkmark & \checkmark & \checkmark & \checkmark & \checkmark & \checkmark \\
HiGH &  & \checkmark & & \checkmark & & \checkmark & \checkmark & \checkmark \\
KINGFISH & \checkmark & \checkmark & \checkmark &  & \checkmark & \checkmark & \checkmark & \checkmark \\
DGS &  \checkmark & \checkmark &  &  & \checkmark & \checkmark & \checkmark & \checkmark \\
\hline
\end{tabular}
\end{table*}

\section{SED fitting}\label{sec:fit}

\subsection{Dust model and multi-ISRF}\label{sec:dustmod}
We used the classical \citet{1990A&A...237..215D} dust model, which consists of three different grain populations: polycyclic aromatic hydrocarbons (PAHs), VSGs, and big silicate grains (BGs). The model assumes that PAHs are grains with radii (0.4-1.2)$\times$10$^{-3}$\mi, VSGs correspond to grains with radii of (1.2-15)$\times$10$^{-3}$\mi\  and that BGs are grains with radii larger than 15$\times$10$^{-3}$\mi. The two-grain size approximation proposed by \cite{2015MNRAS.447.2937H} separates the grain size distribution in small grains with radius $a\,<\,0.03$\,\mi, and large grains with radius $a\,>\,0.03$\,\mi,  therefore small grains for our dust models are PAHs and VSGs, whereas large grains correspond to silicate BGs. We decided to apply \citet{1990A&A...237..215D} dust model for the present study because its simplicity and the small number of free parameters compared to the more sophisticated recent dust models \citep[e.g.][]{2007ApJ...657..810D,2011A&A...525A.103C,2013A&A...558A..62J,2017A&A...602A..46J} allow us a better comparison with the results from simulations. In \citet{2016A&A...595A..43R} we compared the dust models of both \citet{1990A&A...237..215D} and \citet{2011A&A...525A.103C} and found that they both agree in reproducing the relative abundance of VSG and BG grains. We refer the reader to \citet{2021ApJ...912..103C} for a compilation of the most recent dust models and a rigorous study of the variation of dust mass estimates from them. 

The \citet{1990A&A...237..215D} dust model was already applied to study the integrated SED of the DGS and KINGFISH galaxies in \citet{2020A&A...636A..18R} where we derived \stol\ grain mass ratio for each galaxy and compared the results with other galaxy properties as well as with the predictions from simulations by \citet{2018MNRAS.478.4905A} and  \citet{2019MNRAS.485.1727H}. We improve here the study presented in \citet{2020A&A...636A..18R} by introducing a combination of starlight intensities per unit dust mass to describe the observed SED of the galaxy. The methodology is explained in detailed in \citet{2011A&A...536A..88G}. Here we present a summary of the strategy and refer the reader to section 3.2 in \citet{2011A&A...536A..88G} for a full explanation. 

The main idea is to assume that a mass unit of ISM is heated by an interstellar radiation field (ISRF) with a certain spectral shape. We will assume the ISRF shape as the one corresponding to the solar neighbourhood \citep{Mathis:1983p593}, therefore a value for the scale factor $U_0=1$ corresponds to a radiation field of the solar neighbourhood of 2.2$\times\,10^{-5}$\,W\,m$^{-2}$. The conditions under which the dust is heated in the ISM may differ from the conditions in the local solar neighbourhood. To account for this, we assume a distribution of starlight intensities per unit dust mass through the galaxy that can be approximated by a power-law \citep{2001ApJ...549..215D}:   
       \begin{equation}
          \frac{{\rm d} M_{\rm dust}}{{\rm d} U} \propto U^{-\alpha} 
          \mbox{ with } U_{\rm min}\leq U \leq U_{\rm min} + \Delta\,U  
          \label{eq:dale}
        \end{equation}
where $\alpha$ allows to parametrise the possible physical conditions in the ISM (see section 5.5. in \citealt{2001ApJ...549..215D} for a justification of this choice) and $\Delta\,U$ the range of starlight intensities. The total dust mass of the galaxy will be then: 
\begin{equation}
  M_{\rm dust} = \int_{U_{\rm min}}^{U_{\rm min}+\Delta U} 
                 \frac{{\rm d} M_{\rm dust}}{{\rm d} U}{\rm d} U
  \label{eq:Mdust}
\end{equation}
and the distribution of the ISRF for the galaxy can be characterised following \citet{2011A&A...536A..88G} and \citet{2015A&A...582A.121R} by the mass-averaged starlight intensity: 
\begin{equation}
\langle U\rangle = 
  \frac{1}{M_{\rm dust}}
  \int_{U_{\rm min}}^{U_{\rm min}+\Delta U} U\times \frac{{\rm d} M_{\rm dust}}{{\rm d} U}\,{\rm d} U
\end{equation}
and the variance in the starlight intensity distribution: 
\begin{equation}
\sigma^{2} (U) = 
  \frac{1}{M_{\rm dust}}
  \int_{U_{\rm min}}^{U_{\rm min}+\Delta U}\left(U-\langle U\rangle\right)^2
  \times \frac{{\rm d} M_{\rm dust}}{{\rm d} U}\,{\rm d} U
\end{equation}
We explore if a typical ISRF of a young star cluster of 4\,Myr would give better results in our fitting procedure as it was done in \citet{2016A&A...595A..43R} but we found no improvements. Therefore, we finally keep the ISRF of the solar neighbourhood as the one to build up the starlight distribution of the galaxy. 

\subsection{SED fitting methodology}

We fit the observed SEDs for each individual galaxy with the dust model and multi-ISRF strategy presented in Section\,\ref{sec:dustmod}. The free parameters for our fitting strategy are the masses of the different components in the dust model: $\rm M_{\rm PAH}$, $\rm M_{\rm VSG}$, and $\rm M_{\rm BG}$ in M$_{\odot}$, the minimum value of the power-law distribution in Eq.\,\ref{eq:Mdust}, $U_{\rm min}$ in units of  $U_0$; and the exponent $\alpha$ of the power-law distribution. We keep the maximum starlight intensity $U_{\rm max}=U_{\rm min}+\Delta U$ fixed to a value of 10$^{7}\, U_0$, following the same methodology as in previous studies \citep{2019A&A...624A..80N,2020MNRAS.496.3668D}. Additionally, to include the contribution of the old stellar population in the NIR part of the SEDs, we add a black body of T=5,000\,K parameterised by a scale factor $U_{\rm NIR}$. 

Given a set of input parameters $\vec{\theta}$ and a set of observations $\vec{x}$, the posterior probability function can be described as: 
\begin{equation}
p(\vec{\theta} | \vec{x}) \propto p(\vec{\theta})\,p(\vec{x} | \vec{\theta})
\end{equation}
where $p(\vec{\theta})$ is the \textit{prior} distribution, representing the initial distribution of the parameters, and $p(\vec{x} | \vec{\theta})$ is the likelihood function which, under the assumption that the noise follows a normal distribution, is expressed as: 
\begin{equation}
p(\vec{x} | \vec{\theta}) \propto \exp\left( -\frac{1}{2}\chi^2(\theta)\right)
\end{equation}
where, 
\begin{equation}
\chi^2(\theta) = [\vec{F}_\mathrm{obs} - \vec{F}_\mathrm{mod}(\theta)]^T
  \mathbb{C}^{-1}
  [(\vec{F}_\mathrm{obs} - \vec{F}_\mathrm{mod}(\theta)].
\label{eq_chisqr}
\end{equation}
$\vec{F}_\mathrm{mod}(\theta)$ are the band fluxes for each particular model defined by the set of input parameters $\theta$, $\vec{F}_\mathrm{obs}$ are the observed band fluxes, and $\mathbb{C}$ is the covariance matrix (see next section) which takes into account the uncertainties in the photometry and the correlated and uncorrelated uncertainties in the calibration of the data \citep[see][]{2017A&A...601A..55C,2014ApJ...797...85G}. 
 
\subsubsection{Uncertainties in the SED fitting}\label{sec:uncer}
The uncertainties are incorporated in the fitting procedure via the  covariance matrix, $\mathbb{C}$, which is the sum of a diagonal matrix carrying information of the uncertainties in the photometry for each band and a matrix including the errors in the calibration. The uncertainties in the photometry are taken from the values reported in each individual galaxy sample study (see Section\,\ref{sec:samp}). For the calibration uncertainties we follow the methodology of \citet{2014ApJ...797...85G} and  \citet{2017A&A...601A..55C} \citep[see also][]{2012ApJ...748..123S,2012ApJ...756...40S}. The correlated and uncorrelated uncertainties for PACS and SPIRE were taken from \citet{2014ApJ...797...85G} and for IRAC, MIPS 24\,\mi\ and MIPS\,70\mi\ we use the values reported in \cite{2017A&A...601A..55C}. For MIPS 160\mi\ emission we use an absolute calibration uncertainty (correlated noise) of 12\% and a repeatability of 5\% from \citet{2007PASP..119.1038S}. We account for uncorrelated calibration uncertainties for \textit{WISE} as 2.4\%, 2.8\%, 4.5\% and 5.7\% for \textit{WISE}\,3.4\,\mi, 4.6\,\mi\, 12\,\mi\, and 22\,\mi, respectively \citep{2013AJ....145....6J} and 20\% for the IRAS\,60\,\mi\ \citep{2018A&A...609A..37C}. 

\subsubsection{Markov chain Monte Carlo approach}
\begin{figure*}
\includegraphics[width=\textwidth]{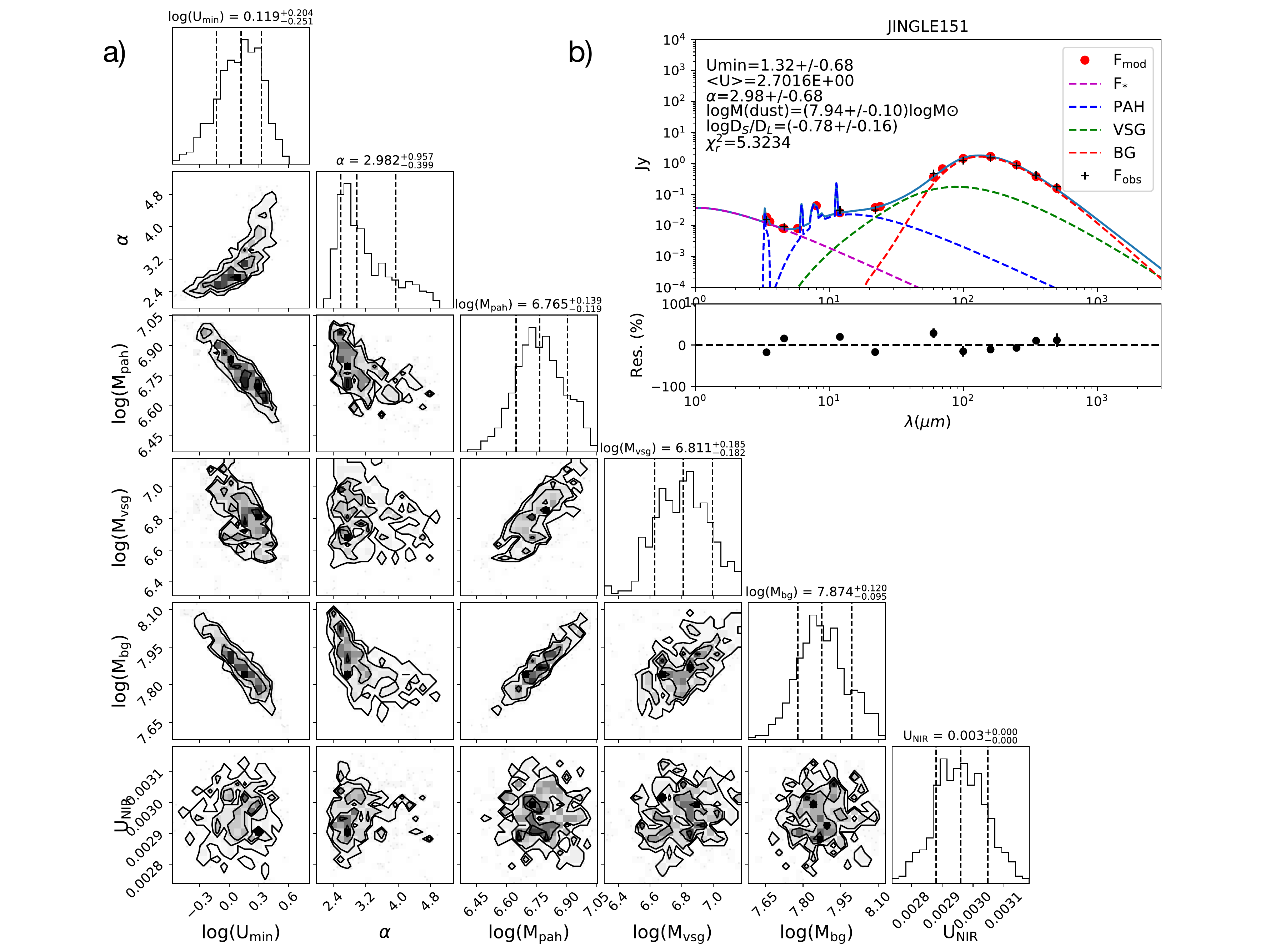}
\caption{(a) Corner plot showing the posterior probability distribution of the free parameters for the galaxy JINGLE\,151. At the top of each panel we show the 50th percentile of the marginalised distribution for each parameter as well as the 16th and 84th percentiles which are taken as estimations of the uncertainties in the best fit parameters. (b) SED with the best fit (blue continuous line) which is the sum of the emission of  PAHs (blue dashed line), VSGs (green dashed line), BGs (red dashed line) and the emission from old stars ($\rm F_{*}$, purple dashed line). Observed data are represented with crosses and red dots correspond to the modelled fluxes for each band. In the bottom panel we show the residuals of the fit in \%.}\label{fig:corner}
\end{figure*}

We apply the Markov chain Monte Carlo ensemble sampler implemented by \citet{2013PASP..125..306F} in the \texttt{emcee} python package\footnote{\url {https://emcee.readthedocs.io}} to sample the posterior probability function. We assume uniform sampling of the \textit{prior} distribution with each input parameter varying within the range presented in Table\,\ref{tab:param}. We initialise 100 chains (\textit{walkers}) with different set of initial parameters and allow for 10,000 steps in each chain to explore the full posterior distribution. We discard the first 5,000 steps in the chains ($\rm N_{burn}$ in the \texttt{emcee} terminology) and record the rest of the chain to study the posterior probability distribution for each parameter.  We checked for convergence estimating the autocorrelation time of the chain, $\tau_{\rm corr}$, for each parameter and taking into account that the length of the chain divided by $\tau_{\rm corr}$ should be higher than 10 . The result of this procedure can be visualised in a \textit{corner} plot that shows the one and two dimensional projections of the posterior probability distribution of the free parameters. We take the best fit value as the median (50th percentile) of the marginalised distribution for each parameter and the 16th and 84th percentiles as an estimation of the corresponding uncertainty.  An example of \textit{corner} plot for one galaxy in our sample is shown in Fig.\,\ref{fig:corner}. The best fit values are given in the top of the distribution for each parameter. 

\begin{table}
\centering
\caption{Range of the free parameters in our SED fitting procedure}
\label{tab:param}
\begin{tabular}{lc}\hline
Parameter & Range \\
\hline 
$\rm\log (M_{\rm PAH}/M_{\odot}) $  & [2, 9]\\
$\rm\log (M_{\rm VSG}/M_{\odot})$  & [2, 9]\\
$\rm\log (M_{\rm BG}/M_{\odot})$ &  [2, 9]\\
$\alpha$ & [1, 5] \\
$\rm\log U_{\rm min}$ &  [-2, 7] \\
$\rm U_{NIR}$ &  [10$^{-3}$, 5] \\
\hline
\end{tabular}
\end{table}

Using the best fit parameters obtained from our MCMC method we generate the best SED model. We show an example of the best fit SED in the right panel of Fig.\,\ref{fig:corner}. Applying the covariance matrix presented in Section\,\ref{sec:uncer} we derive the \chidos\ for our best fit. This is shown in the legend of the panel, giving us an idea of how good the fitting for this galaxy is.  

\section{Results}\label{sec:results}

\begin{figure*}
\includegraphics[width=0.47\textwidth]{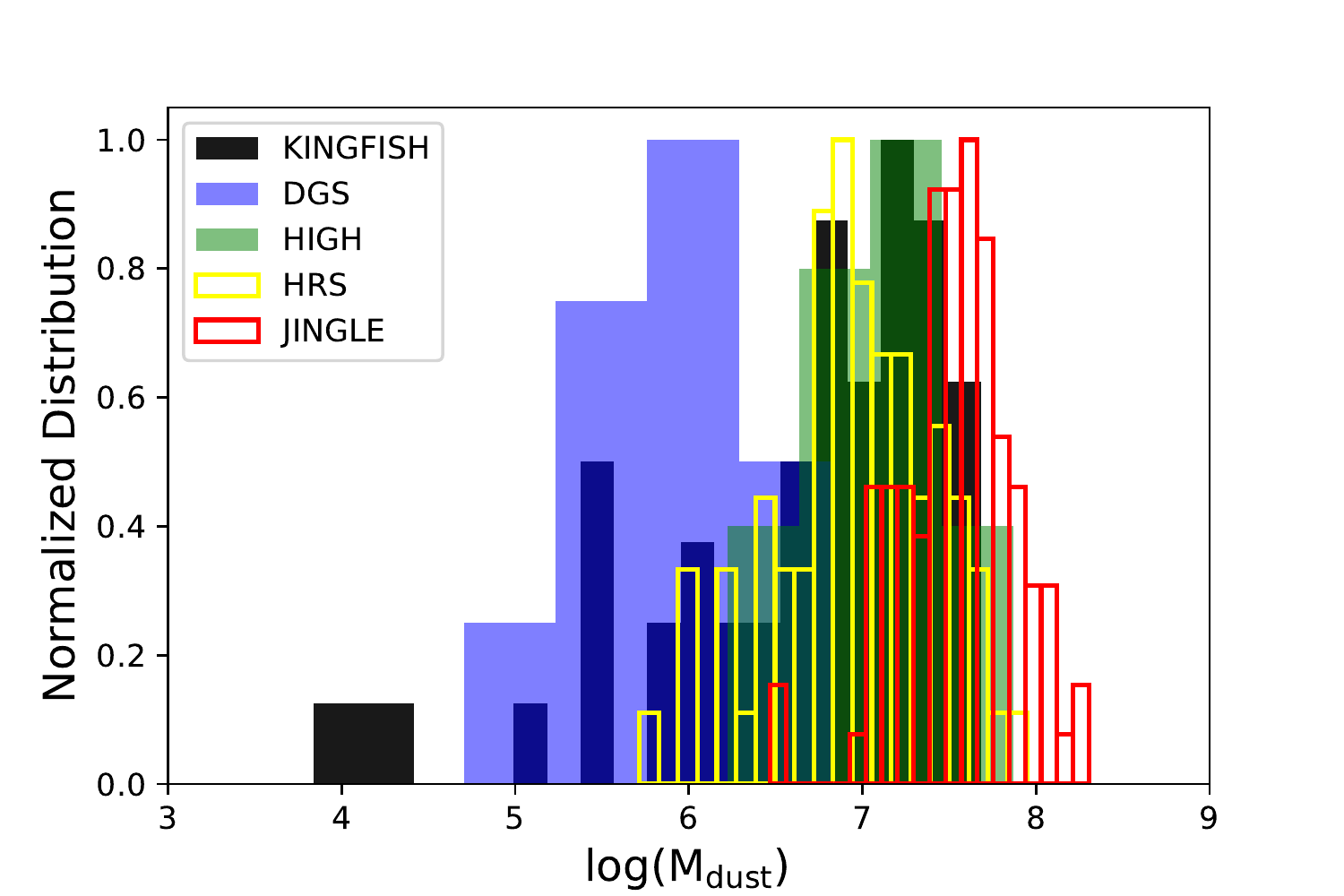}
\includegraphics[width=0.47\textwidth]{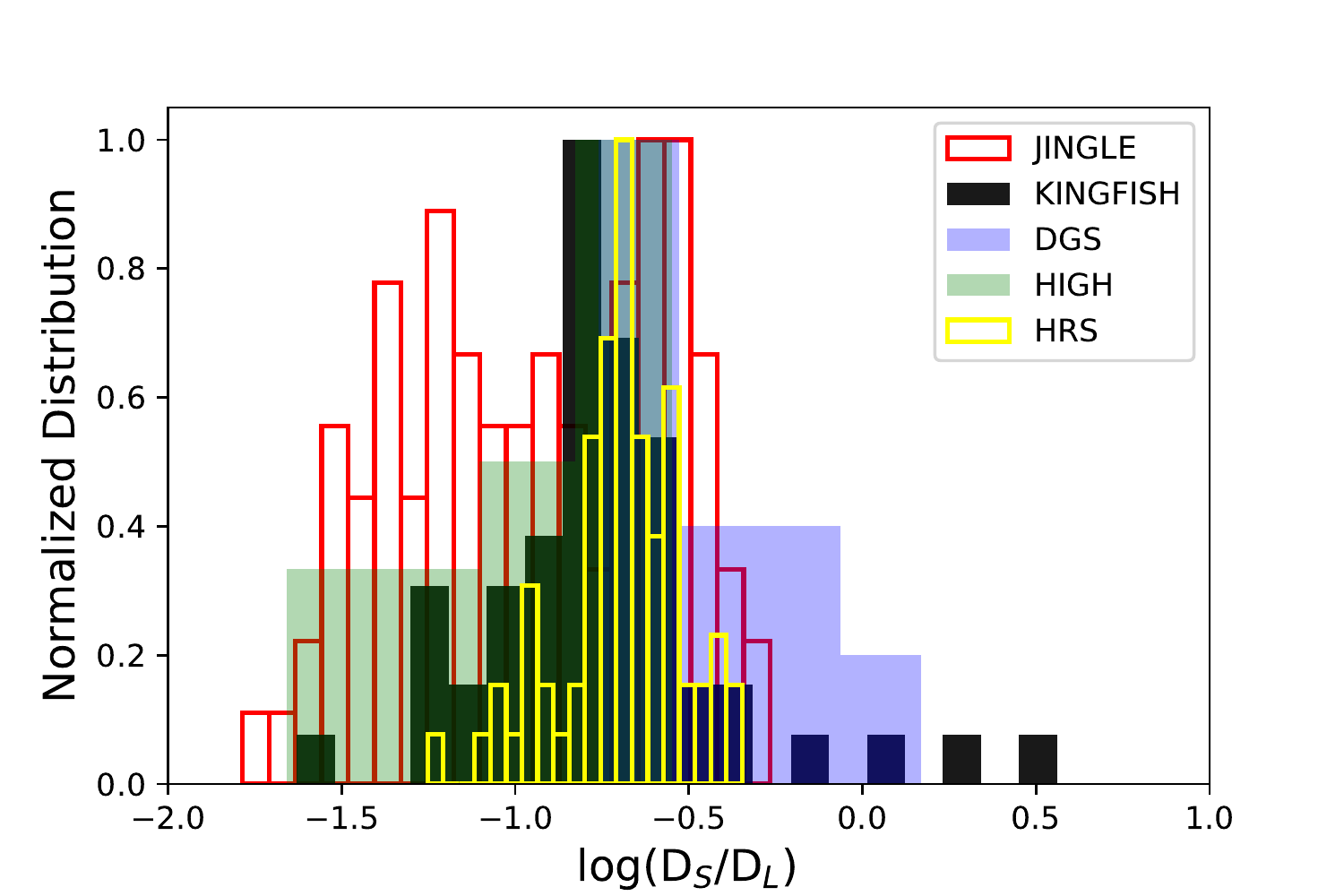}
\caption{Distribution of the dust masses (left) and the \stol\ grain mass ratios (right) derived from our fitting technique. The distributions are normalised to the corresponding maximum. }\label{fig:resultshist}
\end{figure*}

The best fit parameters from our fitting routine provide us with an estimation of the total dust mass of the galaxy, as well as the masses of the different dust components: PAHs, VSGs and BGs. We derive the \stol\ grain mass ratio as the ratio between the total mass of PAHs and VSGs and the mass of BGs. The corresponding error is derived from the estimated uncertainties for the mass of each dust grain component obtained in the SED fitting and using error propagation. In Fig.\,\ref{fig:resultshist} we show the distribution of the dust masses (left panel) and the \stol\ grain mass ratios (right panel) for our final selected galaxy sample. We cover more than 4 orders of magnitude in dust masses and a wide range in \stol\ grain mass ratio from $\log_{10}(D_{S}/D_{L})\sim\,-1.8$ to $\log_{10}(D_{S}/D_{L})\sim\,0.3$. In Table\,\ref{tab:mean} we show the mean values for each galaxy sample. 

\begin{table}
\centering
\caption{Mean values of the logarithm of the \stol\ grain mass ratio ($\log_{10}(D_{S}/D_{L})$) for each galaxy sample included in this study.}
\label{tab:mean}
\begin{tabular}{lc}\hline
Galaxy sample & $\log_{10}(D_{S}/D_{L})$  \\
\hline 
JINGLE & -0.908$\pm$0.004 \\
HRS & -0.676$\pm$0.002 \\
HiGH & -0.805$\pm$0.026\\
KINGFISH& -0.691$\pm$0.007 \\
DGS & -0.33$\pm$0.03  \\
\hline
\end{tabular}
\end{table}

JINGLE galaxies have values in the lower range of the \stol\ grain mass ratio distribution, showing a higher fraction of large grains, as well as higher dust mass content, which as expected shows that most of the dust mass is in the form of large grains. This is not surprising given that JINGLE galaxies were selected among those in $H$-ATLAS with \textit{Herschel} 250 and 350\,\mi\ bands. DGS galaxies are in the opposite side of the dust mass and \stol\ grain mass ratio distributions, presenting high values of \stol\ grain mass ratio and covering the lower regime of dust mass distribution. In general DGS galaxies have lower gas and stellar mass content (see Fig.\,\ref{fig:scarel}) and have in general lower metallicity than JINGLE galaxies, which would translate into a lower dust mass content than the JINGLE sample. Relano et al. 2020,  where a similar SED fitting was done but with a single ISRF,  also found that DGS galaxies exhibit a higher DS/DL than the rest of the galaxy sample used in their study. The dust masses and the \stol\ grain mass ratios for our galaxy sample are presented in Table\,\ref{tab:results}.

\begin{figure*}
\includegraphics[width=0.49\textwidth]{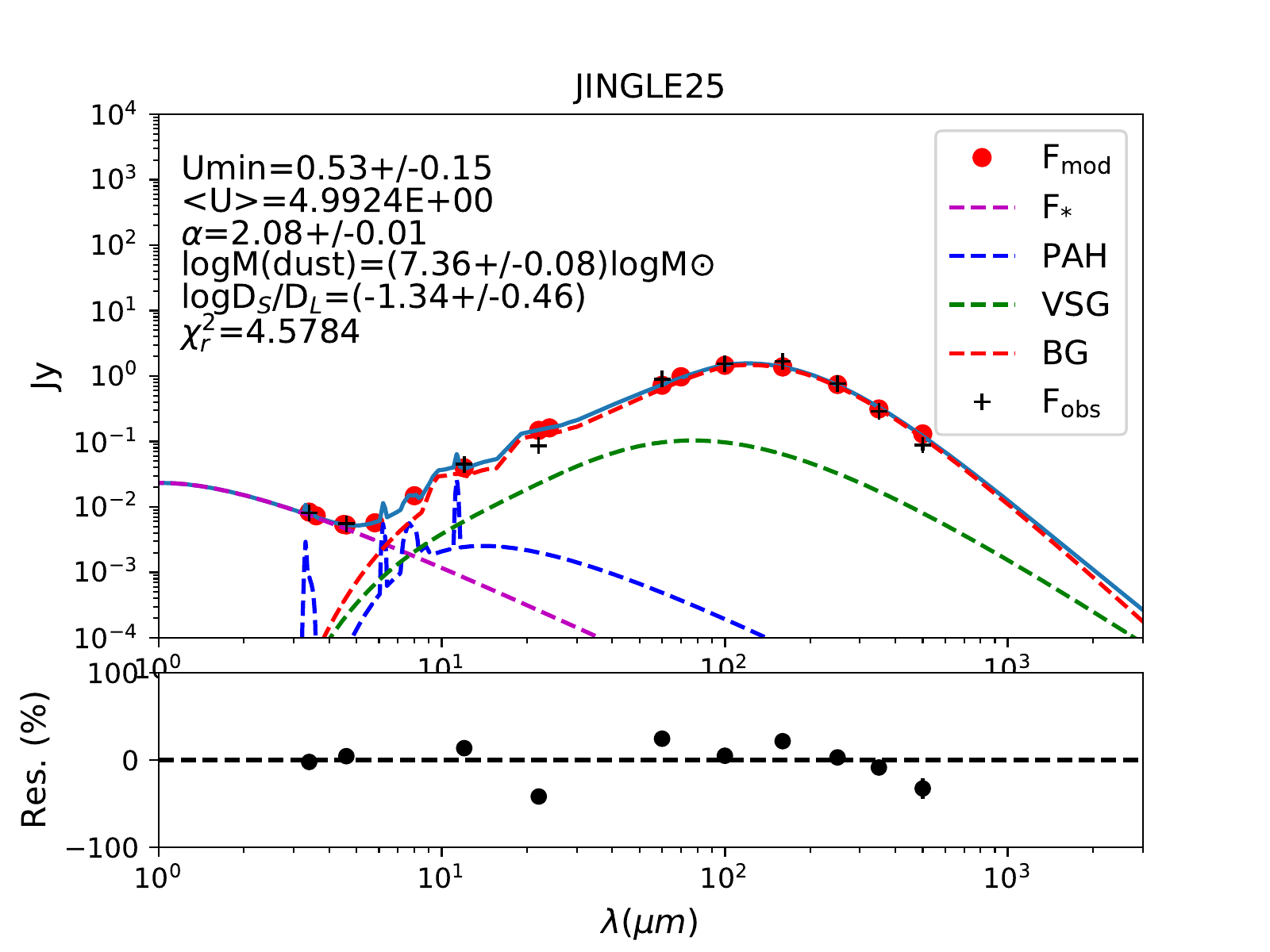}
\includegraphics[width=0.49\textwidth]{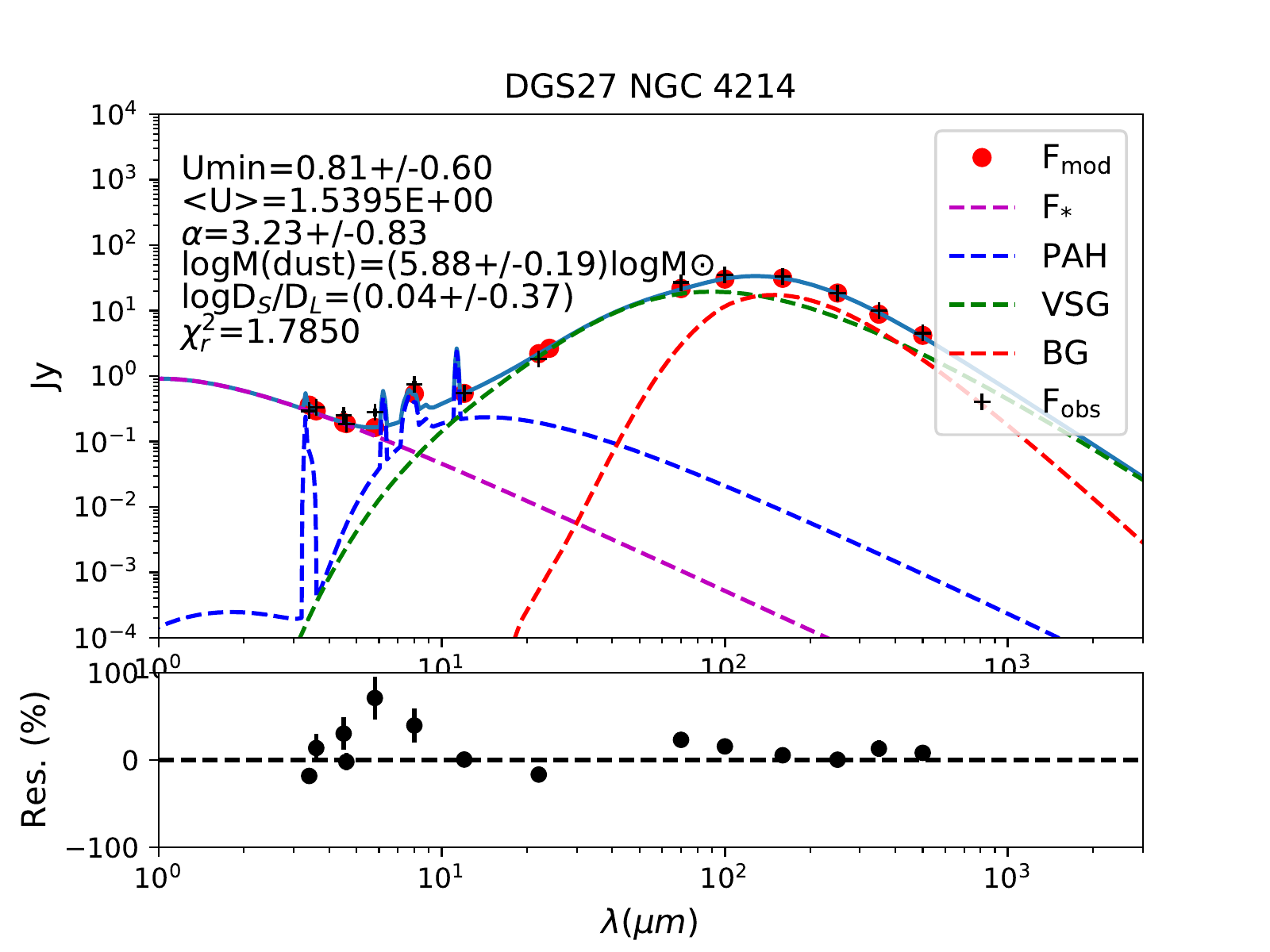}
\caption{Illustration of two galaxies in the extreme range of the \stol\ grain mass ratio distribution. Left: SED fitting for galaxy JINGLE 25 presenting having $\log_{10}(D_{S}/D_{L})=-1.34$. Right: SED fitting for NGC\,4214 from the DGS with $\log_{10}(D_{S}/D_{L})=0.04$. Small grains refer to the combination of PAHs and VSGs (blue and green dashed lines, respectively)}\label{fig:SED_S2L}
\end{figure*}

In Fig.\,\ref{fig:SED_S2L} we show an illustration of the SED fitting of two galaxies representative of extreme values of \stol\ grain mass ratio. JINGLE 25 (left panel) presenting a value of $\log_{10}(D_{S}/D_{L})=-1.34$ and NGC\,4214 (right) from the DGS sample with $\log_{10}(D_{S}/D_{L})=0.04$. In the first case, the dust mass in the form of PAHs and VSGs represents less than 5\% of the total dust mass in the galaxy, while NGC\,4214 has $\sim$\,50\% of its dust mass content in the form of small (PAHs and VSGs) grains. 

The wide \stol\ grain mass ratio range covered by our galaxy sample shows the variety of observed SEDs we fit in this study and allows us to perform a detailed analysis on how the relative dust grain size distribution depends on other physical properties and the evolutionary status of the galaxy. In Section\,\ref{sec:compsim} we will explore how the dust mass and the \stol\ grain mass ratio compare with results from simulations where the evolution of the dust grain size distribution has been taken into account. In the next section, as a robust check of our results, we will study the consistency of the total dust mass and \stol\ grain mass ratios derived here with previous results in the literature.

\subsection{Comparison with previous studies} 
Dust masses have been derived previously for our galaxy samples using different dust models and fitting techniques. In this section we compare our dust mass estimations with those from the literature to show the robustness of our fitting technique.  Dust mass for the DGS and KINGFISH galaxies were obtained by \citet{2015A&A...582A.121R}, while for JINGLE, HiGH, KINGFISH and HRS the dust masses were derived in \citet{2020MNRAS.496.3668D}. In both studies the implementation of a multi-component ISRF heating the dust grains as the one presented here was adopted. The dust model, however, was different in each study: \citet{2015A&A...582A.121R} used two different models, one with graphite grains and the other with amorphous carbon grains; and  \citet{2020MNRAS.496.3668D} used the THEMIS dust model presented in \citet{2017A&A...602A..46J}, which accounts for aromatisation of carbonaceous grains and mantle thickness within a dust evolution context. In Figs.\,\ref{app:compMdRR15} and\,\ref{app:compMdDLooze20} we show the comparison of the dust masses derived in this paper with the results from these studies. We find very good agreement with the amorphous carbon grains dust model used in  \citet{2015A&A...582A.121R} for KINGFISH and DGS samples and we underestimate the dust masses derived from the graphite dust model (see Appendix\,\ref{app:comp}). For the case of JINGLE, HiGH, KINGFISH and HRS samples studied in \citet{2020MNRAS.496.3668D}, we find very good agreement covering four order of magnitudes in dust masses and a wide range of galaxy properties. 

Our fitting procedure relies on the parametrisation introduced by \citet{2001ApJ...549..215D} where the dust mass elements heated by the ISRF are distributed in a power-law form. We have compared the dust masses derived here with those obtained in the literature following the same \citet{2001ApJ...549..215D} prescription, but we also compare our results when a single ISRF is assumed. A comparison between single and multi-ISRF approach has been done in resolved scales for the Magellanic Clouds \citep{2017A&A...601A..55C}. These authors found that the multi-ISRF approach typically improves the fits in the 8-24\,\mi\ wavelength range, independently of the dust model used. In Appendix\,\ref{app:comp} we compare the dust masses derived here with those given in  \citet{2020A&A...636A..18R}, where a single ISRF was used to fit the SED of the KINGFISH and DGS sample. The dust masses agree relatively well for both samples within a factor of two (see Fig.\,\ref{app:compISRFRR15}). 

We have also compared the  \stol\ grain mass ratio when a single and multi-ISRF is assumed. \citet{2018ARA&A..56..673G} 
pointed to a degeneracy between the ISRF distribution and the mass fraction of small grains, in the sense that a certain SED can be equally well fit by a single ISRF with a high mass fraction of small grains as by a multi-ISRF representing hotter environment and a lower mass fraction of small grains (see figures 3b and 3c in \citealt{2018ARA&A..56..673G}). We find an agreement between the \DstoDl\ derived from the two approaches for a high fraction of galaxies (see Fig.\,\ref{app:compISRFRR15}). These galaxies would not be affected by a degeneracy, as we find similar  \DstoDl\ using a  multi-ISRF approach that describes a harder ISRF, and using a single ISRF that only allows to scale the intensity of the \citet{Mathis:1983p593} ISFR. However, some galaxies exhibit large \DstoDl\ when a single ISRF is adopted. For them we are able to fit their SEDs either with a single ISRF and high \DstoDl\ or with a low  \DstoDl\ but a multi-ISRF approach. The fact that those galaxies have in general higher dust temperatures than those in the one-to-one line shows that indeed the ISRF should be harder in these systems. These outliers are  examples of the degeneracy presented in \citet{2018ARA&A..56..673G}. The comparison of the dust masses and  \DstoDl\  reinforces the robustness of our methodology and the results presented further in this paper. 

\section{Comparison with cosmological simulations}\label{sec:compsim}
In this work we aim to compare the small-to-large grain mass ratio and other dust scaling relations with the results from simulations. \citet{2019MNRAS.485.1727H} and \citet{2018MNRAS.478.4905A} performed SPH cosmological simulations with the GADGET3-Osaka N-body/SPH code presented in \citet{2017MNRAS.466..105A} and \citet{2017MNRAS.469..870H}. The initial number of particles were $N$= 2$\times$512$^3$ in a comoving simulation box of 50\,$h^{-1}$\,Mpc. Star formation occurs following the prescription given in Eq.\,2 of \citet{2017MNRAS.466..105A}, which parameterises the SFR in terms of the gas density and the free-fall time. Each newly created star particle is seen as a single stellar population with Chabrier's initial mass function \citep{2003PASP..115..763C} and carries information on the stellar mass, metallicity, and formation time. 

In the simulations dust is produced by SNe and  AGB stars and the metal enrichment is assumed to occur $\sim$\,4\,Myr after the star formation event. The simulations include stellar dust production, SN destruction, grain disruption by shattering in the ISM, astration, coagulation and grain growth by accretion in the dense ISM. Dust grains are additionally destroyed by sputtering in hot gas regions (T$>$10$^{6}$\,K)  in the circumgalactic medium. Each mechanism affects the dust grain size distribution differently:  stellar dust production supply dust grains in the form of large grains, dust destruction affects both large and small grains, while coagulation increases the fraction of large grains. Fragmentation associated with shattering increases the fraction of small grains. Accretion is favoured in the presence of small grains and predominantly increases the total mass of small grains. 
The simulations adopt the two-grain size distribution approximation presented in  
\citet{2015MNRAS.447.2937H}, separating the grain size distribution into small ($a\lesssim$\,0.03\,\mi) and large ($a\gtrsim$\,0.03\,\mi) grains. The separation is justified on the basis of a full calculation of the grain size distribution performed by \citet{2013MNRAS.432..637A}, who showed that the processes dominating the small grain abundance create a bump in the size distribution at small grain sizes, while dust production by stars creates a bump in the large grain size regime. The separation between these two bumps is at $a\simeq0.03\,\mi$. 

Besides, \citet{2019MNRAS.485.1727H} include a simple treatment of AGN feedback in their simulations using a model based on \citet{2014PASJ...66...70O}, where they turn off cooling above certain halo masses, trying to mimic the heating effect by AGN feedback. The result of including AGN feedback is to reduce the metal enrichment and star formation in massive dark matter halos. 

In the following sections we compare the results of our observational analysis for the combined galaxy sample with the predictions of the SPH cosmological simulations of  \citet{2019MNRAS.485.1727H}, which include AGN feedback treatment. The comparison allows us to analyse how the  \stol\ grain mass ratio, which gives information on the main mechanisms affecting the dust evolution, varies as a function of the galaxy properties. In an attempt to find plausible explanations for the deviations we see between observations and simulations, we also compare our observations with the results of isolated galaxy simulations performed by the GADGET4-Osaka code \citep{2022MNRAS.tmp.1342R,2022MNRAS.tmp.1338R}. This code includes a treatment of dust physics similar to that of GADGET3-Osaka \citep{2020MNRAS.491.3844A}, but with a more realistic approach to describe the dense environment where accretion and coagulation take place and incorporates metal diffusion that might enhance the amount of small grains in the ISM (see Appendix\,\ref{app:Rom} for a detailed description of these simulations).

\subsection{Scaling relations}

\subsubsection{Stellar mass, star formation rate and gas mass}\label{sec:scarel}

\begin{figure*}
\includegraphics[width=0.45\textwidth]{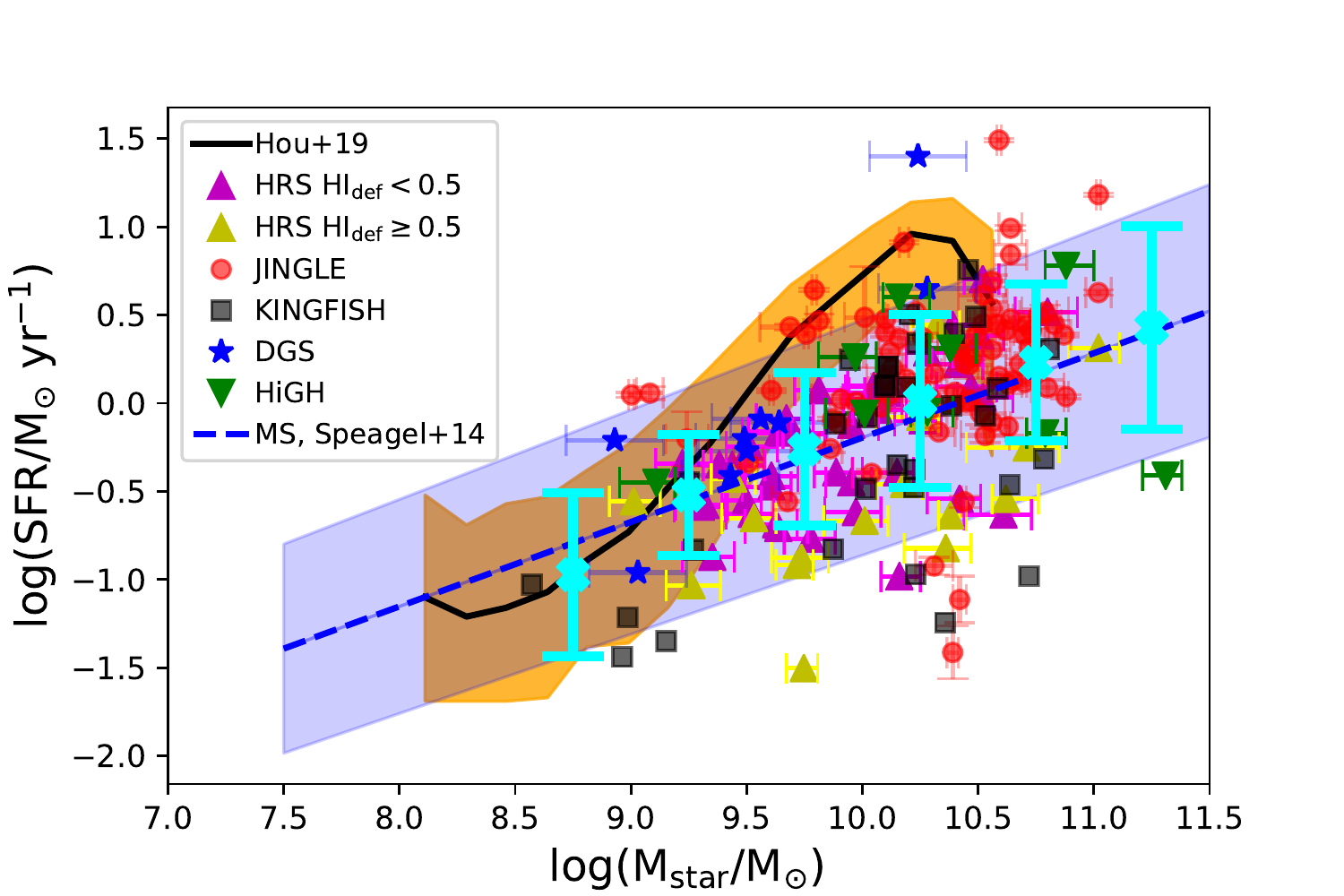}
\includegraphics[width=0.45\textwidth]{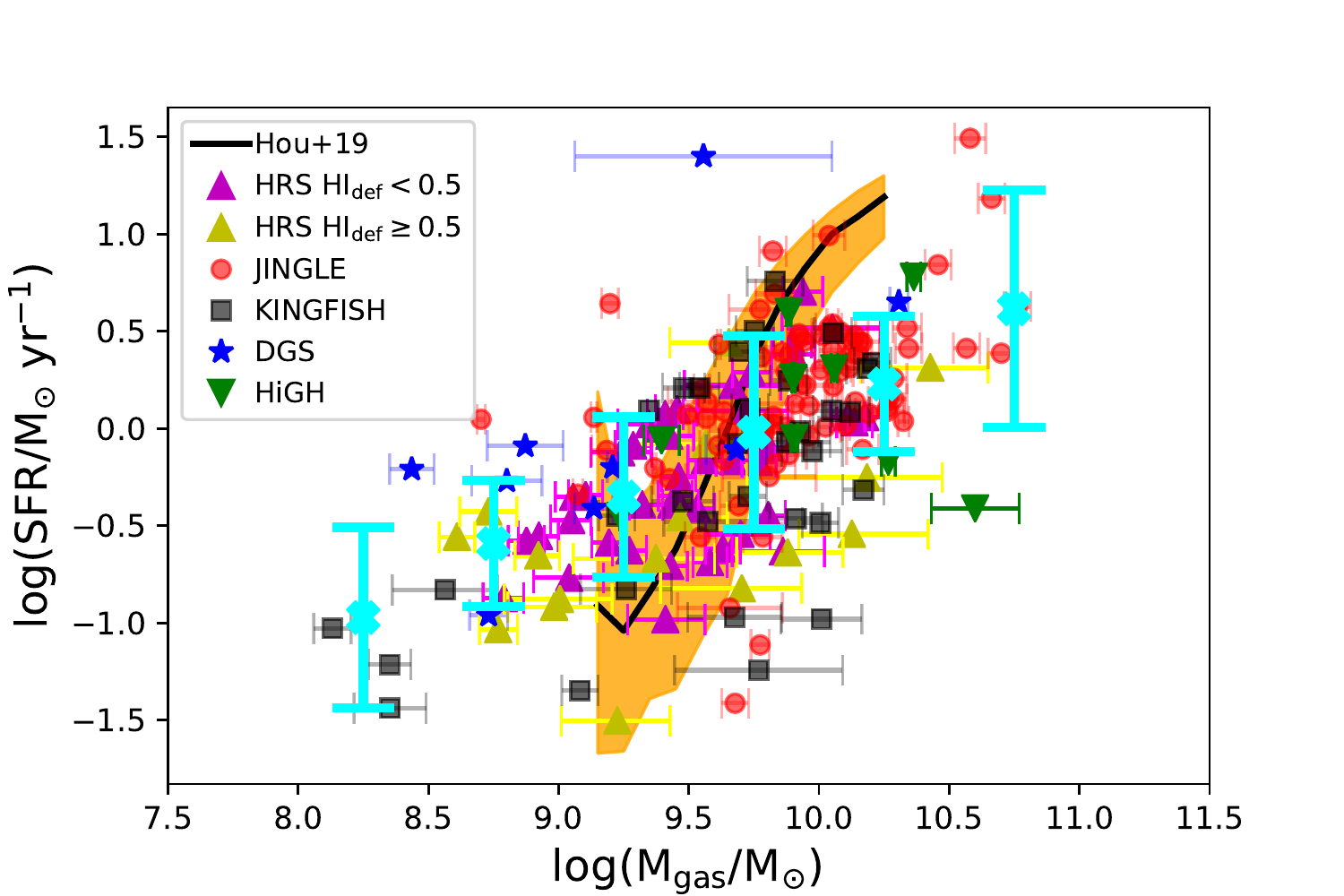}
\caption{SFR versus stellar mass (left) and SFR versus total gas mass (right) predicted by the simulations (continuous black line representing the median of the PDF and orange area enclosing the 16th and 84th percentiles) and obtained from observations (JINGLE  (red points), KINGFISH (black points), DGS (blue stars), HiGH (green stars), HRS with H{\sc{i}}-deficient (H{\sc{i}}$_{\rm def}\geq$\, 0.5) in yellow and non-deficient (H{\sc{i}}$_{\rm def}<$\,0.5) in magenta. The total gas mass of our galaxy sample is the sum of the atomic and molecular gas mass, where the molecular gas mass has been obtained using the relation between $\rm M_{\rm HI}/M_{\rm star}$ and $\rm M_{\rm H_2}/M_{\rm HI}$ obtained by \citet{2020A&A...633A.100C} for galaxies in the Local Universe and a factor of 1.36 has been applied to account for the helium contribution. The dashed blue line and area represent the star formation Main Sequence relation for $z=0$  and its corresponding dispersion derived in \citet{2014ApJS..214...15S} using a compilation of observations from the literature. The cyan crosses and corresponding error bars represent mean values and standard deviations of magnitudes represented in the $y$-axis for bins of 0.5\,dex in the $x$-axis.}\label{fig:scarel}
\end{figure*}

In order to better understand the comparison involving the total dust mass and the relative contribution of the different grain components, we first analyse general scaling relations involving the total stellar mass, star formation rate, as well as the total gas content of the galaxy. In Fig.\,\ref{fig:scarel} we show the relation between the star formation rate and stellar mass (left panel) and gas mass (right panel) for both the simulations and the observations.

The total gas mass includes both atomic and molecular gas masses. Atomic gas mass are taken from the literature which are derived using H{\sc{i}} observations (see Section\,\ref{sec:samp} for each galaxy sample). We do not have CO observations for all the galaxies in our sample, and therefore for homogeneity, we make use of  the correlation between $\rm M_{\rm HI}/M_{\rm star}$ and $\rm M_{\rm H_2}/M_{\rm HI}$ obtained by \citet{2020A&A...633A.100C} for galaxies in the Local Universe  to estimate the molecular gas masses. The reason to prefer an estimation of the molecular gas mass using the relation from \citet{2020A&A...633A.100C} is to include as many galaxies as possible when studying the relations of dust masses and \stol\ grain mass ratio with stellar mass, star formation rate, gas and dust mass. To be conservative, we apply the relation to galaxies in our sample that have $\rm M_{\rm HI}/M_{\rm star}$ within the range where the relation was observationally derived. This ensures we are not inferring molecular gas masses outside its validity range. We explain in Appendix\,\ref{app:compCasaXCOLD} how $\rm M_{\rm H_2}$ have been obtained from \citet{2020A&A...633A.100C} relation and we compare $\log(\rm M_{\rm HI}+M_{\rm H_2})$ obtained here with the values inferred using the scaling relation presented in \citet{2022arXiv220200690S} based on xCOLD GASS data. We furthermore compare $\log(\rm M_{\rm HI}+M_{\rm H_2})$ for those galaxies for which estimates of $\rm M_{\rm H_2}$ can be done from CO observations. Except for the H{\sc i}-deficient(H{\sc i}$_{\rm def}$\footnote{H{\sc i} deficiency parameter, H{\sc i}$_{\rm def}$, is defined as the difference in logarithmic scale between the expected and the observed H{\sc i} mass of a galaxy \citep{1984AJ.....89..758H}.}$\geq$\, 0.5) HRS galaxies, the mean differences between the prescription given in \citet{2020A&A...633A.100C} and the one provided in \citet{2022arXiv220200690S} are within 0.2\,dex.

The observed data broadly agree with the predicted SFR-$\rm M_{\rm star}$ relation by the simulations (continuous black line and orange area in left panel in Fig.\,\ref{fig:scarel}). The observations follow the star formation main sequence (SFMS) relation for star-forming galaxies derived by \citet{2014ApJS..214...15S} at $z=0$ (blue dashed line with blue area). The H{\sc i}-deficient HRS galaxies fall below the SFMS. These galaxies exhibit a lack of atomic  hydrogen gas, probably removed due to interactions with other galaxies, which would produce a decrease of SFR. Therefore, the location of these galaxies in the SFR-$\rm M_{\rm star}$ diagram is expected to be below the SFMS relation of star-forming galaxies. There is a trend for the cosmological simulations in \citet{2019MNRAS.485.1727H} to deviate from the observational SFMS relation obtained in  \citet{2014ApJS..214...15S} and from the relation found for our galaxy sample for stellar masses within $\log(\rm M_{star})\sim9.7-10.5$. Indeed the trend shown for our observations agrees better with the fitted relation by \citet{2014ApJS..214...15S} at $z=0$ than with the result from simulations. The discrepancy, which is relatively small if we take into account the dispersion of the observations and the simulations, cannot be produced by differences of the initial mass function: the cosmological simulations from \citet{2019MNRAS.485.1727H} as well as \texttt{MagPhys} \citep{2008MNRAS.388.1595D}, which is the code to derive the SFR, both used the \citet{2003PASP..115..763C} IMF. 

\citet{2019MNRAS.485.1727H} already added AGN feedback in their analysis and lowered the star formation efficiency with respect to previous simulations of the same group \citep{2018MNRAS.478.4905A}. A more sophisticated AGN feedback treatment or SNe feedback in high-resolution simulations \citep{2022arXiv220100970O}, could eventually reduce further the SFR in the high stellar mass regime and would produce a better agreement than what is observed here. Another possible cause of the discrepancy between simulations and observations could be related with the star formation prescription assumed by the simulations in \citet{2019MNRAS.485.1727H}. 
In these simulations only gas particles with density and temperature above a certain threshold (n=0.1\,cm$^{-3}$ and T$\leq$10$^{4}$\,K) are able to produce stars at a rate parameterised by a constant star formation efficiency, $\epsilon_{*}$=0.01. 
The discrepancy between simulations and observations is not constant for the whole stellar mass range but it happens at high stellar masses and metallicities. Simulations at higher resolutions including a more sophisticated parameterisation of the star formation rate might be able to alleviate the discrepancies seen in SFMS relation, as well as those seen in the SFR-M$_{\rm gas}$  relation (right panel of Fig.\,\ref{fig:scarel}). We note here that the discrepancy between observations and simulations in the SFMS relation also occurs in the stellar mass$--$metallicity relation, as we show in the next section. 

\subsubsection{Metallicity derived from observations}\label{sec:MsZ}

Metallicity for each galaxy in our sample was inferred using the oxygen abundance\footnote{Metallicity, normalised to solar metallicity $Z/Z_{\odot}$, was obtained using the oxygen abundance and the relation: $Z/Z_{\odot} = 10^{[\rm O/H]}$, where $[\rm O/H]=\rm \log(O/H)-\log(O/H)_{\odot}$. As in the simulations we assume $Z_{\odot}=0.02$, which corresponds to 12+log(O/H) = 8.93.} derived from observations. Derivation of oxygen abundances from spectroscopic emission lines is commonly done using either the direct method where information from electron temperature is available or via empirical calibration of strong emission lines. While the first method would give a more accurate value of the oxygen abundance, observations of temperature sensitive emission lines is not always possible. We rely on oxygen abundances derived from empirical calibrations in this study. Due to the diversity of the galaxy sample we were not able to use a single calibration to derive oxygen abundances. However, we try to be as consistent as possible and adopt for our galaxies the following calibrations among those available in the literature. For JINGLE, HiGH, HRS and KINGFISH we use oxygen abundances derived from the O3N2 calibration of \citet{2004MNRAS.348L..59P}, reported previously in the literature \citep[][respectively]{2018MNRAS.481.3497S,2017MNRAS.471.1743D,2013A&A...550A.115H,2019A&A...623A...5D}. We favoured oxygen abundance derived from O3N2 calibration of \citet{2004MNRAS.348L..59P}, as this last one empirical calibration is less accurate than O3N2 in the metallicity range where both calibrations can be applied \citep[see][for a recent comparison of different metallicities calibrations available in the literature]{2021MNRAS.500.2359Z}. Oxygen abundances were reported in \citet{2015A&A...582A.121R} for DGS  using the PT05 calibration \citep{2005ApJ...631..231P}. \citet{2013PASP..125..600M} compares the metallicities for DGS derived from PT05 with those derived from the direct method and found small differences of $\sim$0.1\,dex. We did not attempt to perform a conversion between PT05 and O3N2 calibrations for the DGS galaxies as the validity of the metallicity range for the O3N2 calibration of \citet{2004MNRAS.348L..59P} falls outside of most of the low metallicities galaxies in DGS. 

\begin{figure}
\includegraphics[width=0.5\textwidth]{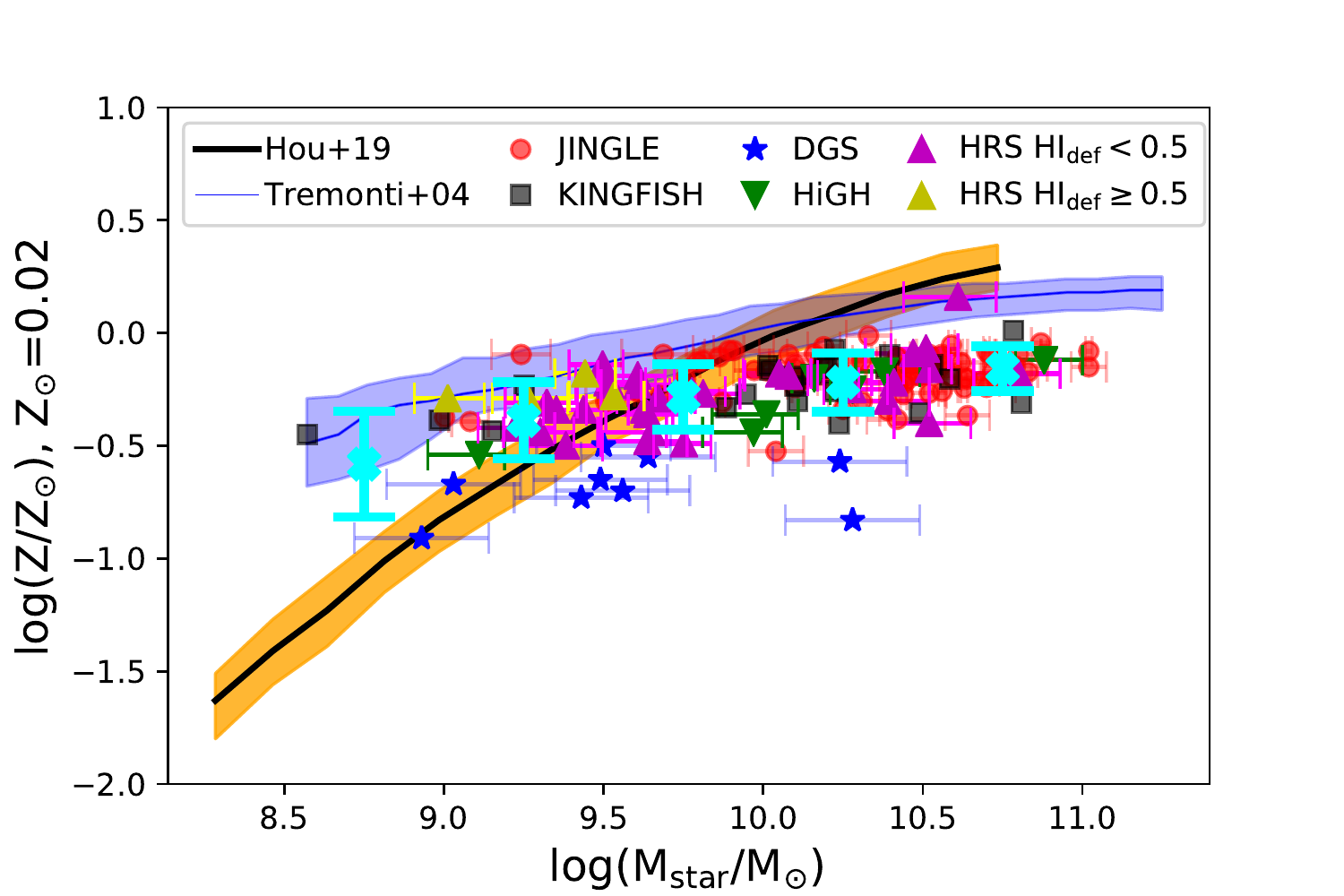}
\caption{$M_{\rm star}-Z$ relation for our galaxy sample: JINGLE (red points), KINGFISH (black points), DGS (blue stars), HiGH (green stars) and HRS (magenta stars). Blue continuous line and dashed black line correspond to the $M_{\rm star}-Z$ relation presented in \citet{2004ApJ...613..898T} and in the simulations from \citet{2019MNRAS.485.1727H}, respectively. The blue and orange shaded areas enclose the 16th and 84th percentiles of the PDF and give an estimate of the uncertainties in the $M_{\rm star}-Z$ relation derived in each work.}\label{fig:Mstar-Z}
\end{figure}

With these calibrations our galaxy sample follows a continuous $M_{\rm star}-Z$ relation\footnote{Metallicities derived using N2 calibrator give anomalous high values of Z in the high stellar mass end of the $M_{\rm star}-Z$ relation.}, albeit shifted to $\sim$\,0.3\,dex lower metallicities than the relation found by \citet{2004ApJ...613..898T} (see Fig.\,\ref{fig:Mstar-Z}). The difference in oxygen abundance is related to the metallicity calibration used in this paper in comparison with the methodology used in \citet{2004ApJ...613..898T} to estimate oxygen abundances based on photoionisation models. As it was shown in \citet{2008ApJ...681.1183K} the  O3N2 calibration of \citet{2004MNRAS.348L..59P} predicts lower oxygen abundances than those estimated in \citet{2004ApJ...613..898T}. The differences are minimal in the lowest stellar mass regime and increases towards higher stellar masses, in agreement with the trend seen in this paper. When comparing to the simulations from  \citet{2019MNRAS.485.1727H} we find that the trend predicted by the simulations deviates from the behaviour shown in the observations, especially in the high stellar mass regime, where the simulations give higher metallicity values than those derived from observations. This is consistent with the deviation of the simulations in the SFMS relation (left panel in Fig.\,\ref{fig:scarel}) shown in previous section. The higher SFR obtained by the simulations at stellar masses of $\log(\rm M_{star})\gtrsim\,10$ would imply more stars forming per unit time and therefore an increase in the metal enrichment of the ISM, as it is seen in Fig.\,\ref{fig:Mstar-Z}. 

\subsubsection{Dust-to-gas ratio}\label{sec:D2G}

\begin{figure*}
\includegraphics[width=0.45\textwidth]{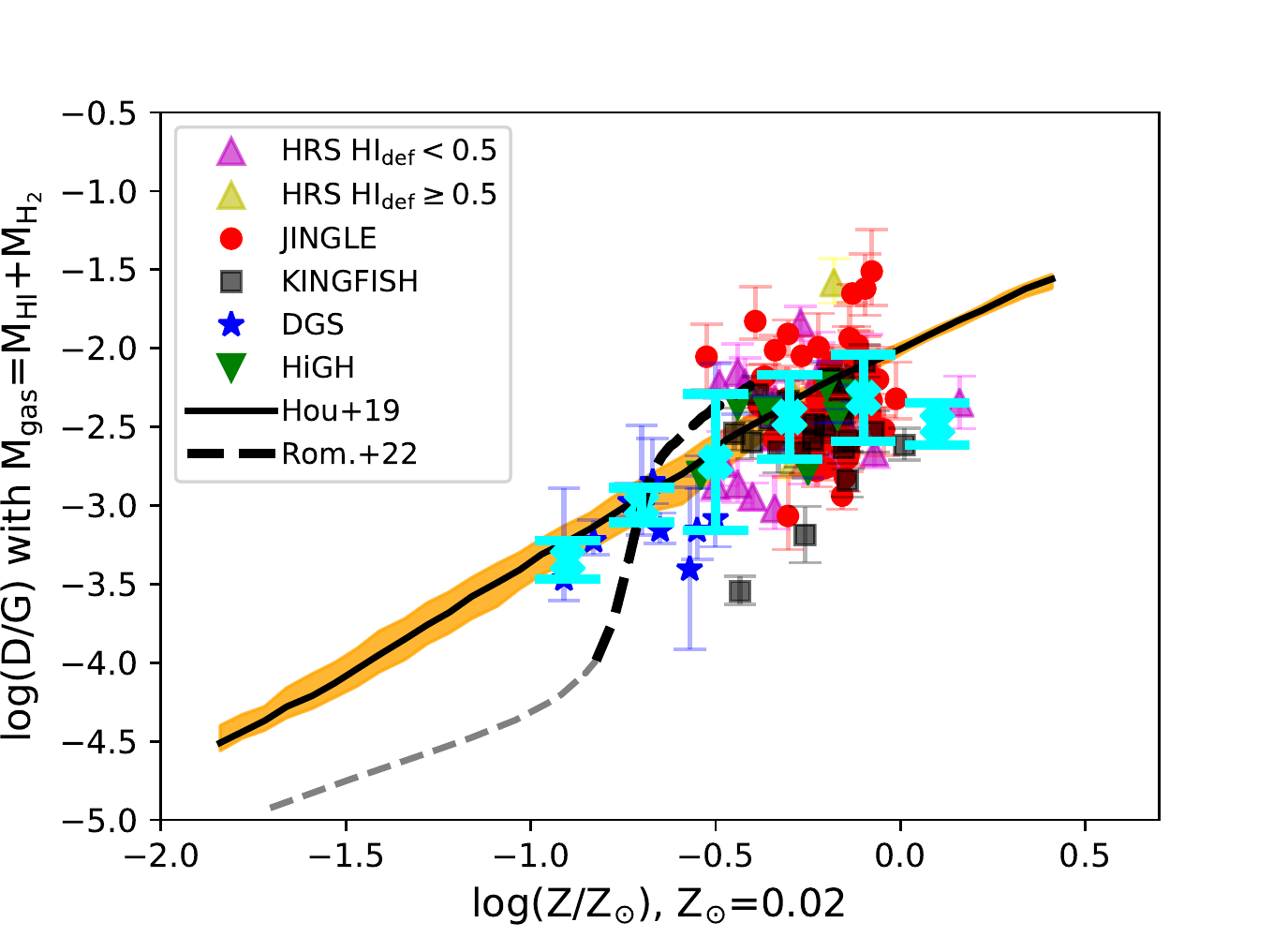}
\includegraphics[width=0.45\textwidth]{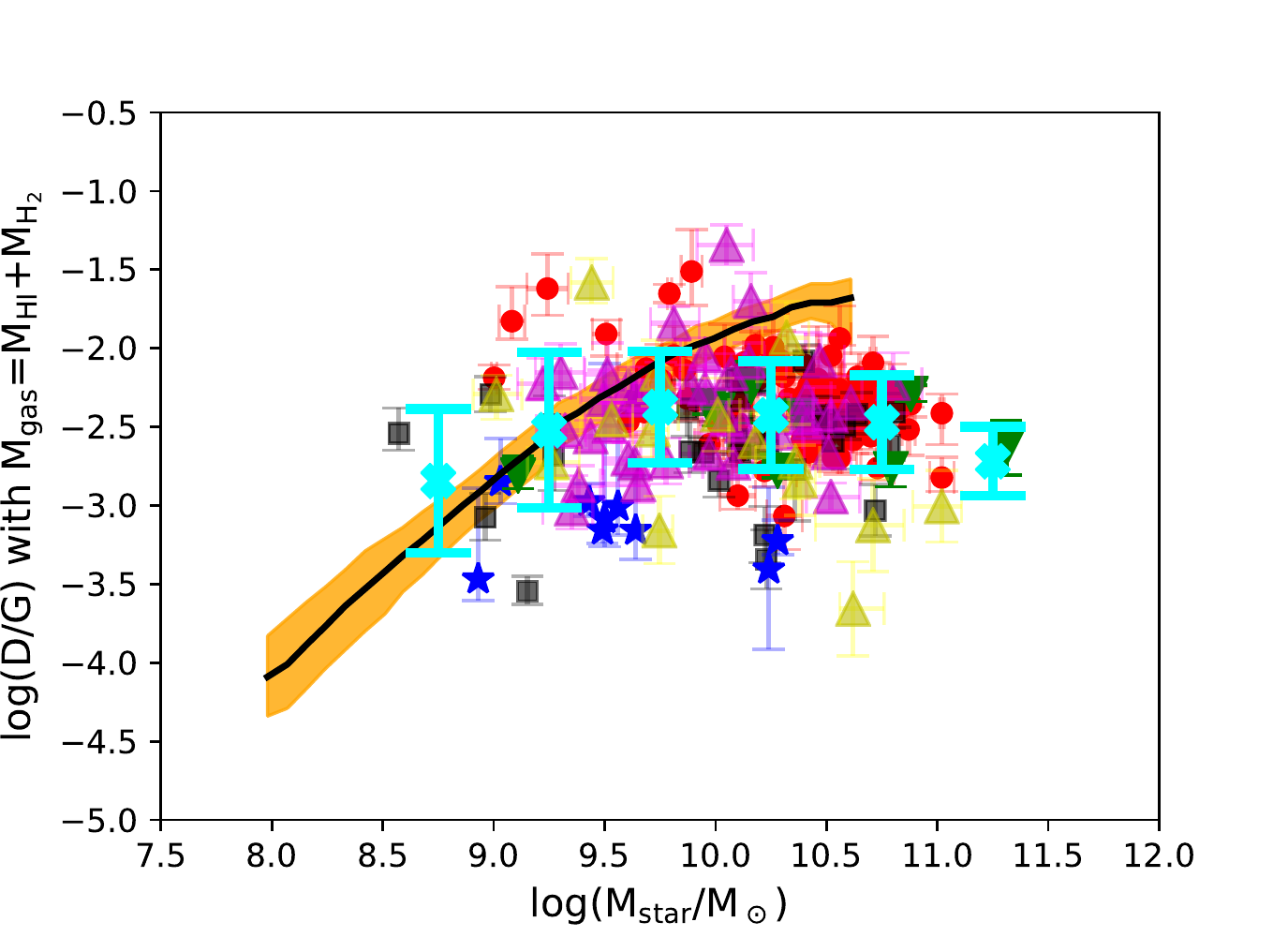}
\includegraphics[width=0.45\textwidth]{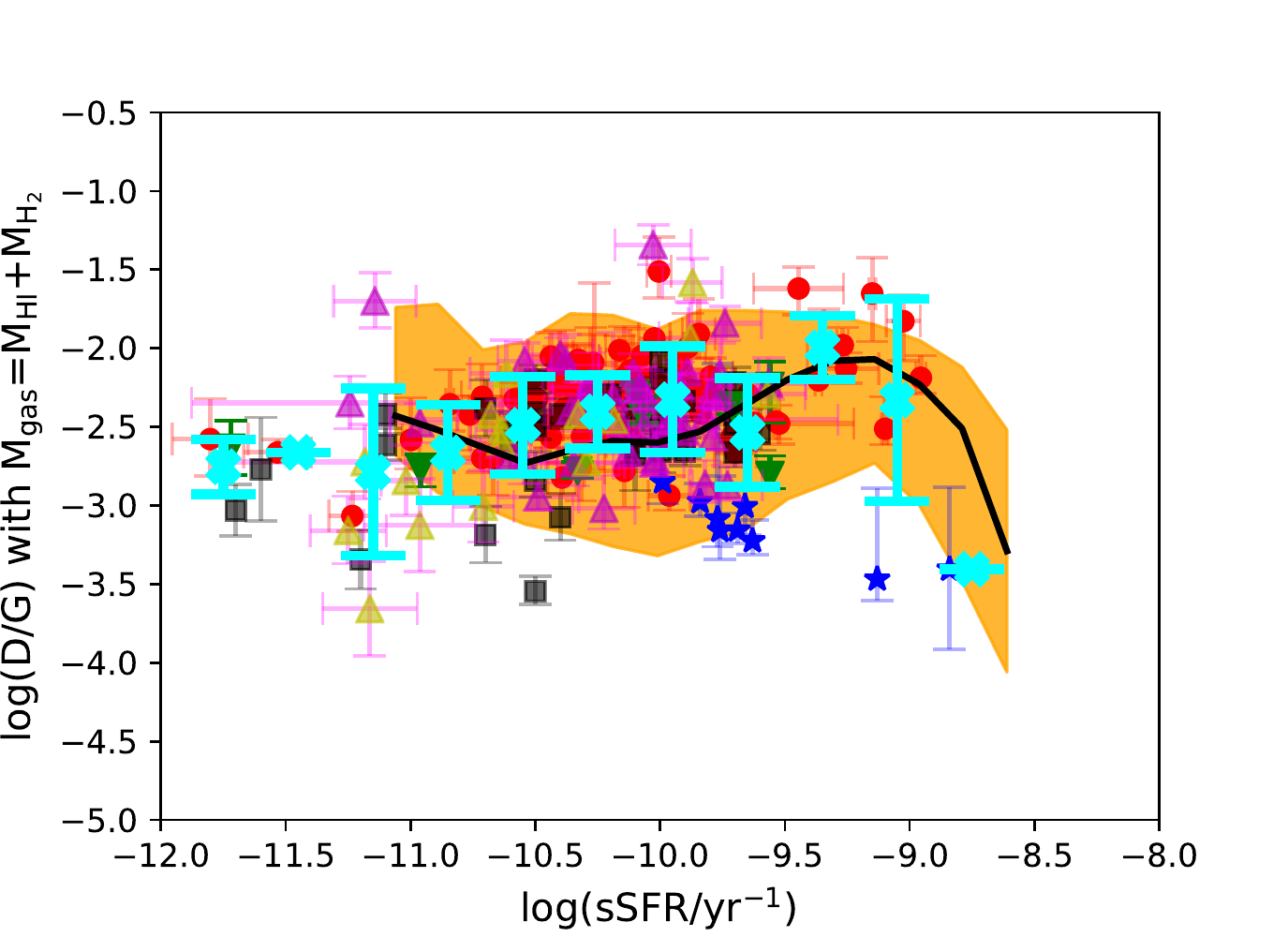}
\includegraphics[width=0.45\textwidth]{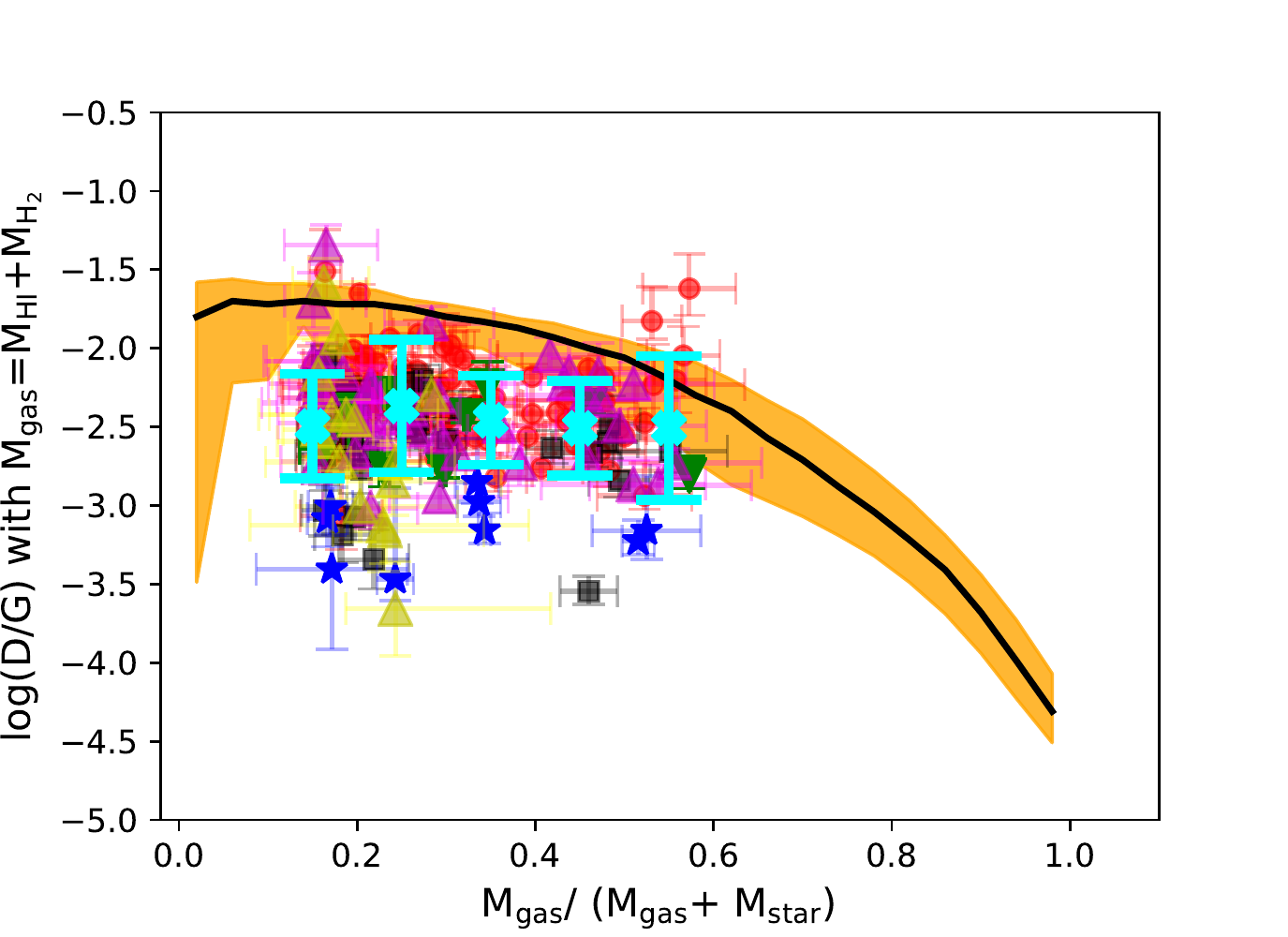}
\caption{Comparison of the dust-to-gas mass ratio (D/G) for our galaxy sample with hydrodynamic simulations. The black continuous line shows the 50th percentile of the simulated galaxy distribution at $z$=0 in the GADGET3-Osaka cosmological simulation \citep{2019MNRAS.485.1727H} and the yellow colour represents the area within the 16th and 84th percentiles. The black dashed line shows the evolutionary track of an isolated galaxy simulation by GADGET4-Osaka code with the thinner line showing the earlier phases at t\,$\lesssim$\,0.5\,Gyr (see Appendix\,\ref{app:Rom}).
Top-left: D/G versus metallicity, top-right: D/G versus stellar mass, bottom-left: D/G versus sSFR, and bottom-right: D/G versus   gas mass fraction, $\rm M_{gas}/(M_{gas}+M_{star})$ with $\rm M_{gas}=1.36\,(M_{HI}+M_{H_{2}})$. Since we do not have CO observations for all the galaxies in our sample we derive molecular gas masses using the relation found by \citet{2020A&A...633A.100C} for galaxies in the Local Universe, as explained in Appendix\,\ref{app:compCasaXCOLD}. We apply this relation to galaxies satisfying the validity range where the relation was derived ($-2<\log(\rm M_{\rm HI}/M_{\rm star})<0$), therefore in the bottom-right panel we do not have galaxies with gas mass fraction above $\sim$\,0.6. The cyan crosses and corresponding error bars represent mean values and standard deviations of magnitudes represented in the $y$-axis for bins in the $x$-axis. The predicted D/G is above the observations in the high stellar mass regime indicating an overprediction of the dust mass in simulations (see Section\,\ref{sec:D2G}).}\label{fig:D2G}
\end{figure*}

In Figs.\,\ref{fig:D2G} we show the dust-to-gas (D/G) ratio for our galaxy sample as a function of other physical properties. As we mentioned in Section\,\ref{sec:scarel}, gas masses include atomic (obtained in the literature using H{\sc i} observations) and molecular gas phases. For homogeneity, and since we do not have CO observations for all the galaxies in our sample, we estimate molecular gas masses in these figures using the prescription found in \citet{2020A&A...633A.100C}, as explained in Appendix\,\ref{app:compCasaXCOLD}.

All the galaxies are spread widely in the four diagrams in Figs.\,\ref{fig:D2G} allowing us a good comparison over the physical parameter ranges covered by the simulations. In general, D/G versus metallicity agrees relatively well with the results of the simulations represented by the black continuous line. However, the observations seem to be more spread than the yellow area representing the dispersion in the simulations, which implies that our observations contains a larger variety of galaxies types with different properties and star formation histories that the simulations fail to describe. 

In the top-right panel of Fig.\,\ref{fig:D2G}  we show the relation between D/G ratio and the stellar mass. The simulations reproduce the observational trend up to a stellar mass of $\log\rm M_{star}\sim\,9.5$. However, toward the high end of stellar mass there is a significant number of galaxies exhibiting lower D/G ratios than the values predicted by the simulations. This discrepancy was also reported by \citet{2019MNRAS.485.1727H} but for a smaller sample of galaxies. In the high stellar mass regime, either the dust mass is over predicted or the simulations show a lack of gas mass. Following the $\rm M_{star}-$metallicity relation in Fig.\,\ref{fig:Mstar-Z}, the simulations predict higher metallicities than the observations in the high stellar mass range. This enhanced chemical enrichment would translate into a higher dust mass content in these massive galaxies. Indeed, the dust mass function at $z=0$ simulated by \citet{2019MNRAS.485.1727H} overpredicts the observed dust mass function in the high dust mass regime. Therefore, an overestimation of the dust mass in massive galaxies seems a plausible explanation for the higher D/G values given by the simulations in comparison with the observed values in massive galaxies. Interestingly, the predictions for the D/G ratios versus metallicity (left-panel in Fig.\,\ref{fig:D2G}) seems to describe well the observed trend (except for some galaxies at high metallicity that are above the relation). This shows that both correlations, D/G$-\rm M_{star}$ and D/G$-$metallicity are related through the $\rm M_{star}-$metallicity relation where in the massive end some tension between observations and simulations are found. 

The other possibility to explain the high values of D/G in massive galaxies in the D/G$-\rm M_{star}$ relation is that the simulations predict a low gas content in massive galaxies, which would make the D/G ratios higher than the observed in our sample. In Appendix\,\ref{app:compfmolSPH} we compare the gas mass fractions predicted by simulations with those obtained for our galaxy sample. The results of the simulations fall within the observed data and traces relatively well the observed trend in the data, which shows that the simulations seems to reproduce the gas mass content in galaxies, and therefore the discrepancies observed in the D/G is mainly due to an overestimation of the dust mass in galaxies by the simulations.

One possible reason for the discrepancy in massive galaxies could be that the AGN feedback might not be suppressing star formation enough and therefore the chemical enrichment would still be high in massive galaxies, with the corresponding increase in the dust mass content. A more sophisticated AGN feedback prescription should be included in the simulations in order to explore these possibilities. Other reasons could be related to the simplification of keeping a constant star formation efficiency for all the galaxies in the simulations, as it was mentioned in previous section. Finally, we note here that the total dust mass in a galaxy depends on the time-scales for the different mechanisms affecting the dust evolution and dust condensation/destruction efficiencies which are still in debate (see Section\,\ref{sec:intro}).

In the bottom-left panel of Fig.\,\ref{fig:D2G} we show the D/G ratio versus the specific SFR, sSFR. Most of the galaxies fall within the area covered by the predictions of the simulations. Both observed data and simulations present a large dispersion with a light trend of increasing D/G ratios for higher sSFR. 
In the bottom-right panel in Fig.\,\ref{fig:D2G} we see the D/G ratio versus the gas mass fraction, where the molecular gas mass has been estimated following the prescription given in  \citet{2020A&A...633A.100C}. The D/G ratios predicted by the simulations are higher for galaxies with low gas mass fractions, which tend to be the most massive ones. This agrees with the trend in the top-right panel where the most massive galaxies have observed D/G that are lower than those predicted by the simulations.

\subsubsection{Dust-to-star ratio}\label{sec:D2S}

The dust-to-star mass (D/S) ratio (sometimes called specific dust mass)  traces the amount of dust per stellar mass that survives the dust destruction and removal processes in the ISM. D/S ratio tends to be constant (to a value close to D/S\,$\sim$\,10$^{-3}$, e.g. Edmunds 2001, \citealt{2019MNRAS.483.4968V}) if there is no dust growth, and decreases to lower values when either stellar dust formation decreases or dust destruction processes are in place. D/S ratio strongly depends on the star formation history of the galaxy \citep{2008A&A...479..669C,2017MNRAS.465...54C}, therefore,  the relation between the D/S ratio and the stellar mass of the galaxy allows us to describe the evolution status of the galaxy. 

\begin{figure*}
\includegraphics[width=0.45\textwidth]{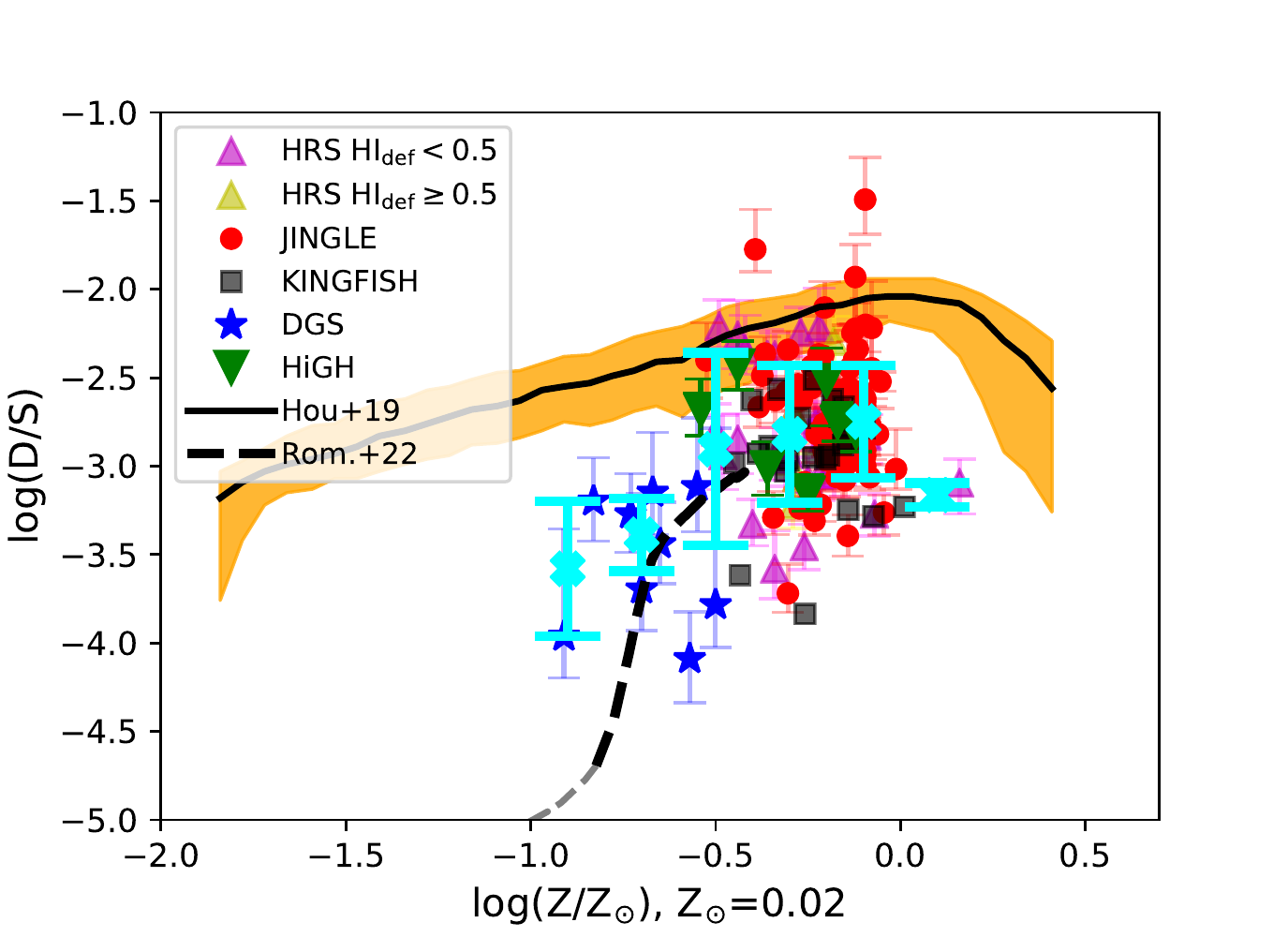}
\includegraphics[width=0.45\textwidth]{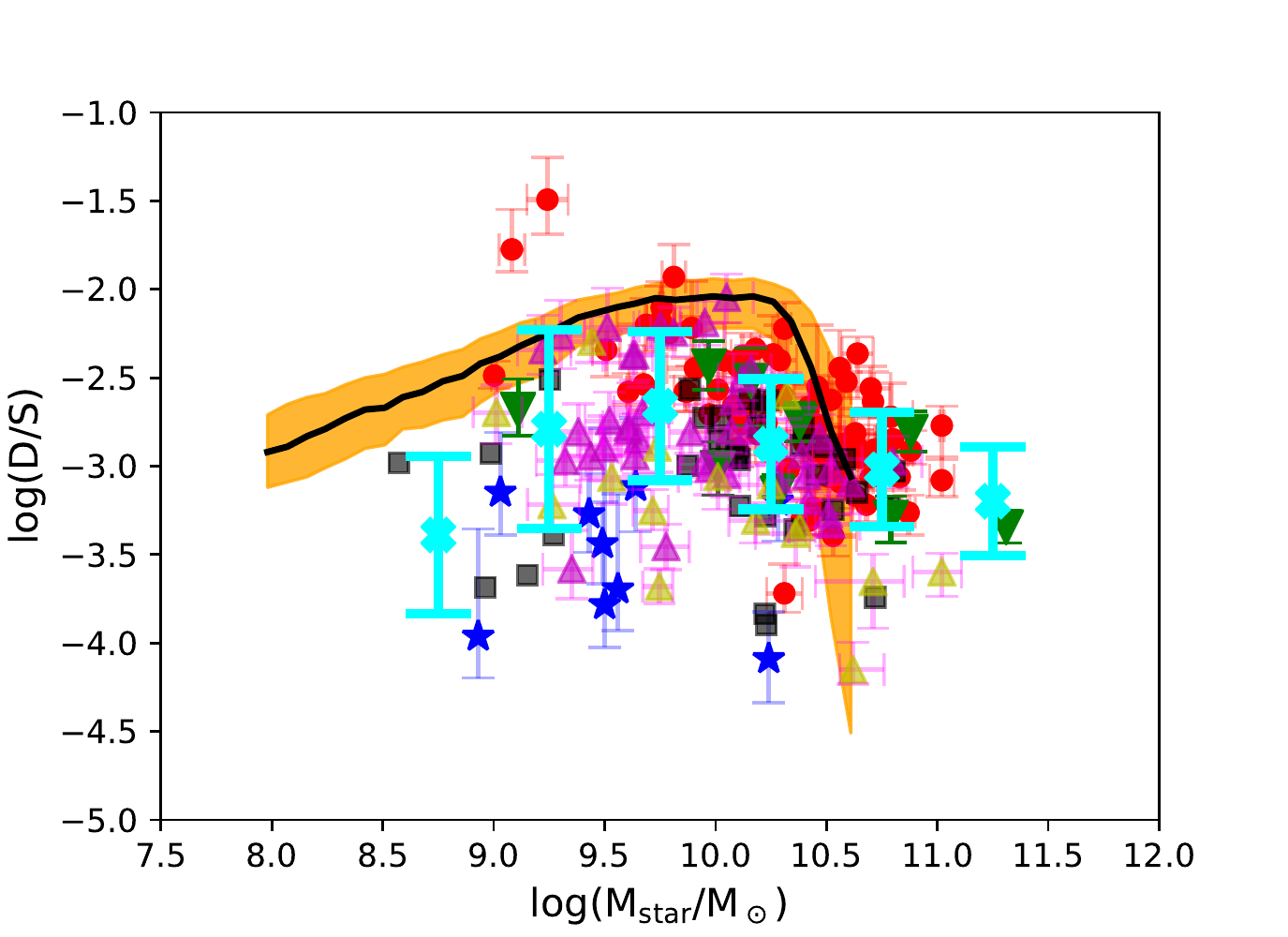}
\includegraphics[width=0.45\textwidth]{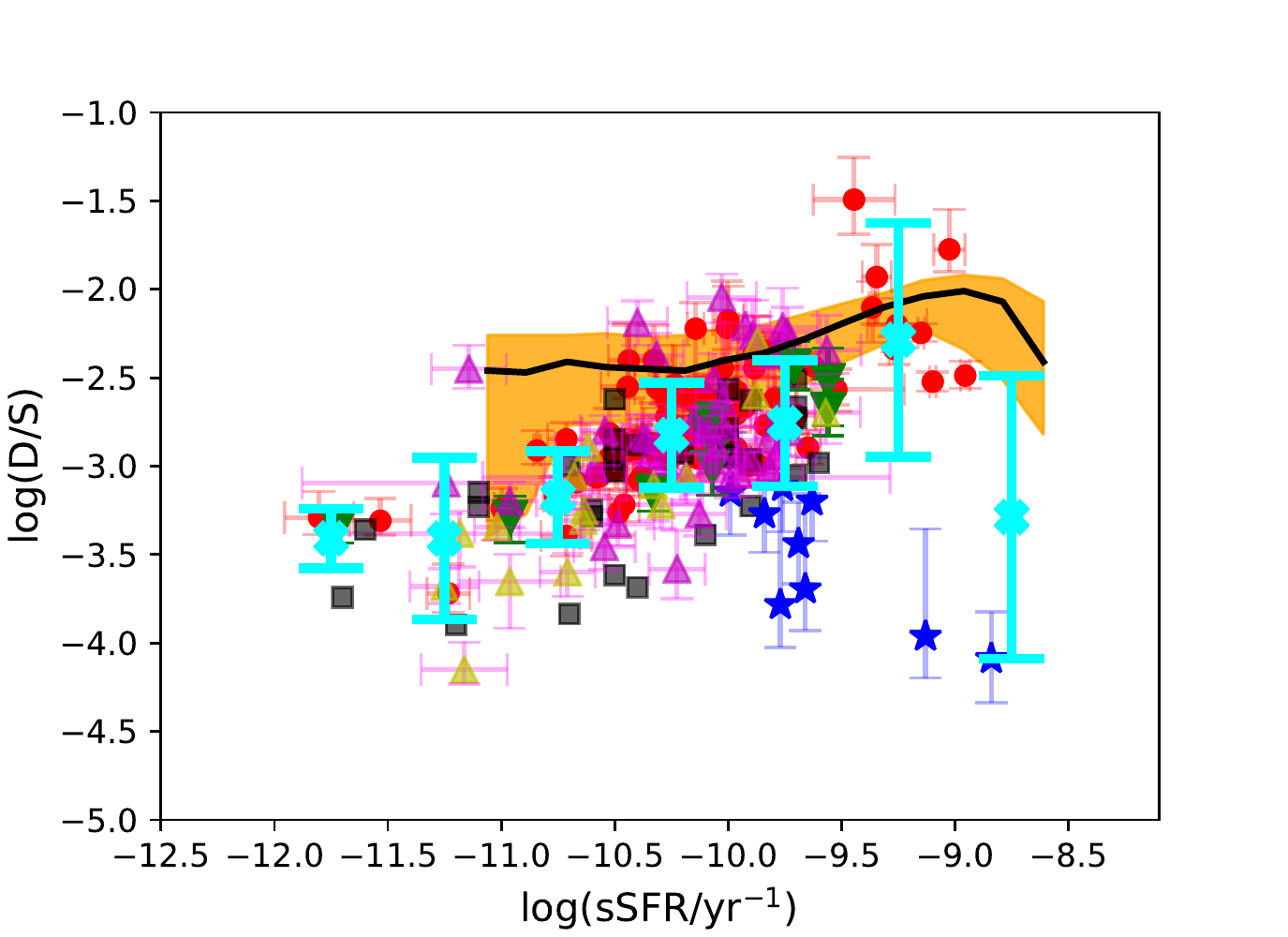}
\includegraphics[width=0.45\textwidth]{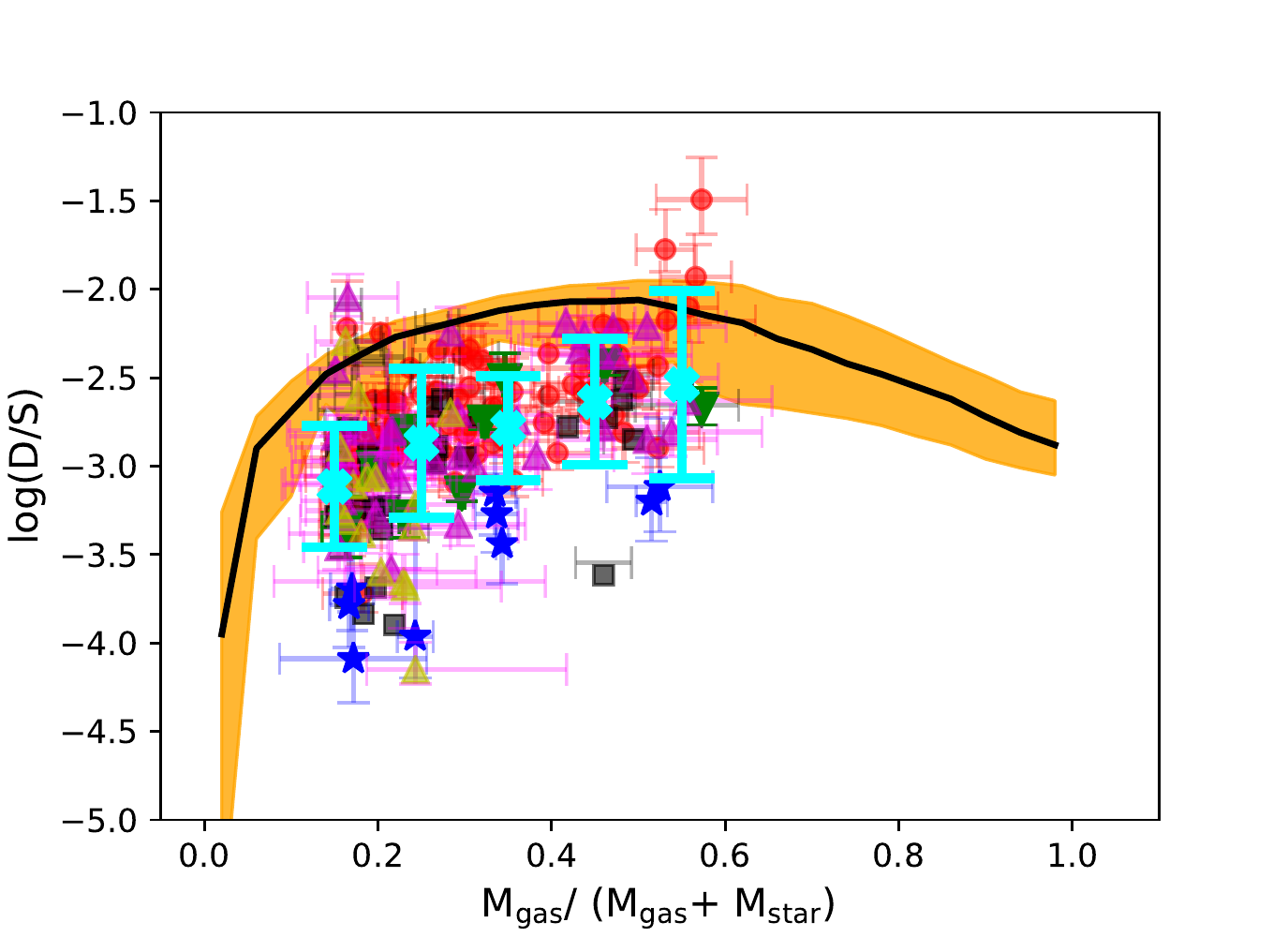}
\caption{Comparison of the dust-to-star ratio (D/S) for our galaxy sample with the hydrodynamic simulations. The black continuous line with the yellow colour, and the black dashed lines represent the same simulations as those in Fig.~\ref{fig:D2G}.Top-left: D/S versus metallicity, top-right: D/S versus stellar mass, bottom-left: D/S versus sSFR, and bottom-right: D/S versus gas mass fraction. The cyan crosses and corresponding error bars represent mean values and standard deviations of magnitudes represented in the $y$-axis for bins in the $x$-axis. The dust masses are in general overpredicted by the simulations (see Section\,\ref{sec:D2S}).}\label{fig:D2S}
\end{figure*}

In general, the observed D/S ratio decreases with the stellar mass in local galaxies \citep[e.g.][]{2020MNRAS.496.3668D}. A declining D/S ratio as a function of stellar mass has also been observed in galaxies up to $z\sim$5 \citep[e.g.][]{2020A&A...644A.144D,2021ApJ...921...40K}. We also see a declining D/S ratio with stellar mass for our galaxy sample (top-right panel of Fig.\,\ref{fig:D2S}). The observational behaviour of the D/S ratio with stellar mass is shaped not only by the SFMS relation and the $M_{\rm star}-Z$ relation shown in Sections\,\ref{sec:MsZ} and \ref{sec:scarel}, but also by the dust evolution processes in the ISM. Low mass galaxies are characterised by a low SFR and low metal content which would then give a low dust mass content. However, the balance between the dust production and destruction in the ISM of these galaxies makes the specific dust mass higher in low mass galaxies. 

The comparison of the D/S ratio between observations and simulations could give us more information about the overestimation of the dust mass for massive galaxies that is inferred from the 
D/G trends shown in Fig.\,\ref{fig:D2G}. In the top-left panel of Fig.\,\ref{fig:D2S} we explore the D/S-metallicity relation. Almost all of our galaxies present lower values of observed D/S ratios compared with the predictions from simulations within the whole metallicity range, which shows that dust masses would be overestimated for all our galaxies if the stellar mass would be accurately reproduced by the simulations. The fact that the simulations predict relatively well the behaviour of $\rm f_{\rm gas}$ as a function of stellar mass (see Appendix\,\ref{app:compfmolSPH} and Fig.\,\ref{fig:MgMs_Ms}) reassures that in a general way, this is the case.  In contrast to simulations from \citet{2019MNRAS.485.1727H}, the D/S versus metallicity predictions from \citet{2022MNRAS.tmp.1342R}, which correspond to an individual isolated galaxy with metal diffusion (see Appendix\,\ref{app:Rom}), reach D/S values comparable with those observed in our galaxy sample.

In the top-right panel in Fig.\,\ref{fig:D2S} we show the D/S-stellar mass relation. Although the observed D/S ratio is lower than the expected from simulations in the $\log\rm (M_{star}/M_\odot)\sim$\,9-10 regime, it is well reproduced by the simulations for most massive galaxies at $\log\rm (M_{star}/M_\odot)\sim$\,10.5. Thus, an overestimation of the dust masses in galaxies with  $\log\rm (M_{star}/M_\odot)\sim$\,10.5  would mean an overestimation of the stellar mass for these galaxies as well.  Indeed, the galaxy stellar mass function predicted by \citet{2019MNRAS.485.1727H} shows a bump at log(M$_{\rm star}\gtrsim$10.5) that is not seen in the observations (see fig.\,1 in \citealt{2019MNRAS.485.1727H}). In this regime, not only the dust mass is overestimated but also the stellar mass is higher than what is observed, and therefore the D/S ratio comes into an agreement with observations in massive galaxies. The simulations also tends to produce a higher D/S ratio than the observed values for galaxies with low gas mass fraction (bottom-right panel of  Fig.\,\ref{fig:D2S}), however, the observed increasing trend of D/S with gas mass fraction is well described by the simulations.

At intermediate stellar masses ($\log\,\rm M_{star}\sim9.5-10$), the simulations shows a constant D/S ratio that reproduces the behaviour of chemical and dust evolution models of MW type galaxies where there is a continuous replenishment of dust due to a prolonged star formation activity but the dust destruction and formation processes give a relatively constant D/S ratio (see top-right panel in figure 3 of \citealt{2017MNRAS.465...54C}). At higher stellar masses AGN feedback is suppressing star formation and the simulations show a declining trend of D/S with stellar mass in agreement with observations. Our observations show a declining D/S ratio for galaxies with stellar mass  
$\log\,\rm M_{star}\gtrsim9.5$. However, not all the galaxies in our sample are affected by AGN and a significant number of them are spiral disk galaxies with continuous star formation. Moreover, dust can be removed from the ISM by outflows due to intense star formation and SN events, which will also affect the evolution of the D/S ratio \citep{2015MNRAS.449.3274F,2019A&A...632A..43M}. Interestingly, the simulations in \citet{2017MNRAS.468.1505M}, which use the star formation model proposed by \citet{2003MNRAS.339..289S} and incorporate AGN feedback, find a declining trend of D/S with stellar mass matching relatively well the trend from observations. However, \citet{2017MNRAS.468.1505M} simulations fail to reproduce the observed D/G versus metallicity trend, which is crucial to describe properly the evolution of the interstellar dust. The evolution models in \citet{2017MNRAS.465...54C} for galaxies formed by rapid collapse of gas that triggers an intense star formation event and evolve to a passive phase are able to describe the observed declining trend of the D/S ratio with stellar mass. All this shows that D/S--$\rm M_{star}$ relation is intimately related to feedback and star formation history of the galaxy.

The observed trend for the D/S--sSFR relation is in better agreement with the results from simulations than the D/S--Z or D/S--$\rm M_{star}$ relations. D/S ratio increases with increasing sSFR both for simulations and observations. An increase of the D/S ratio with sSFR has also been reported in the literature \citep[e.g.][]{2010MNRAS.403.1894D,2015A&A...582A.121R,2020MNRAS.496.3668D}  and it can be explained as a consequence of the different mechanisms that affect the chemical and dust evolution in galaxies. Galaxies with high sSFR tend to have a high fraction of gas, form stars at a very high rate and the dust mass increases as a consequence of the intense star formation. As the galaxy evolves the gas mass (and consequently star formation as well) decreases while the stellar mass rises, causing the sSFR to decline. Dust mass also decreases as star formation drops, producing a declining trend of the D/S ratio at low sSFR. This behaviour has been successfully reproduced with chemical and dust evolution models for different galaxy types \citep[e.g.][]{2015A&A...582A.121R,2020A&A...641A.168N}.

\subsection{Small-to-large grain ratio}

The \stol\ grain mass ratio reflects the relative importance of the mechanisms affecting the dust grain size distribution. As it has been predicted by simulations \citep[e.g.][]{2019MNRAS.485.1727H,2018MNRAS.478.4905A}, for low-metallicity galaxies, log(Z/Z$_{\odot}$)\,$\sim-2$, dust is mainly produced by stars and shattering is the only source of small grains, while at higher metallicities (-2\,$\leq$\,log(Z/Z$_{\odot}$)\,$\leq$\,-1), accretion would become efficient and \DstoDl\ would start to increase significantly. At -1\,$\leq$\,log(Z/Z$_{\odot}$)\,$\leq$\,-0.5, coagulation would become efficient enough to produce a balance between the amount of small grains created by accretion and shattering, and the large grains created via coagulation, giving as a result a constant \DstoDl. At even higher metallicities, log(Z/Z$_{\odot}$)\,$\geq$\,-0.5, coagulation dominates against accretion and shattering and \DstoDl\ decreases with increasing Z.

\begin{figure*}
\includegraphics[width=0.45\textwidth]{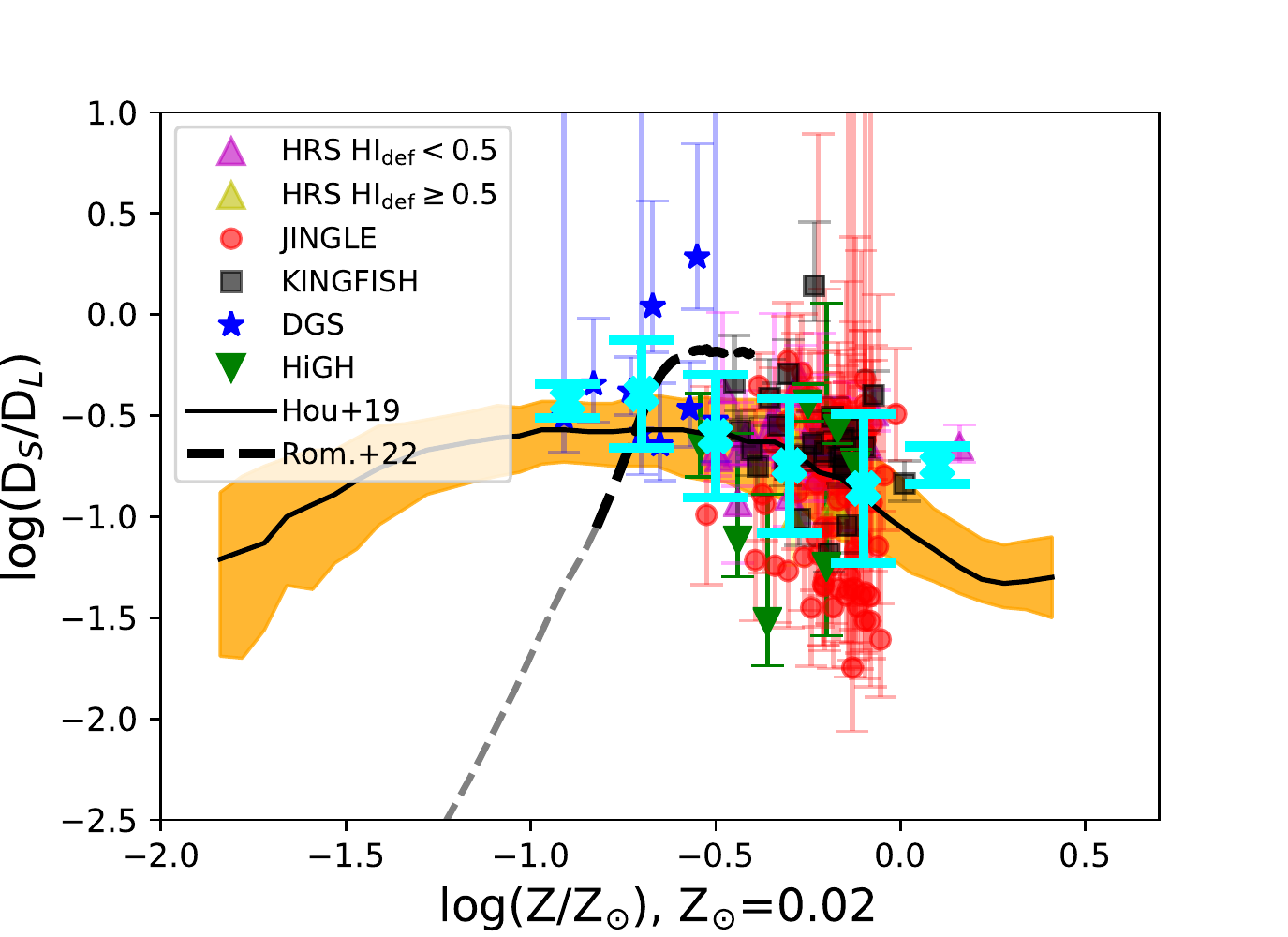}
\includegraphics[width=0.45\textwidth]{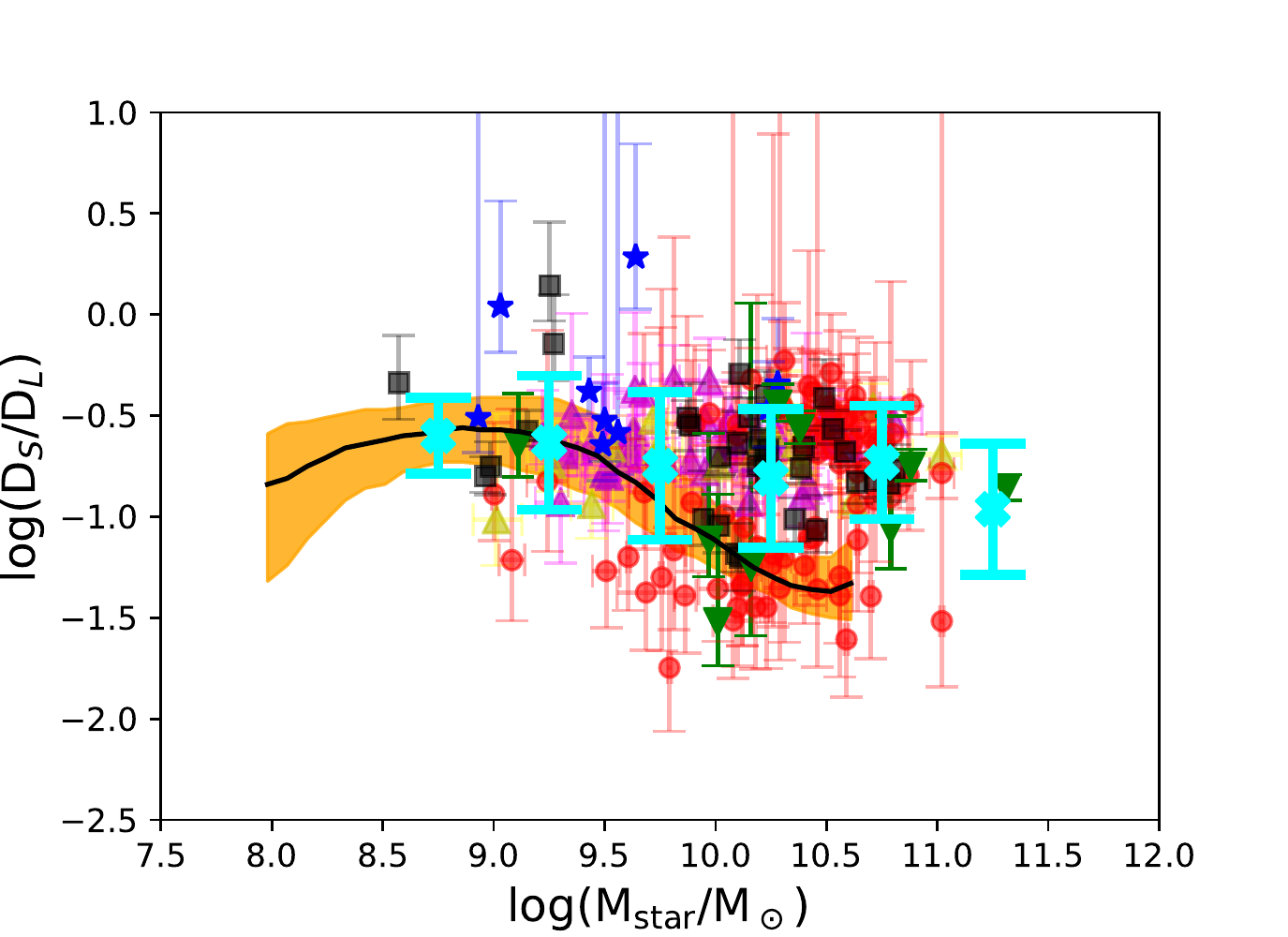}
\includegraphics[width=0.45\textwidth]{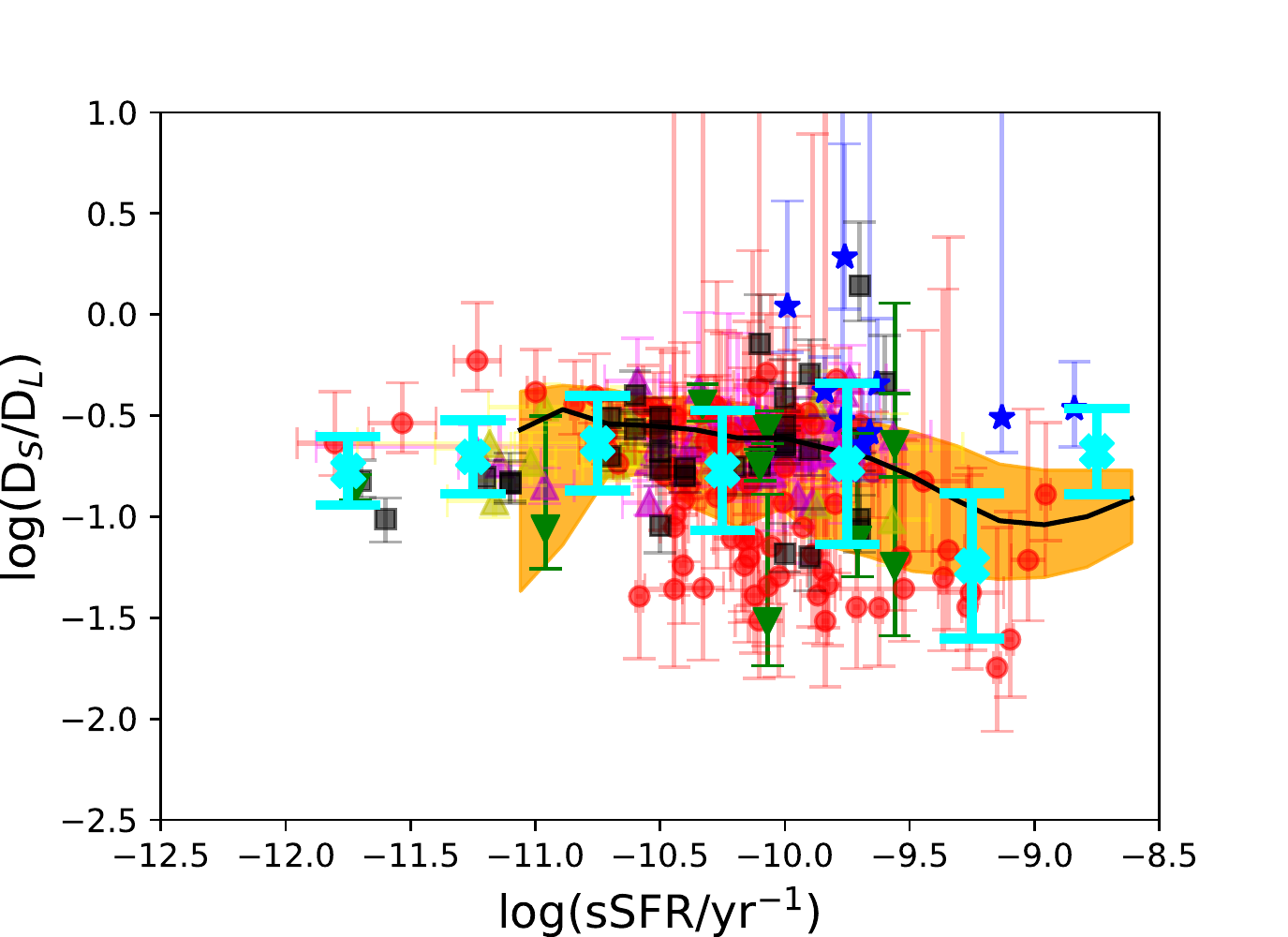}
\includegraphics[width=0.45\textwidth]{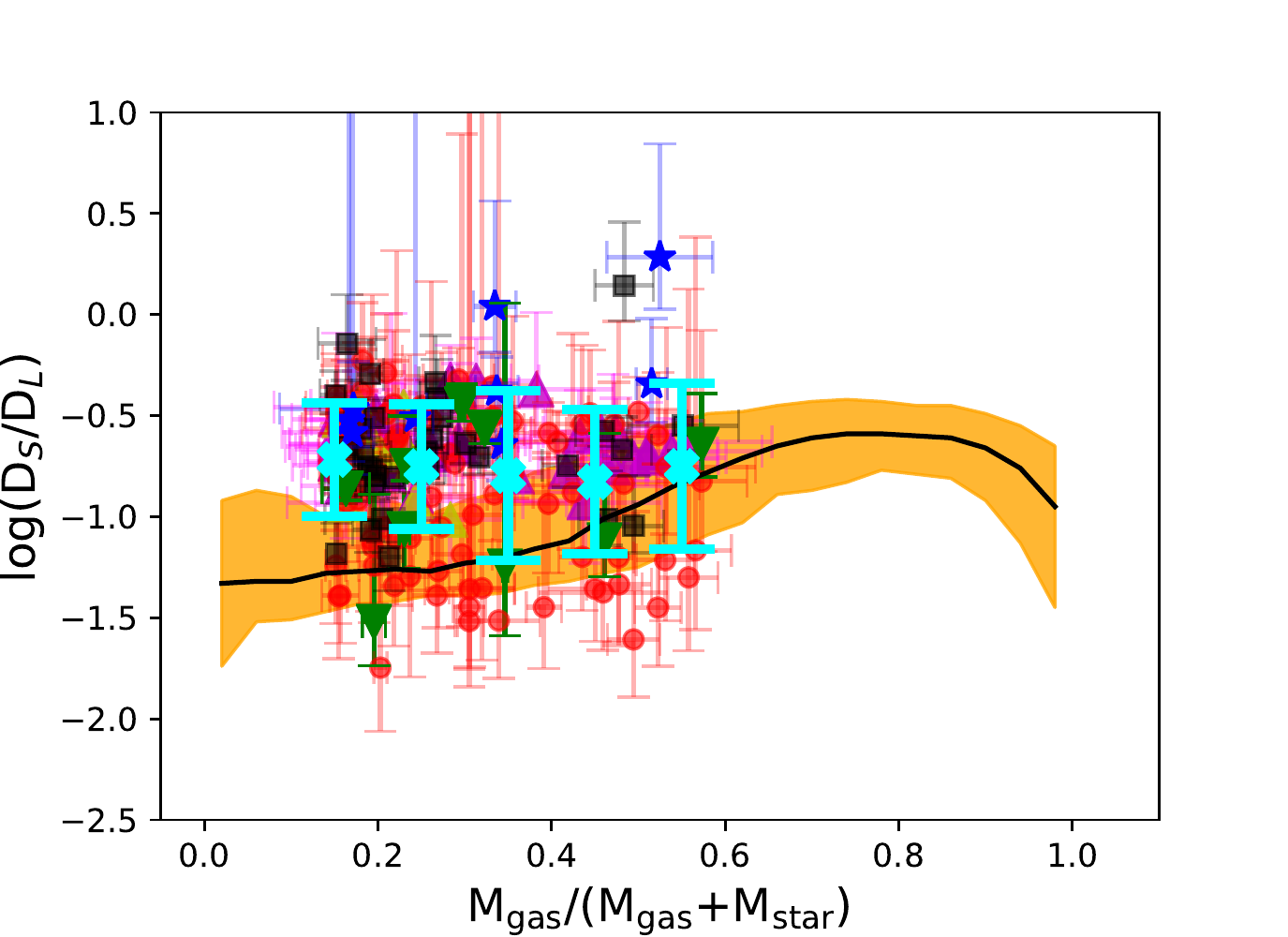}
\caption{Comparison of the \stol\ grain mass ratio (\DstoDl) with results of the simulations. In each panel we show the trend of \DstoDl\ with different galaxy properties: metallicity (top-left), stellar mass (top-right), sSFR (bottom-left) and gas mass fraction (bottom-right).  The black continuous line with the yellow colour, and the black dashed lines represent the same simulations as those in Fig.~\ref{fig:D2G} and \ref{fig:D2S}. The cyan crosses and corresponding error bars represent mean values and standard deviations of magnitudes represented in the $y$-axis for bins in the $x$-axis. In massive galaxies, observed \DstoDl\ tend to be higher than the values prediction from simulations.}\label{fig:S2L}
\end{figure*}

In Fig.\,\ref{fig:S2L} we show the \stol\ grain mass ratio (\DstoDl) versus different galaxy properties with the same colour code for each galaxy sample as in previous figures. The top-left panel describes very nicely the different steps in the evolution of the \DstoDl\ with metallicity. Most of the galaxies in our sample shows \DstoDl\ values consistent with the phase where a  balance between accretion and shattering producing small grains and coagulation giving large grains is occurring. At $\log_{10}(Z/Z_{\odot})\geq$-0.3  a significant fraction of galaxies (mainly from JINGLE sample and a few from KINGFISH and HiGH) shows lower values of \DstoDl\ than those expected from the simulations in this high metallicity regime. The late phase of the simulations of an individual galaxy with metal diffusion \citep{2022MNRAS.tmp.1342R} agrees with the observations as well. 

In the top-right panel of Fig.\,\ref{fig:S2L} we show the \DstoDl\ versus stellar mass. We see two trends here: $i)$ Most galaxies from the KINGFISH, HiGH, HRS and JINGLE show \DstoDl\ within a narrow range of values ($\log_{10}(D_{S}/D_{L})\sim-0.5$), and $ii)$ the rest of galaxies presents a declining trend with stellar mass similar to the predictions from simulations. The behaviour of \DstoDl\ versus sSFR (bottom-left panel in Fig.\,\ref{fig:S2L}) shows a large dispersion with a small fraction of galaxies with low \DstoDl\ falling outside the predictions of the simulations. A large dispersion is also seen in the \DstoDl\ versus f$_{\rm gas}$ panel (bottom-right panel in Fig.\,\ref{fig:S2L}), in this case the galaxies with low values of  \DstoDl\ are within the range of predictions from simulations. 

\begin{figure}
\includegraphics[width=0.45\textwidth]{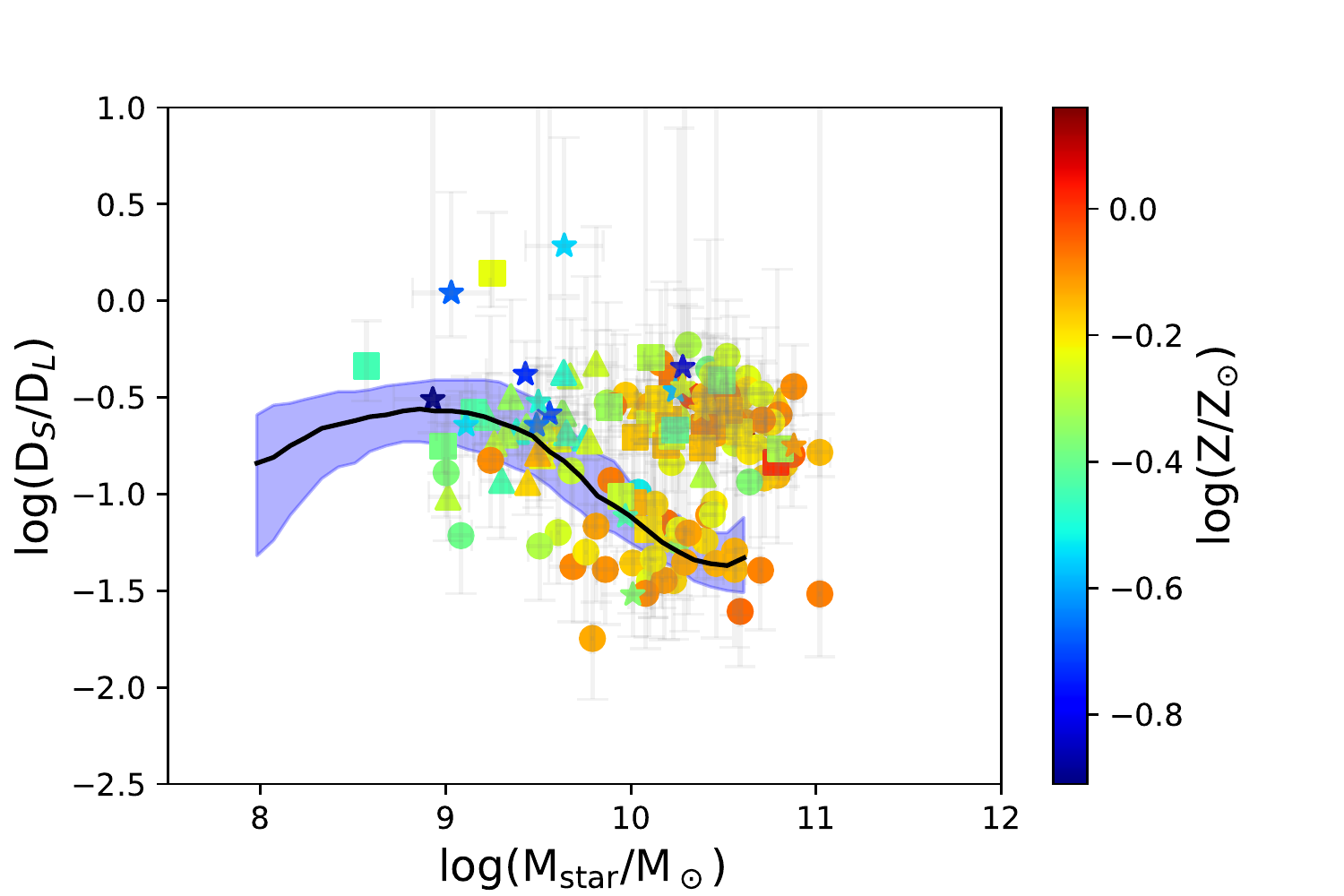}
\includegraphics[width=0.45\textwidth]{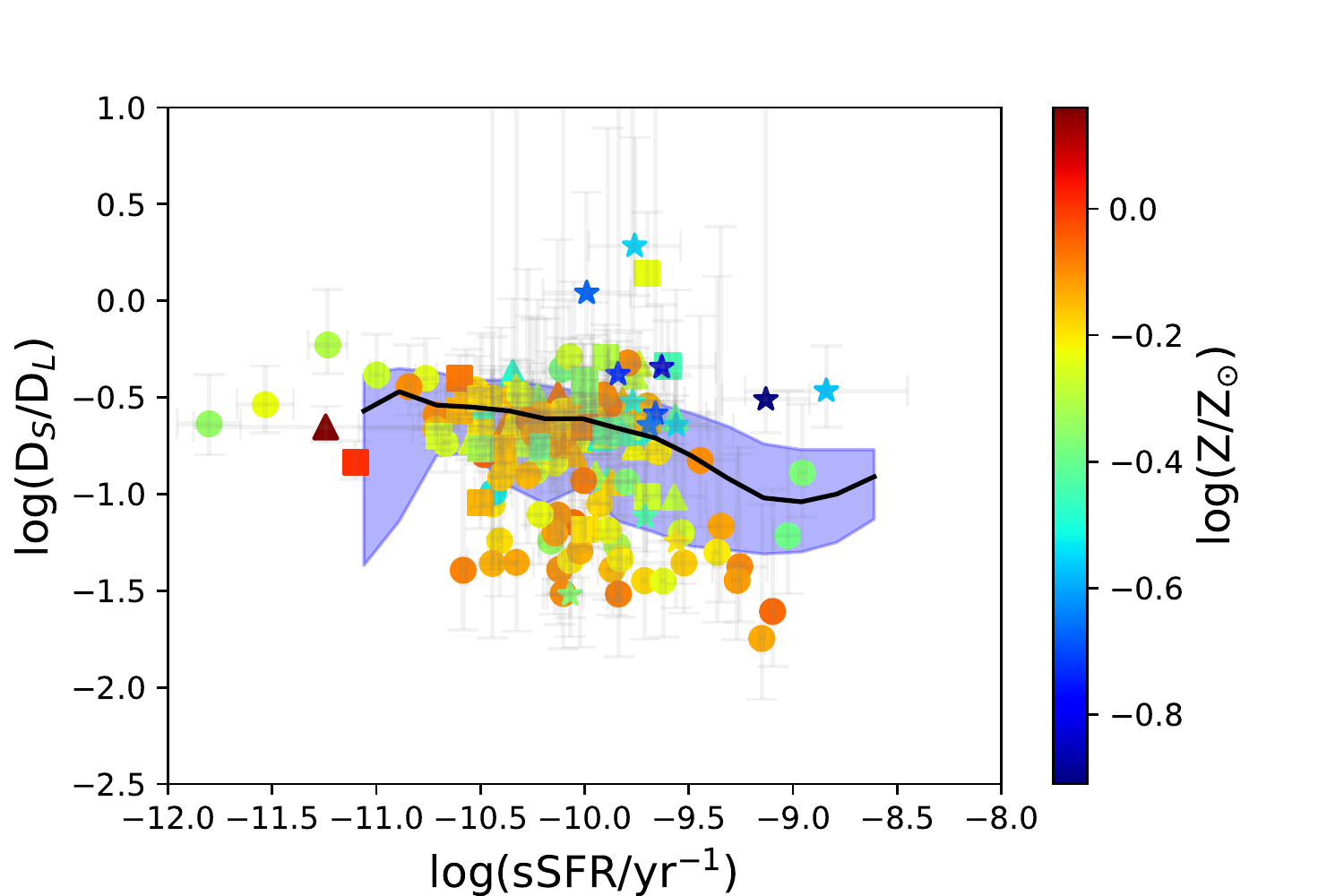}
\includegraphics[width=0.45\textwidth]{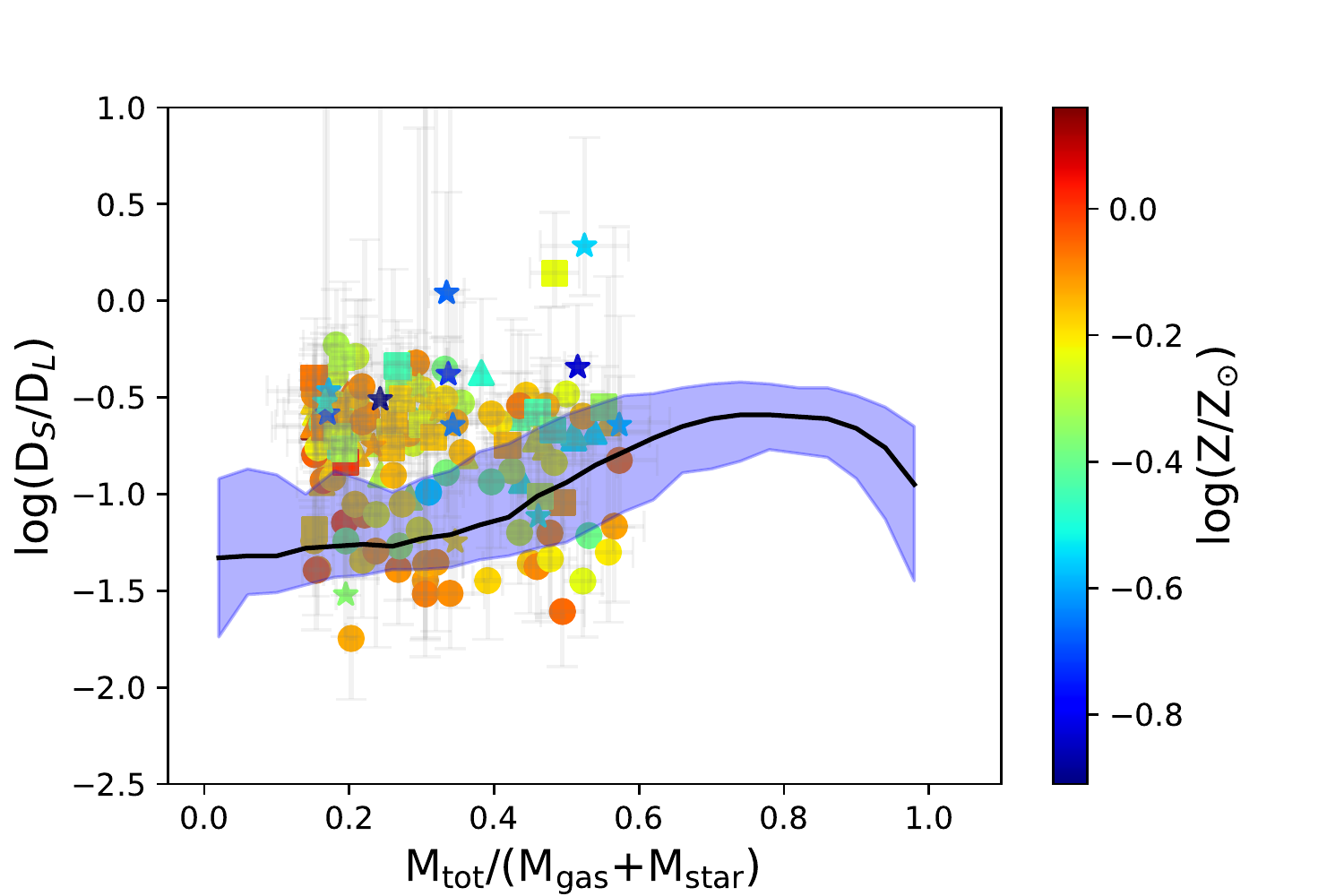}
\caption{Distributions of the \stol\ grain mass ratio (\DstoDl) versus stellar mass (top), sSFR (middle) and gas mass fraction (bottom) for our galaxy sample colour coding with metallicity. The black continuous line shows the 50th percentile of the particle distribution in the simulations from \citet{2019MNRAS.485.1727H} and the blue colour represents the area within the 16th and 84th percentiles.}\label{fig:S2L:metal}
\end{figure}

\begin{figure}
\includegraphics[width=0.45\textwidth]{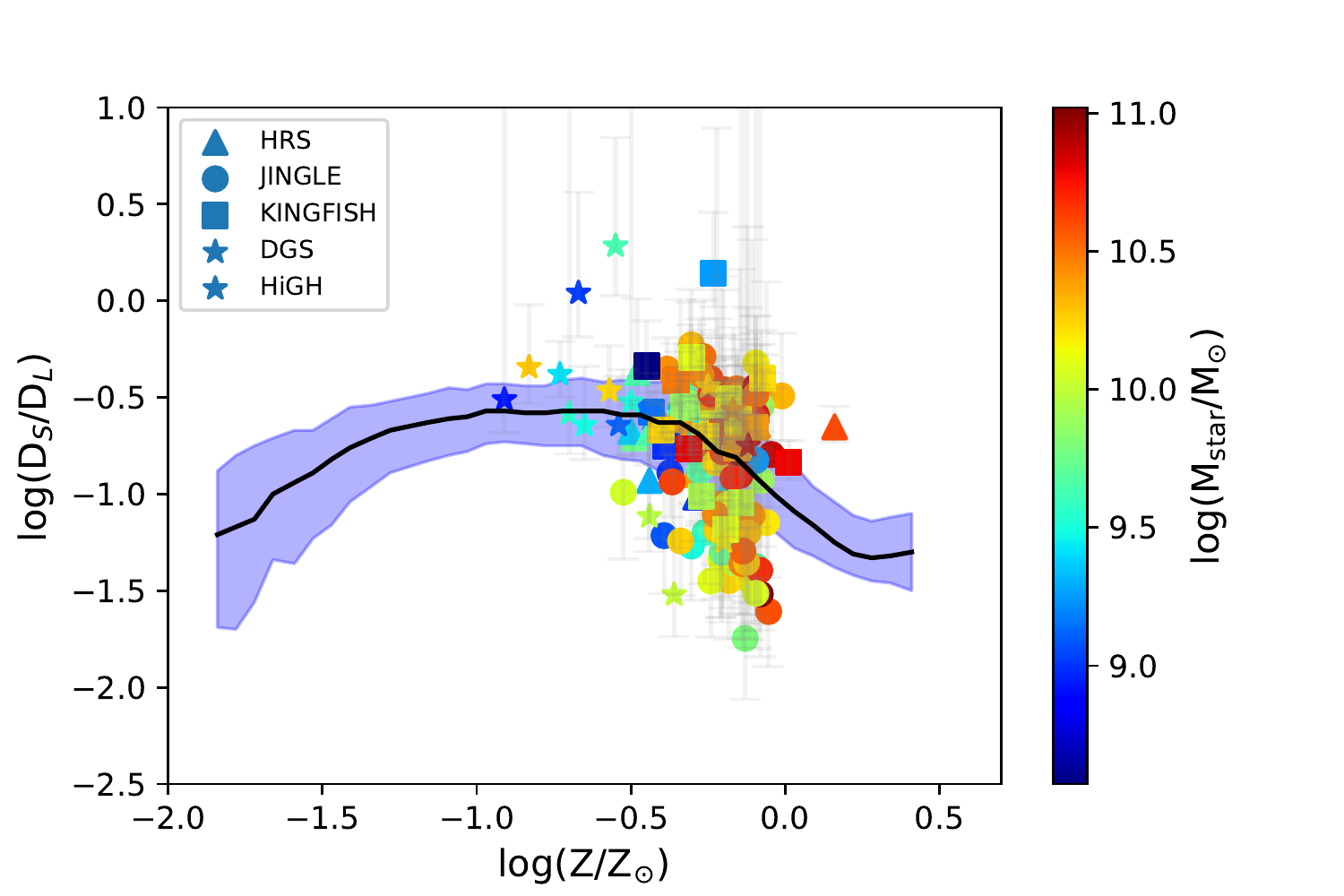}
\includegraphics[width=0.45\textwidth]{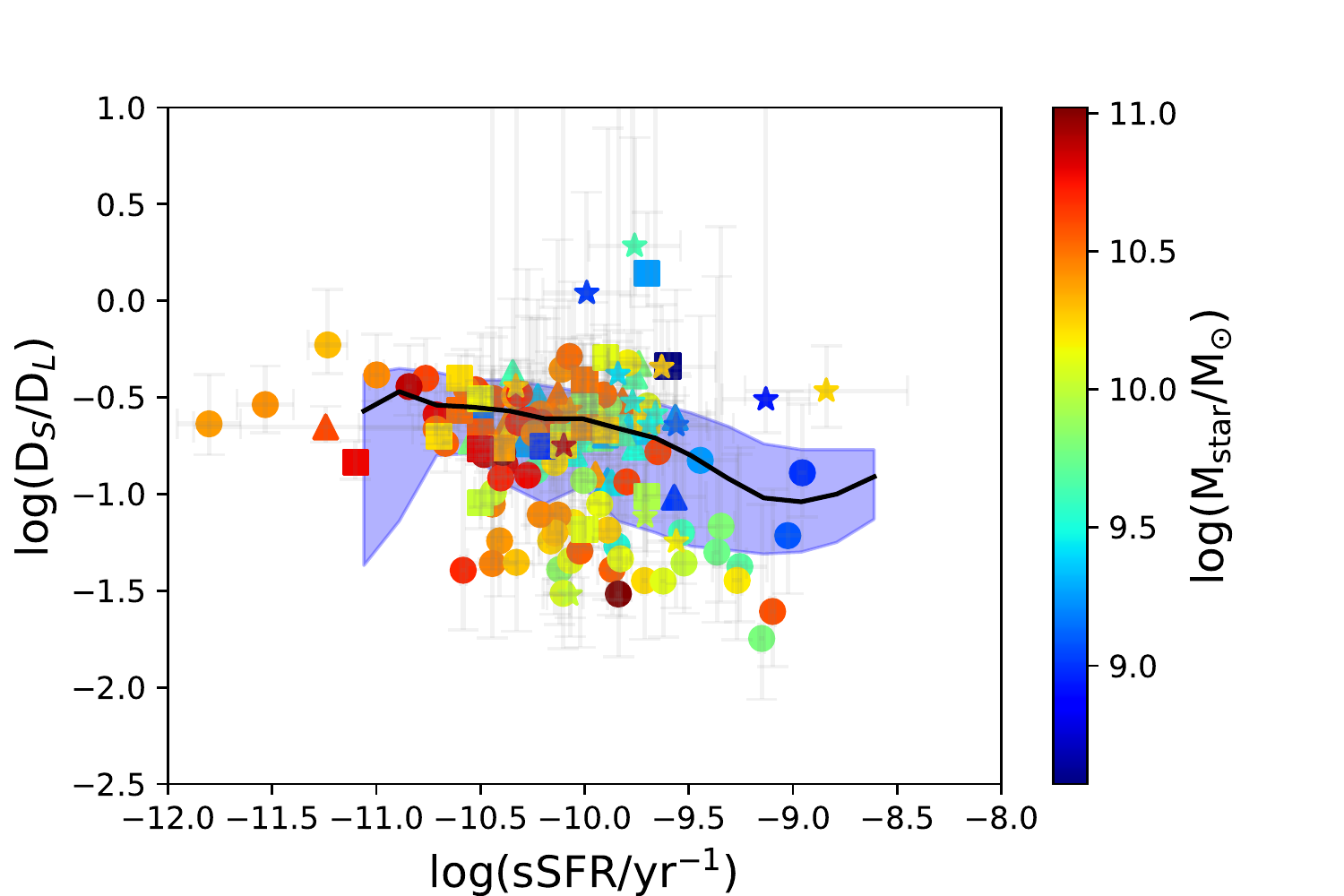}
\includegraphics[width=0.45\textwidth]{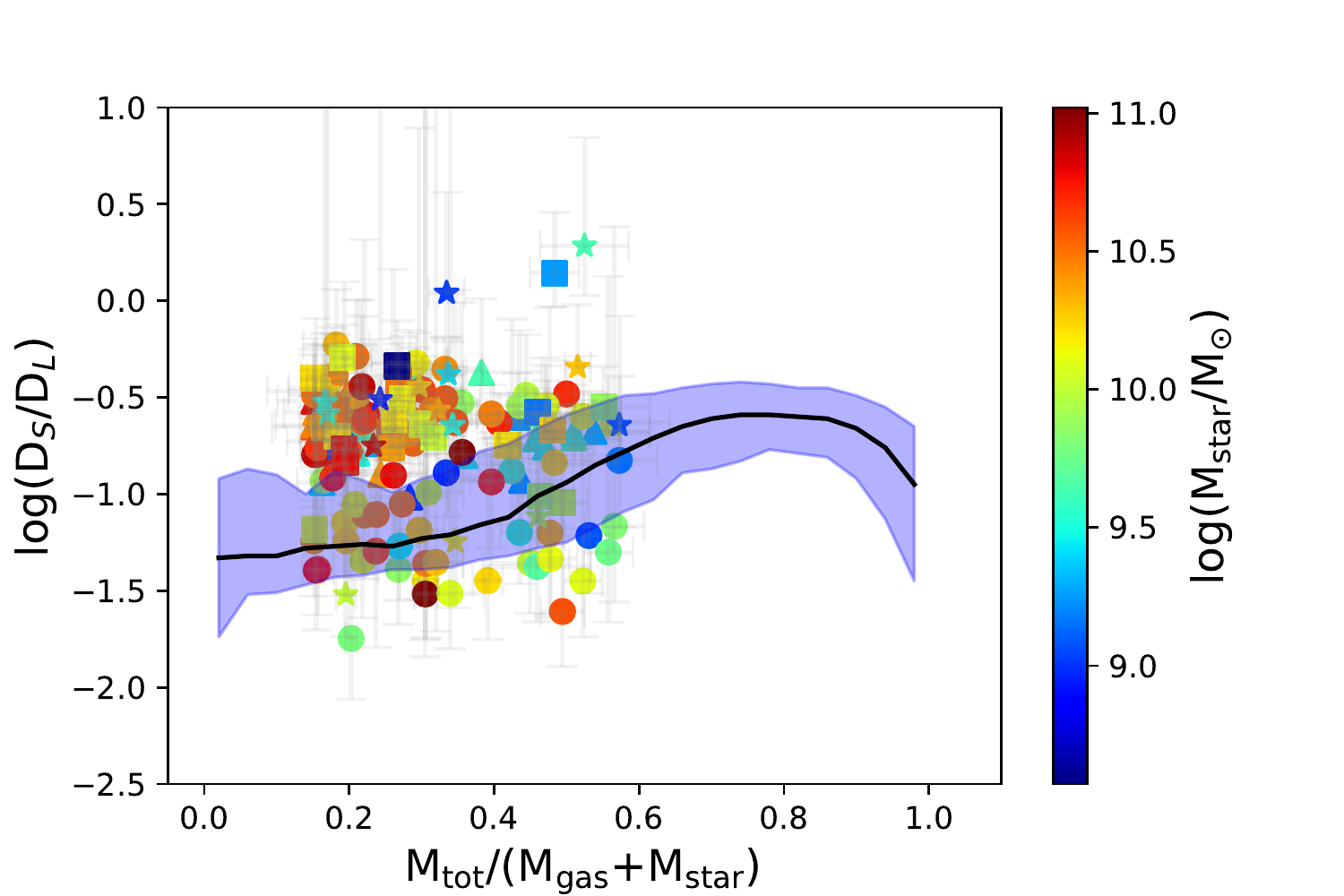}
\caption{Distributions of the \stol\ grain mass ratio (\DstoDl) versus metallicity (top), sSFR (middle) and gas mass fraction (bottom) for our galaxy sample with a colour code indicating the stellar mass for each galaxy. The black continuous line shows the 50th percentile of the particle distribution in the simulations from \citet{2019MNRAS.485.1727H} and the blue colour represents the area within the 16th and 84th percentiles.}\label{fig:S2L:Ms}
\end{figure}

We study in more detail the distributions presented in Fig.\,\ref{fig:S2L} in order to characterise the two trends in the \DstoDl-stellar mass distribution: the galaxy subsample that follows the predictions of the simulations and the other subsample with $\log(D_{S}/D_{L})\sim-0.5$ in the high stellar regime.  In Fig.\,\ref{fig:S2L:metal} we plot the  trends as in Fig.\,\ref{fig:S2L} but now colour coding our galaxy sample with metallicity. The two trends are clearer seen in the top panel of Fig.\,\ref{fig:S2L:metal}. One subsample of galaxies follows the behaviour of the simulations within the whole stellar mass range ($\log\rm (M_{star}/M_\odot)\sim$\,8$-$10.5), covering a large range of metallicities; and another subsample exhibits \DstoDl\ values within a relatively narrow range ($\log(D_{S}/D_{L})\sim-0.5$) and at slightly higher metallicities. This last subsample is outside the area covered by the predictions of the simulations in the \DstoDl-stellar mass distribution. The two samples are also differentiated in the \DstoDl-f$_{\rm gas}$ distribution (bottom panel of Fig.\,\ref{fig:S2L:metal}), the sample with constant \DstoDl\ again not being  consistent with the results from the simulations. Interestingly, in the \DstoDl-sSFR distribution (middle panel in Fig.\,\ref{fig:S2L:metal}) the set of galaxies that fall within the area covered by the simulations is the sample with galaxies having $\log(D_{S}/D_{L})\sim-0.5$, being the other sample not in agreement with the simulated \DstoDl-sSFR relation. As an illustration, we present in Fig.\,\ref{fig:S2L:Ms} the distributions coloured coded with stellar mass, showing that the galaxies that follow the prediction of the simulations cover a wide range in stellar mass (see top and middle panels of Fig.\,\ref{fig:S2L:Ms}). 
 
\subsubsection{Small-to-large grain ratio and dust mass}

\begin{figure}
\includegraphics[width=0.45\textwidth]{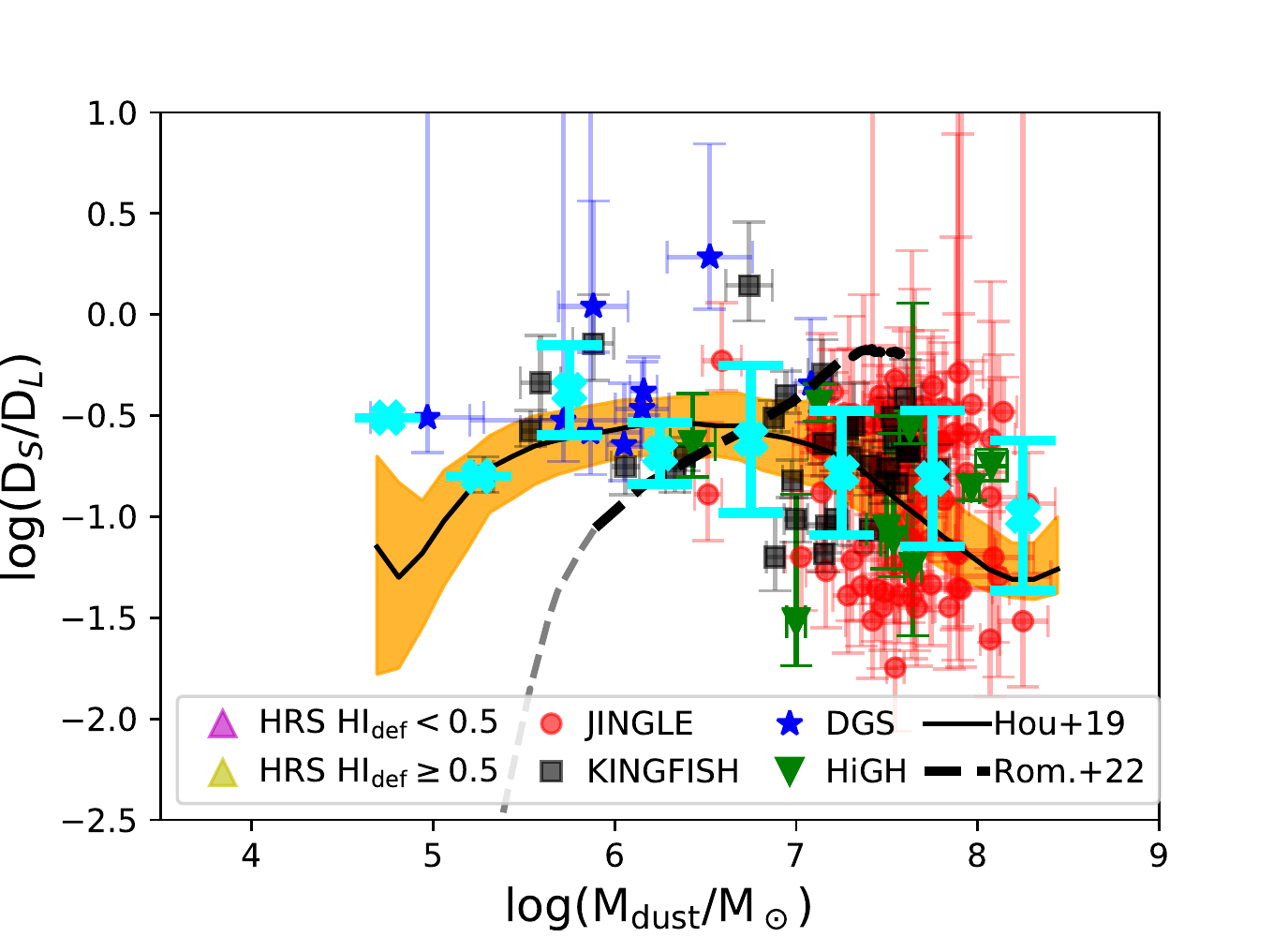}
\caption{Small-to-large grain mass ratio \DstoDl\  versus the total dust mass obtained with our fitting procedure for our galaxy sample. Colour code is the same as previous figures. The total dust mass has been obtained as a sum of the mass of the different dust components.  The black continuous line with the yellow colour, and the black dashed lines represent the same simulations as those in Fig.~\ref{fig:D2G} and \ref{fig:D2S}. The cyan crosses and corresponding error bars represent mean values and standard deviations of magnitudes represented in the $y$-axis for bins in the $x$-axis.}\label{fig:S2L:Md}
\end{figure}

In Fig.\,\ref{fig:S2L:Md}, we show \stol\ grain mass ratio versus the total dust masses derived from our fitting. The total dust mass has been obtained adding the mass of the different dust grains components. DGS galaxies are those with lower dust masses and higher  \DstoDl. The observed data present a hint of decreasing  \DstoDl\ for galaxies with high dust masses, although in the high dust regime there is a significant dispersion in the data, as some galaxies present the same \DstoDl\  ratio and others follow the declining trend of \DstoDl\  at high dust mass predicted by the simulations. 

The dashed black line represents the simulation of an individual isolated galaxy  \citep[][see Appendix\,\ref{app:Rom}]{2022MNRAS.tmp.1342R} that includes a parametrisation for the mass fraction in the form of dense clouds in cold and dense gas particles, called $f_{\rm dense}$ in the simulation studies of \citet{2017MNRAS.466..105A} and \citet{2019MNRAS.485.1727H}. $f_{\rm dense}$ is fixed in these simulations ($f_{\rm dense}=0.1$ in \citet{2019MNRAS.485.1727H} and $f_{\rm dense}=0.5$ in \citet{2017MNRAS.466..105A}), and is directly linked to the accretion and coagulation timescales that can affect the estimation of the dust mass content in a galaxy. Therefore, the value of $f_{\rm dense}$ assumed in the simulations can influence the D/G, D/S and \DstoDl\ ratios (see fig.\,12 in \citealt{2017MNRAS.466..105A} for a comparison of results with $f_{\rm dense}=0.5$ and $f_{\rm dense}=0.1$). 

The idea behind the parametrisation of $f_{\rm dense}$ in the simulation of an individual isolated galaxy proposed in \citet{2022MNRAS.tmp.1338R}  is to reach more realistic global dense gas fractions of $\sim$20\,\%, typical of MW type galaxies, than the global values predicted by the simulations of \citet{2017MNRAS.466..105A} and \citet{2019MNRAS.485.1727H}. Indeed, if we assume that the molecular gas mass fraction is a proxy of $f_{\rm dense}$ (the mass fraction in the form of dense clouds in cold and dense gas particles in simulations) we can compare the molecular gas mass fraction of our galaxies with the global values of $f_{\rm dense}$ obtained in the simulations. The comparison is done in Appendix\,\ref{app:compfmolSPH}, where we show that our galaxies have higher molecular gas mass fraction than the global values of $f_{\rm dense}$ predicted in the simulations of \citet{2019MNRAS.485.1727H} (see Fig.\,\ref{fig:fmol_MgMsXcold}).

It is important to note that in the case of the simulations of an isolated galaxy from \citet{2022MNRAS.tmp.1342R}, at late ages when the dust mass is reaching its maximum value, the \DstoDl\ ratio reaches values comparable with the highest values observed in massive galaxies. These simulations incorporate metal diffusion in the dust evolution treatment which produces an efficient way of  transport of large grains into the diffuse medium where they can be shattered into small grains. They are able to better reproduce the observational behaviour of those galaxies presenting a constant value for the \DstoDl\ ratio. While the model with diffusion might explain the high values of \DstoDl, it remains unclear whether the model can simultaneously explain the subset of galaxies with low \DstoDl. Cosmological simulations using the model by \citet{2022MNRAS.tmp.1342R} are needed in order to provide a more conclusive sample in this regard.

\section{Discussion}\label{sec:discussion}

In this section we discuss in detail the implications of the comparison of the \stol\ grain mass ratio derived from observations and  simulations. Our goal is to characterise observationally under which physical properties the different mechanisms (accretion, coagulation and shattering)  dominate the dust evolution in galaxies. This will provide some help to constrain the initial conditions and model parameters of future simulations treating the evolution of the dust grain size distribution. 

\subsection{Relation to molecular gas mass fraction}\label{sec:molgas}

The \stol\ grain mass ratio reflects the relative importance of the mechanisms affecting the dust grain size distribution. In particular, accretion and coagulation predominantly occur in the dense and cold gas. Such a gas phase is well represented by molecular clouds \citep{2000PASJ...52..585H,2009MNRAS.394.1061H}.
In contrast, shattering is efficient in an opposite condition -- the warm and diffuse phase.
Thus, we assume that the molecular gas mass fraction is an indicator of the dense gas fraction, which regulates the balance among the above various interstellar processing mechanisms. Under this assumption, we explore here how the main trends observed in previous sections are described in combination with the molecular gas content in the galaxy. 

\begin{figure*}
\includegraphics[width=0.45\textwidth]{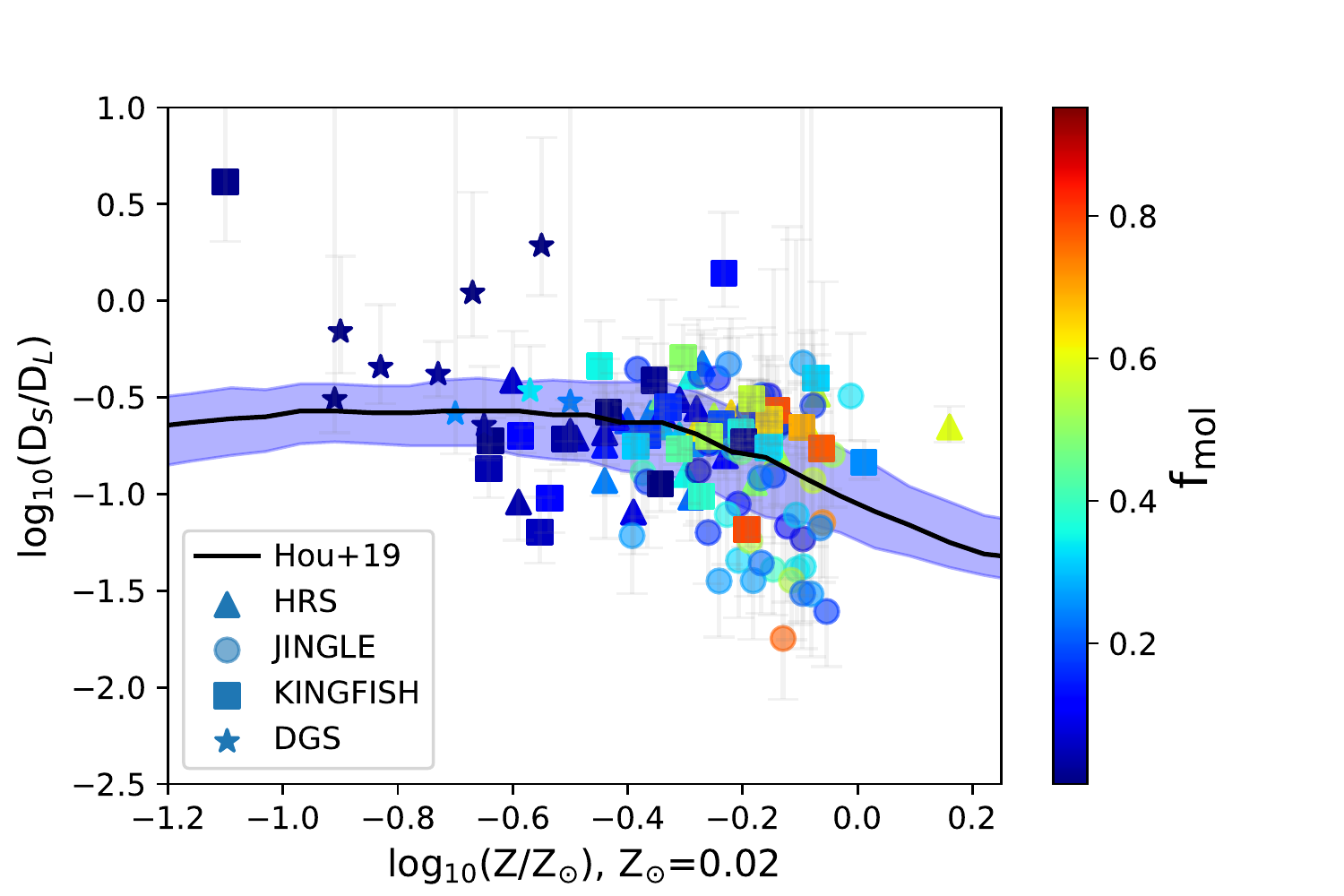}
\includegraphics[width=0.45\textwidth]{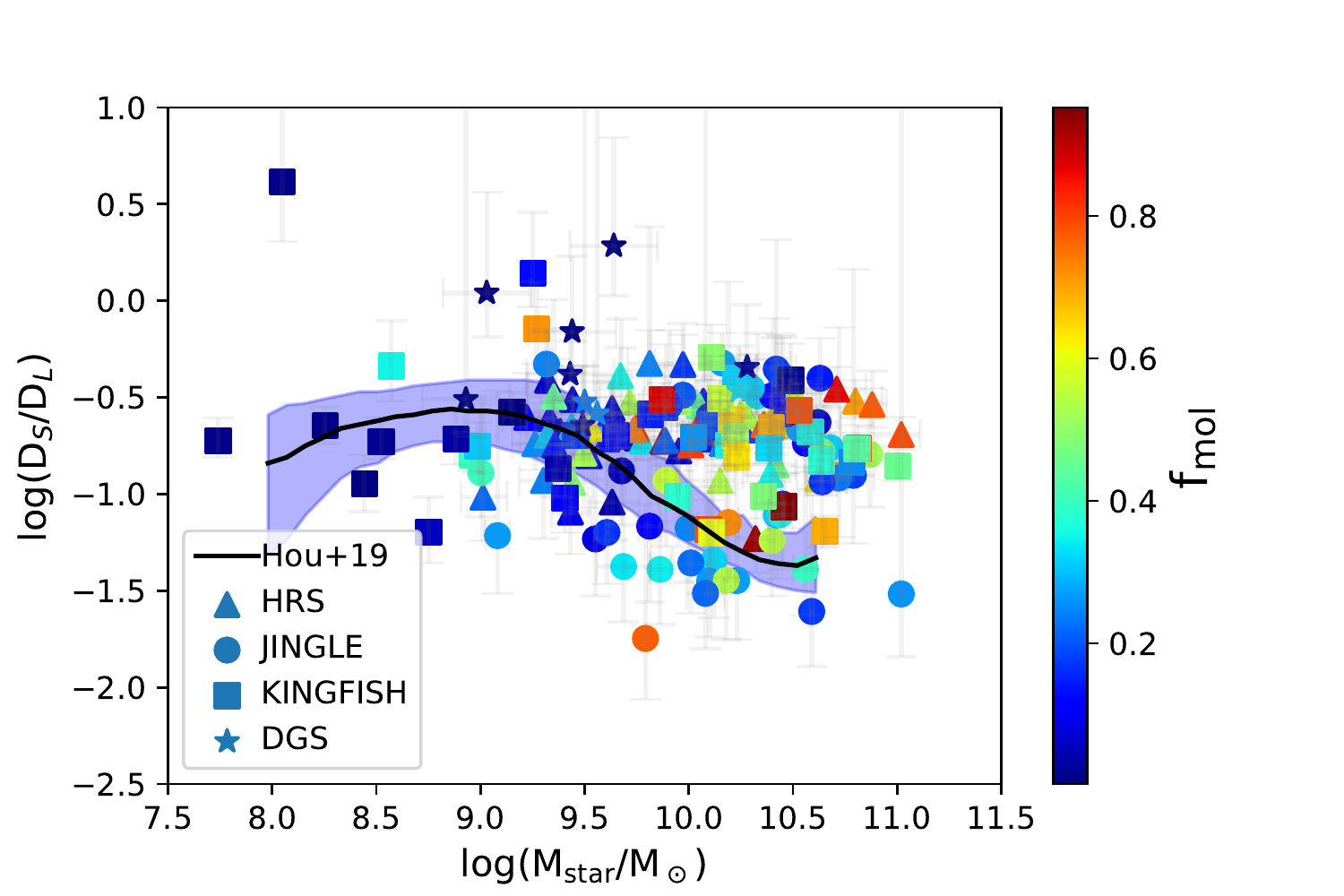}
\caption{\DstoDl\ versus metallicity (left) and stellar mass (right) colour coded with molecular gas mass fraction for our galaxy sample: JINGLE (circles), KINGFISH (squares), DGS (stars), HiGH (upside down triangles) and HRS (triangles). We only include galaxies with molecular gas mass measurements in the literature. Median values for the errors in $\log(D_{S}/D_{L})$ of the galaxies having $-0.6<\log(D_{S}/D_{L})<-0.4$ are: $+$0.09 and $-$0.17\,dex. In these galaxies coagulation is not an important mechanism affecting the dust evolution.}\label{fig:S2L_metfmol}
\end{figure*}

In Fig.\,\ref{fig:S2L_metfmol} we show \DstoDl\ versus metallicity (left) and stellar mass (right) colour coded with the molecular gas mass fraction ($\rm f_{\rm mol} = M_{H_{2}}/(M_{HI}+M_{H_{2}})$. We do not have molecular gas mass estimates (those derived from CO observations) for all our galaxy sample, therefore in these figures only those galaxies with CO observations reported in the literature have been included. There are no estimates of molecular gas masses for HiGH galaxies, therefore this sample has not been included in these plots. All the molecular gas masses have been obtained using a Milky Way CO-to-H$_{2}$ conversion factor ($X_{\rm CO}=2.0\times10^{20}\rm cm^{-2} (K\,km\,s^{-1})^{-1}$, \citealt{2013ARA&A..51..207B}). Those galaxies with molecular gas mass estimates in the literature with other $X_{\rm CO}$ factors were recalculated with the Milky Way $X_{\rm CO}$ to obtain an homogeneous data set. 

In the left panel of Fig.\,\ref{fig:S2L_metfmol} we see that in general there is trend of declining \DstoDl\ ratio with high values of f$_{\rm mol}$ (and high metallicity). The DGS galaxies with low metallicity show high \DstoDl\ values and low molecular gas mas fractions (f$_{\rm mol}$). For galaxies with high metallicity, high molecular gas mass fractions and low values of \DstoDl\ ratio, coagulation might be the dominant process affecting the dust evolution as the coagulation time scale depends on the inverse of the dense gas mass fraction \citep[e.g.][]{2017MNRAS.466..105A}. There are however some galaxies at high metallicities having all a similar value of $\log(D_{S}/D_{L})\sim-0.5$. These galaxies are clearly separated in the \DstoDl\ stellar mass relation (right panel of Fig.\,\ref{fig:S2L_metfmol}). They have high metallicity, high molecular gas mass fractions, and {\it high} \DstoDl\ values. In these galaxies, either accretion or shattering (or both mechanisms) might be dominating over coagulation despite their high molecular gas mass fractions. Alternatively, metal diffusion could enhance the amount of large grains in the diffuse medium and, therefore, increase the efficiency of shattering and the total amount of small grains. It is interesting to note that applying a metallicity dependence of the CO-to-H$_{2}$ conversion factor \citep{2020A&A...643A.180H}, which would change our molecular gas mass estimates, we obtain the same results as those presented here. In Appendix\,\ref{app:XCO} we can see the same figures but with a metallicity dependence of $X_{\rm CO}$ from \citet{2020A&A...643A.180H}. 

\subsection{ISRF heating the dust}\label{sec:isrf}

We have seen in Fig.\,\ref{fig:S2L:Md} that there is a significant dispersion in the \stol\ grain mass ratios for dusty galaxies. Some galaxies follow the predictions from simulations while others tend to have a similar \DstoDl\ ratio. We explore in this section if the ISRF heating the dust could have an impact in this trend. An intense ISRF, related to massive star formation, would heat the dust to higher temperatures. The massive star formation could also lead to a higher rate of SNe with the corresponding dust processing either in the form of dust grain destruction or shattering of large grains into small ones. 

\begin{figure*}
\includegraphics[width=0.45\textwidth]{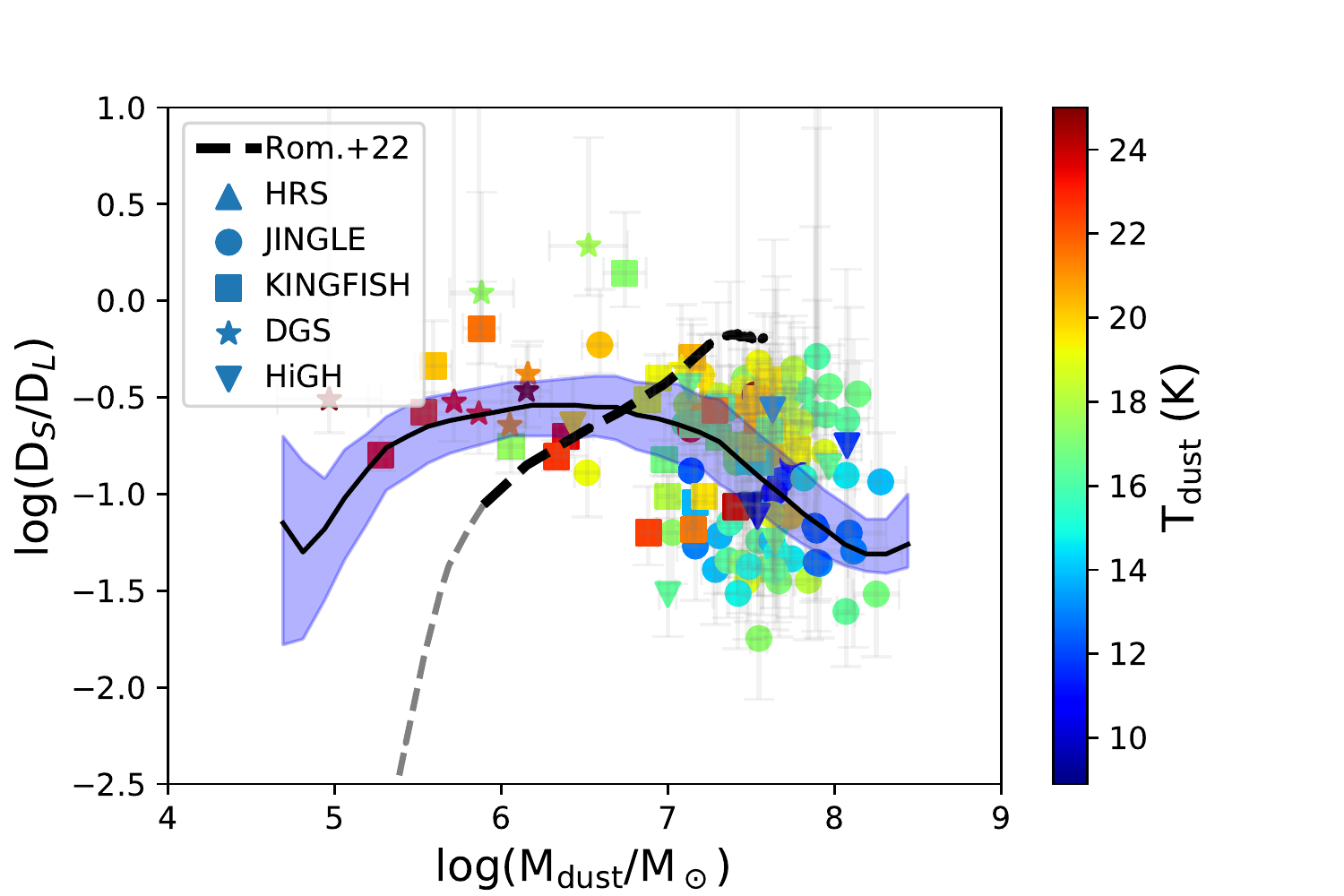}
\includegraphics[width=0.45\textwidth]{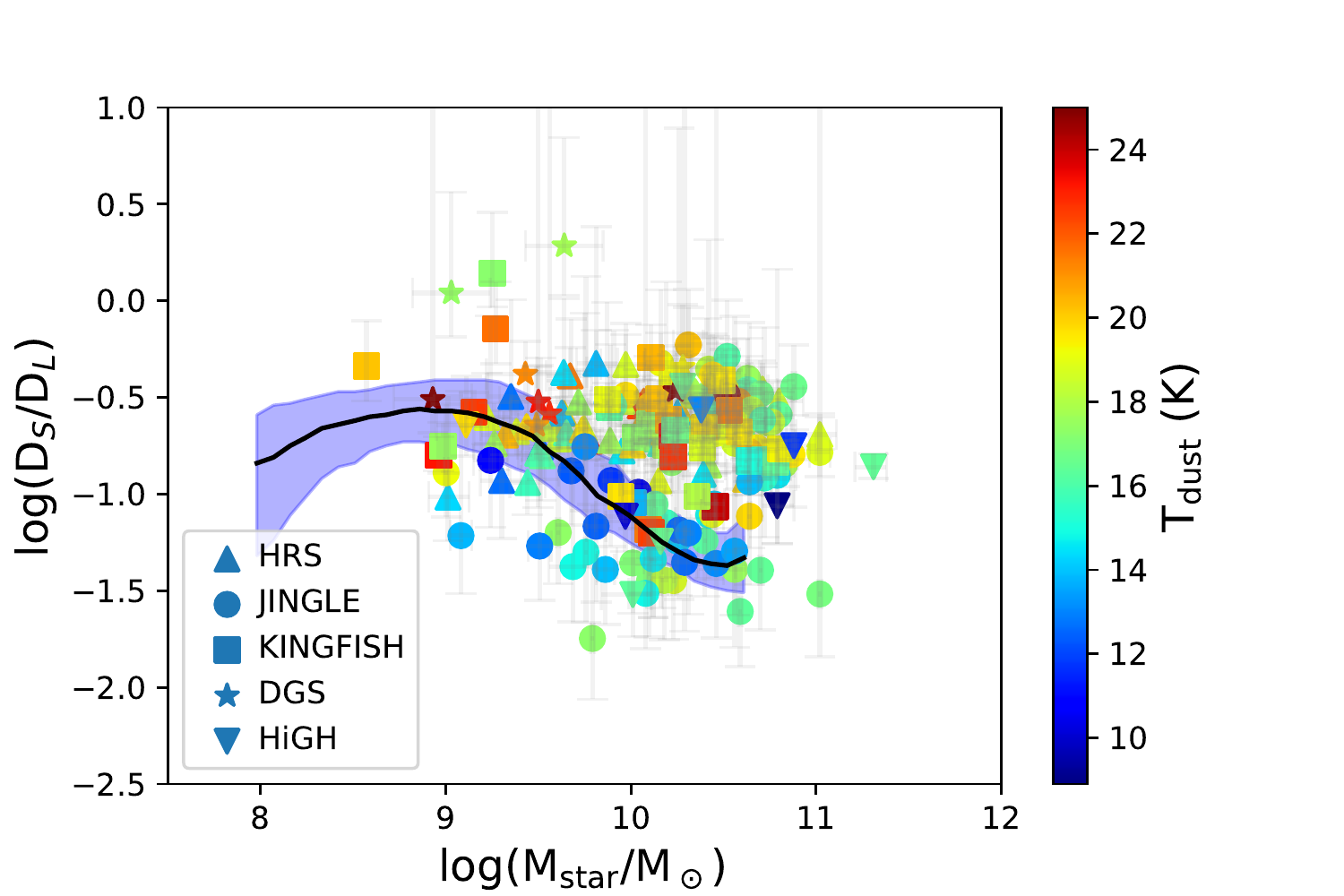}
\caption{\DstoDl\ versus dust mass (left) and stellar mass (right) colour coded with an estimation of the dust temperature, $T_{\rm dust}$. This has been derived using the relation: $T_{\rm dust}=18\times\ U_{\rm min}^{1/(4+\beta)}$, with $\beta$\,=\, 2 and the normalisation of \citet{2014ApJ...780..172D} for $U$ = 1 (see Section\,\ref{sec:isrf}). Symbols are the same as in previous figures. Galaxies with high values of $\log(D_{S}/D_{L})$ also have high T$_{\rm dust}$. The relative increase of small grain mass fraction in these galaxies could be due to a combined effect of a more intense SFR, traced by the high T$_{\rm dust}$ in these systems, and metal diffusion (see Section\,\ref{sec:isrf}).}\label{fig:S2LvMd_Td}
\end{figure*}

In the left panel of Fig.\,\ref{fig:S2LvMd_Td} we show the \stol\ grain mass ratio versus the dust mass colour coded with an indication of the dust temperature estimated using the values of $\rm U_{\rm min}$ obtained from the SED fitting ($T_{\rm dust}=18\times\ U_{\rm min}^{1/(4+\beta)}$, with $\beta$\,=\, 2). Galaxies with low dust masses tend to have higher dust temperatures. In the high dust mass regime there is a spread in dust temperature, with some galaxies presented low temperatures, T$_{\rm dust}\sim\,12-14$\,K, and other having T$_{\rm dust}\sim\,18-20$\,K . In the right panel of Fig.\,\ref{fig:S2LvMd_Td} we show the \stol\ grain mass ratio versus the stellar mass. The separation in the dust temperature is more visible in this panel. Galaxies having $\log(D_{S}/D_{L})\sim-0.5$ tend to have T$_{\rm dust}$ typically above 20\,K, while galaxies with lower \DstoDl\ values and following the predictions of the simulations present in general  lower dust temperatures (T$_{\rm dust}\sim\,15-18$\,K).

If we assume that dust temperature is related to star formation\footnote{In Fig.\,\ref{appfig:sfms} we show that the galaxies having high  \DstoDl\ tend to be in the upper area of the SFR$-$stellar mass relation.}, this separation in temperatures agrees with the scenario proposed above where in galaxies with high dust masses and high dust temperatures shattering is important. Together with accretion, shattering can balance coagulation even in galaxies with high molecular gas mass fractions, giving as a result higher values of \DstoDl.
The simulations of an isolated galaxy from \citet{2022MNRAS.tmp.1342R} incorporating metal diffusion, which has the effect of transporting large grains from the dense, star-forming regions into the diffuse medium where they can be efficiently shattered into small grains, reproduce very well the high  \DstoDl\ ratios observed here. Indeed, \citet{2022MNRAS.tmp.1342R} simulations of an individual isolated galaxy predict two separated branches in \DstoDl$-$metallicity distribution, one at high \DstoDl\ ratio, corresponding to the disk of the galaxy, and another branch corresponding to dust in the circumgalactic medium with lower values of $\log(D_{S}/D_{L})\lesssim-1.0$. We see here that low \DstoDl\ ratios are related to low dust temperatures, which suggests that the dust in the circumgalactic medium would be somewhat colder than the dust in the galaxy disk. 

Finally, we note that in general high T$_{\rm dust}$ and high \DstoDl\ ratio tend to enhance the emission at short wavelengths, therefore there is also the possibility that our fitting procedure could give somehow a possible degeneracy between \DstoDl\ ratio and the ISRF, that might produce an artificial relation between \DstoDl\ ratio and T$_{\rm dust}$. We believe this is not the case as we do not see a relation between the \DstoDl\ ratio and T$_{\rm dust}$ in Fig.\,\ref{app:compISRFRR15} when a multi-ISRF approach is used.

\section{Conclusions}\label{sec:conc}

We have fitted the spectral energy distribution of a sample of 247 local galaxies separating the emission of three different dust components: PAHs, very small grains and large grains. The galaxy sample has been extracted from a set of galaxy surveys covering a wide range of physical properties. The mass of the PAHs and the very small grains in our dust model have been added to represent the total mass of small grains. With this definition we have derived \stol\ grain mass ratios  (\DstoDl) for each galaxy of our sample and we have compared our results with the predictions from simulations. The comparison has helped us to analyse the different mechanisms that dominate the evolution of interstellar dust. The main conclusions of this study are: 

\begin{itemize}
\item[--]  Cosmological simulations from \citet{2019MNRAS.485.1727H} reproduce well the dust-to-gas (D/G)$-$metallicity relation observed for our galaxy sample. However, at high stellar masses the D/G ratio obtained from the simulations is above the observed values. We show that the overestimation of the D/G in the high mass regime indicates that the dust mass is overestimated by the simulations in this mass range, as gas masses in the simulations seem to reproduce well the observed values. An overestimation of the dust mass by simulations is supported by the larger dust-to-star (D/S) values predicted from the simulations in comparison with the observed D/S ratios.  

\item[--] The values of \DstoDl\ obtained from the observed SED fitting are in general within the predictions of the simulations. The relation between \DstoDl\ and stellar mass reveals a set of galaxies with high stellar mass showing \DstoDl\ within a narrow range of values ($\log_{10}(D_{S}/D_{L})\sim-0.5$). These galaxies deviate from the results of simulations which predict lower \DstoDl\ values at these high stellar masses, while the rest of the sample follows nicely the behaviour of \DstoDl\ with stellar mass and also with dust mass. 

\item[--] We analyse further the sub-sample of galaxies with similar values of $\log_{10}(D_{S}/D_{L})\sim-0.5$. These galaxies tend to have high metallicity and {\it high} molecular gas mass fractions in comparison with the rest of the galaxy sample. Despite of the high molecular gas mass fraction, coagulation does not seem to be an important mechanism affecting the dust evolution  in these galaxies. Either accretion or shattering (or both mechanisms) might be dominating over coagulation to produce the high \DstoDl\ ratios. 

\item[--] Using a proxy for the dust temperature based on the scale of the ISRF derived from our SED fitting we find that dust seems to be hotter in the galaxies having similar values of $\log_{10}(D_{S}/D_{L})\sim-0.5$ than in the other objects of our sample. We suggest that in these galaxies shattering might be an efficient mechanism to convert large grains into smaller ones. The simulations of an isolated galaxy from \citet{2022MNRAS.tmp.1342R} including metal diffusion, which is able to transport large grains from dense regions to a more diffuse medium where they can be easily shattered, gives  \DstoDl\ ratios in agreement with the high values observed for this sub-sample of galaxies, however a larger sample of simulated galaxies will be needed in order to check whether or not their model can explain the full range of observed galaxies. 
\end{itemize}

We have presented here a comparison of the  \stol\ grain mass ratio inferred from observed SEDs and those obtained from simulations that include a treatment for the evolution of the dust grain size distribution. The comparison allows us to explore the magnitude of the different mechanisms in our galaxy sample. Based on this comparison we also highlight here some possible additions that could be taken into account in future simulations. 

A detailed AGN feedback prescription and a more sophisticated prescription for the SFR might eventually be useful to obtain results that agree better with the observed SFMS relation and the mass$-$metallicity relation, but also with the relations presented here. A more accurate calibration of the dense gas mass fraction in the simulations would be needed in order to accurately describe the trend of the \stol\ grain mass ratio with other galaxy properties obtained from observations. Finally, extra mechanisms such as metal diffusion might be very helpful to reproduce the \stol\ grain mass ratio even in places where the dense and cold medium that might favour the formation of large grains. We highlight here the use of the \DstoDl\ ratio to infer the mechanisms that shape the evolution of the interstellar dust in galaxies with different physical properties.

\section*{Acknowledgements}
MR and UL acknowledge support from project PID2020-114414GB-100, financed by MCIN/AEI/10.13039/501100011033, project AYA2017-84897-P from the Spanish Ministerio de Economía y Competitividad, project P2000334 and A-FQM-510-UGR20 financed by the Junta de Andalucía and from FEDER/Junta de Andalucía-Consejería de Transformación Económica, Industria, Conocimiento y Universidades. IDL acknowledges support from ERC starting grant 851622 DustOrigin. We are grateful to Volker Springel for providing the original version of GADGET-3, on which the GADGET3-Osaka code is based. Our numerical simulations and analyses were carried out on the XC50 systems at the Center for Computational Astrophysics (CfCA) of the National Astronomical Observatory of Japan (NAOJ), Octopus at the Cybermedia Center, Osaka University, and Oakforest-PACS at the University of Tokyo as part of the HPCI System Research Project (hp200041, hp210090). This work is supported in part by the JSPS KAKENHI Grant Number  JP17H01111, 19H05810, 20H00180 (KN). KN acknowledges the travel support from the Kavli IPMU, World Premier Research Center Initiative (WPI), where part of this work was conducted. HH thanks the Ministry of Science and Technology (MOST) for support through grant MOST 107-2923-M-001-003-MY3 and MOST 108-2112-M-001-007-MY3, and the Academia Sinica
for Investigator Award AS-IA-109-M02. YG acknowledges funding from National Key Basic Research and Development Program of China (Grant No. 2017YFA0402704). IL acknowledges support from the Comunidad de Madrid through the Atracción de Talento Investigador Grant 2018-T1/TIC-11035.

\section*{Data Availability}
The datasets were derived from sources in the public domain: \url {http://www.star.ucl.ac.uk/JINGLE/data.html}, \url{http://egg.astro.cornell.edu/alfalfa/data/}, \url{https://cdsarc.cds.unistra.fr/viz-bin/cat/J/PASP/122/261},  \url {https://irsa.ipac.caltech.edu/applications/Scanpi/}, \url{https://cdsarc.cds.unistra.fr/viz-bin/cat/J/A+A/609/A37}, \url {https://cdsarc.cds.unistra.fr/viz-bin/cat/J/A+A/582/A121}, \url{https://cdsarc.cds.unistra.fr/viz-bin/cat/J/ApJS/233/22}.

%%%%%%%%%%%%%%%%%%%% REFERENCES %%%%%%%%%%%%%%%%%%
\bibliographystyle{mnras}
\bibliography{references_v1} % 
%%%%%%%%%%%%%%%%%%%%%%%%%%%%%%%%%%%%%%%%%%%%%%%%%%

%%%%%%%%%%%%%%%%% APPENDICES %%%%%%%%%%%%%%%%%%%%%

\appendix
\section{Comparison of dust masses with previous studies and results from a single ISRF}\label{app:comp}

\begin{figure}
\includegraphics[width=0.5\textwidth]{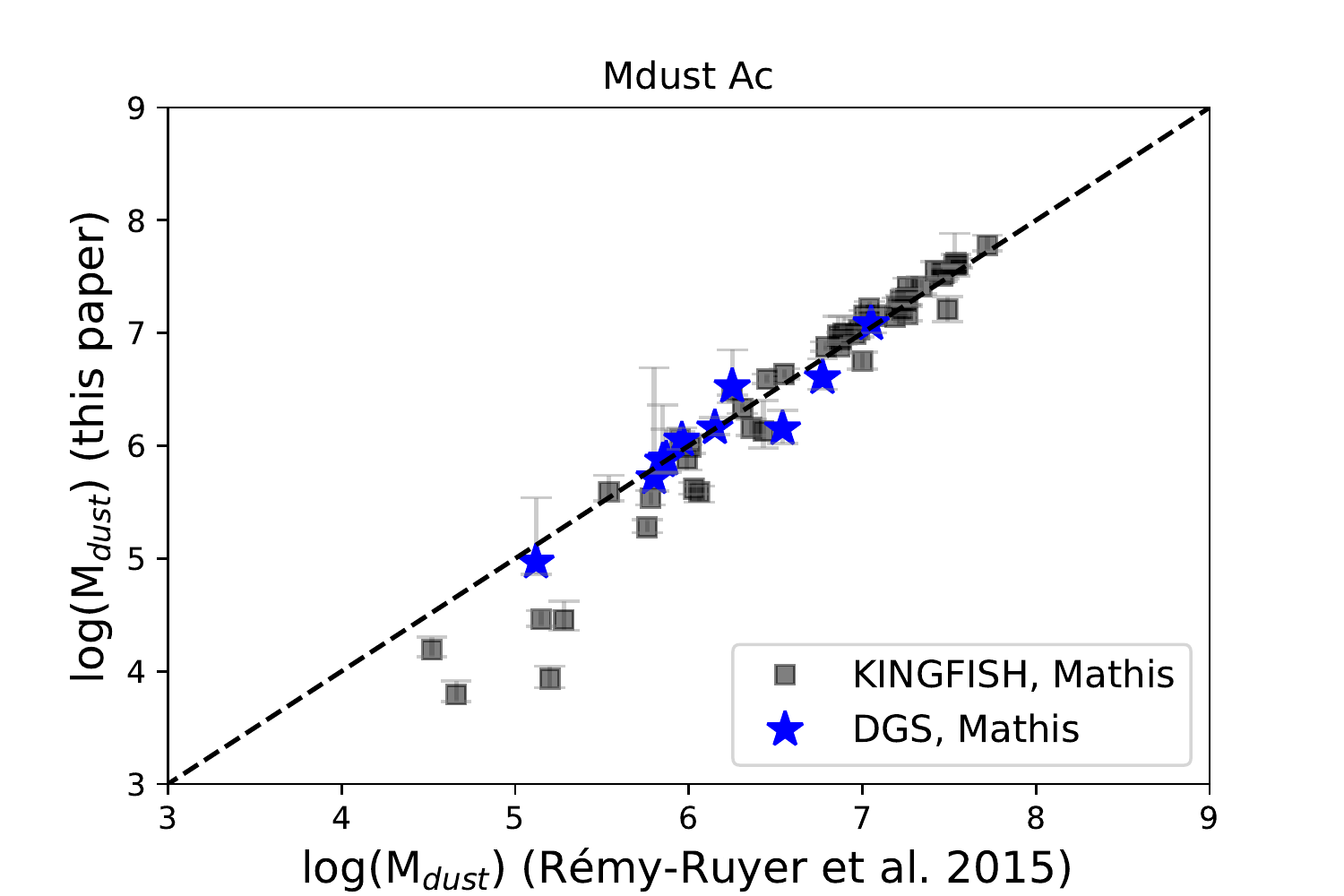}
\includegraphics[width=0.5\textwidth]{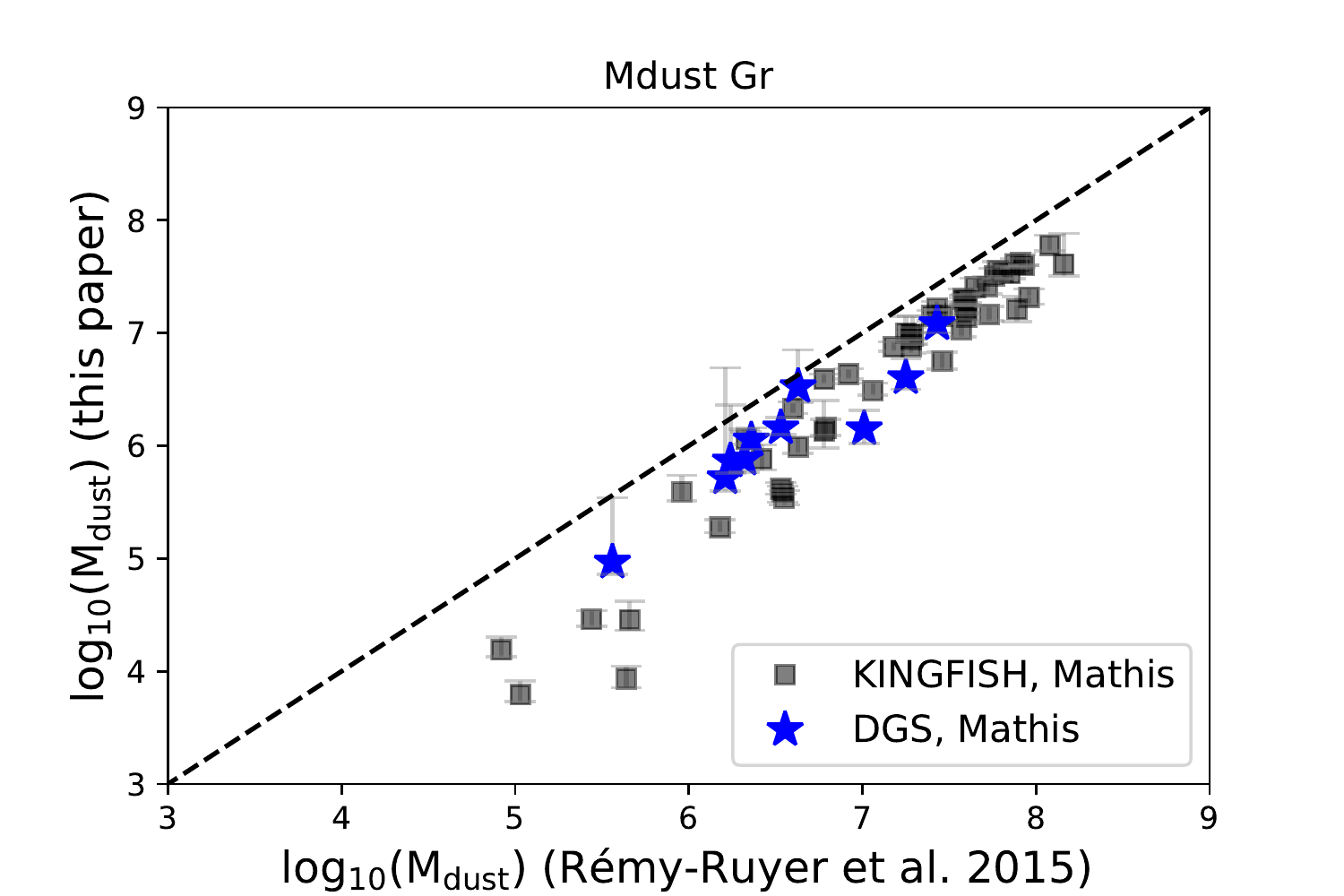}
\caption{Comparison of the dust masses derived in this paper with the two dust models, graphite ($Gr$) and amorphous carbon ($Ac$) grains used in \citet{2015A&A...582A.121R}. Black squares correspond to KINGFISH galaxies while blue stars represents the DGS galaxies considered in our sample. The mean values of the differences in the dust masses derived from both $Gr$($Ac$) dust models are 0.50(0.09)\,dex and 0.45(0.04)\,dex for KINGFISH and DGS galaxy samples, respectively.}\label{app:compMdRR15}
\end{figure}

In this section we compare our dust mass estimations with those from the literature. For DGS and KINGFISH galaxies we compare with results in \citet{2015A&A...582A.121R}. The dust masses were obtained using different dust models and the same fitting approach as the one  presented here. The dust mass is distributed in different mass elements heated by a certain ISRF following the prescription from \citet{2001ApJ...549..215D}. The authors adopted the same strategy with two different dust models: one with a grain composition made of silicates, carbon grains in the form of graphite and PAHs (BARE-GR-S dust model in \citealt{2004ApJS..152..211Z}) where the relative contribution of the PAH component is varied while the graphite-to-silicate ratio is kept fixed; and a second one where the graphite grains are changed to amorphous carbon grains with optical properties obtained from \citet{1996MNRAS.282.1321Z}.  
In Fig.\,\ref{app:compMdRR15} we show the dust masses derived with our dust model for the DGS and KINGFISH galaxies with the estimates from \citet{2015A&A...582A.121R} for the two dust models $Gr$ (based on graphite) in the bottom panel and $Ac$ (based on amorphous carbon grains) in the top panel. We find very good agreement over 4 orders of magnitude in dust masses with the results obtained using the $Ac$ dust model in \citet{2015A&A...582A.121R}. However, we underestimate the dust masses derived from the $Gr$ dust model. As pointed in \citet{2015A&A...582A.121R} these authors found a factor of $\sim 2-2.5$ difference between the dust masses derived from the dust models, with $Gr$ dust masses being systematically higher. The differences are due to the fact that amorphous carbon dust is more emissive in the submillimetre wavelength range and therefore less dust amount is needed to match the same IR luminosity. In our case, allowing the dust mass of the VSG to vary gives a better agreement with the $Ac$ dust model in \citet{2015A&A...582A.121R}. We note that variations of 1.4 in dust mass estimates from different dust models is normally expected \citep{2021ApJ...912..103C}.   

\begin{figure}
\includegraphics[width=0.5\textwidth]{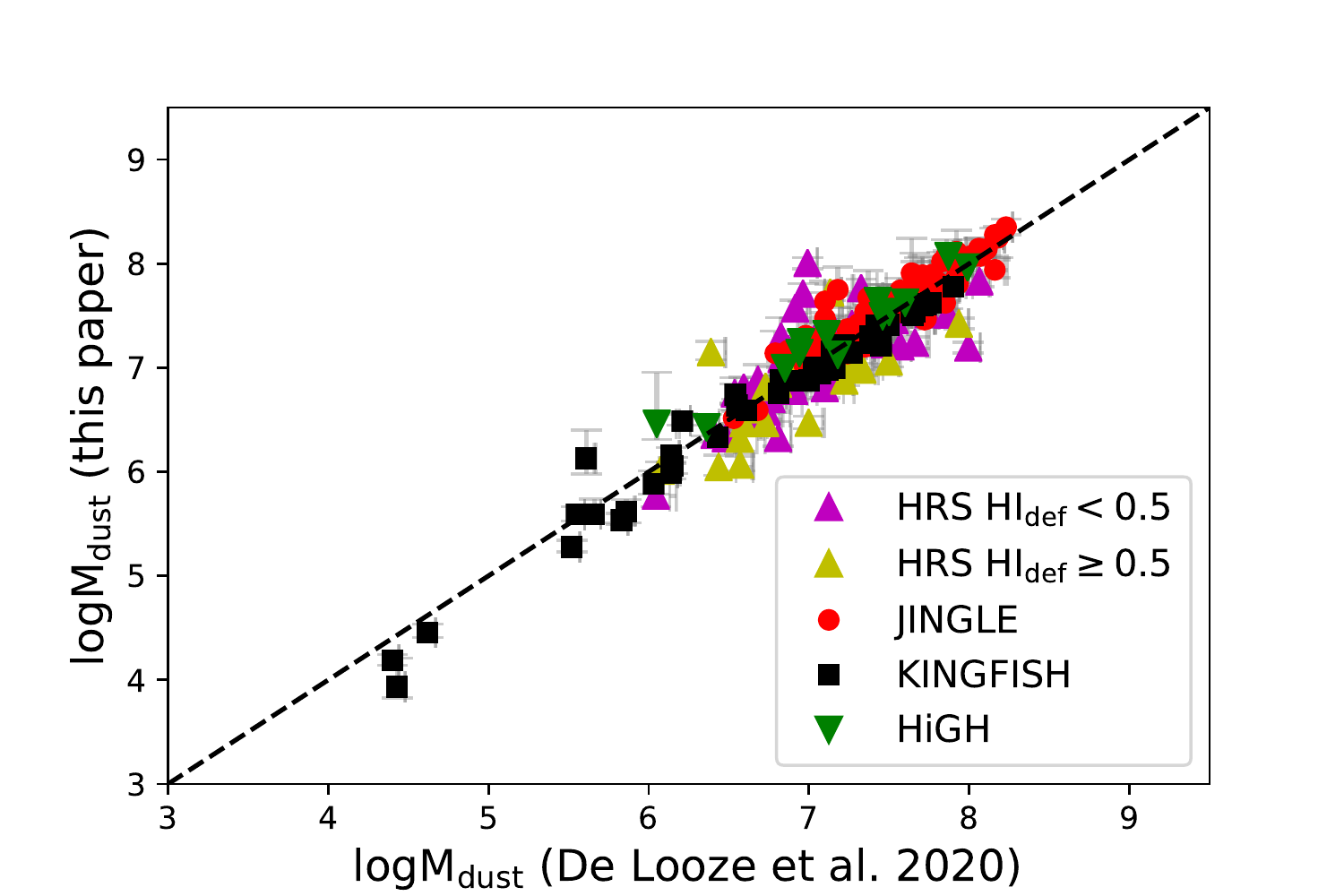}
\caption{Comparison of the dust masses for our galaxy selection from JINGLE (red dots), HiGH (green upside down triangles), KINGFISH (black squares) and HRS (magenta and triangle stars) derived in this study with those obtained in \citet{2020MNRAS.496.3668D}. The mean values of the differences in the dust masses derived by  \citet{2020MNRAS.496.3668D} and those obtained in this studty are  0.25, 0.12, 0.13, 0.15\,dex for HRS, JINGLE,  KINGFISH and HiGH galaxy samples, respectively. \citet{2020MNRAS.496.3668D} did not include the DGS in their sample.}\label{app:compMdDLooze20}
\end{figure}

Dust masses for JINGLE, HiGH and KINGFISH were derived in \citet{2020MNRAS.496.3668D} with the similar strategy of a multi-component ISRF heating the dust grains as the one presented here. The dust model used by these authors is the THEMIS dust model presented in \citet{2017A&A...602A..46J}. In Fig.\,\ref{app:compMdDLooze20} we compare the dust masses derived here with those derived in \citet{2020MNRAS.496.3668D}. We find very good agreement for all the galaxy samples considered in this paper covering four order of magnitudes in dust masses and a wide range of galaxy properties.   

\begin{figure}
\includegraphics[width=0.5\textwidth]{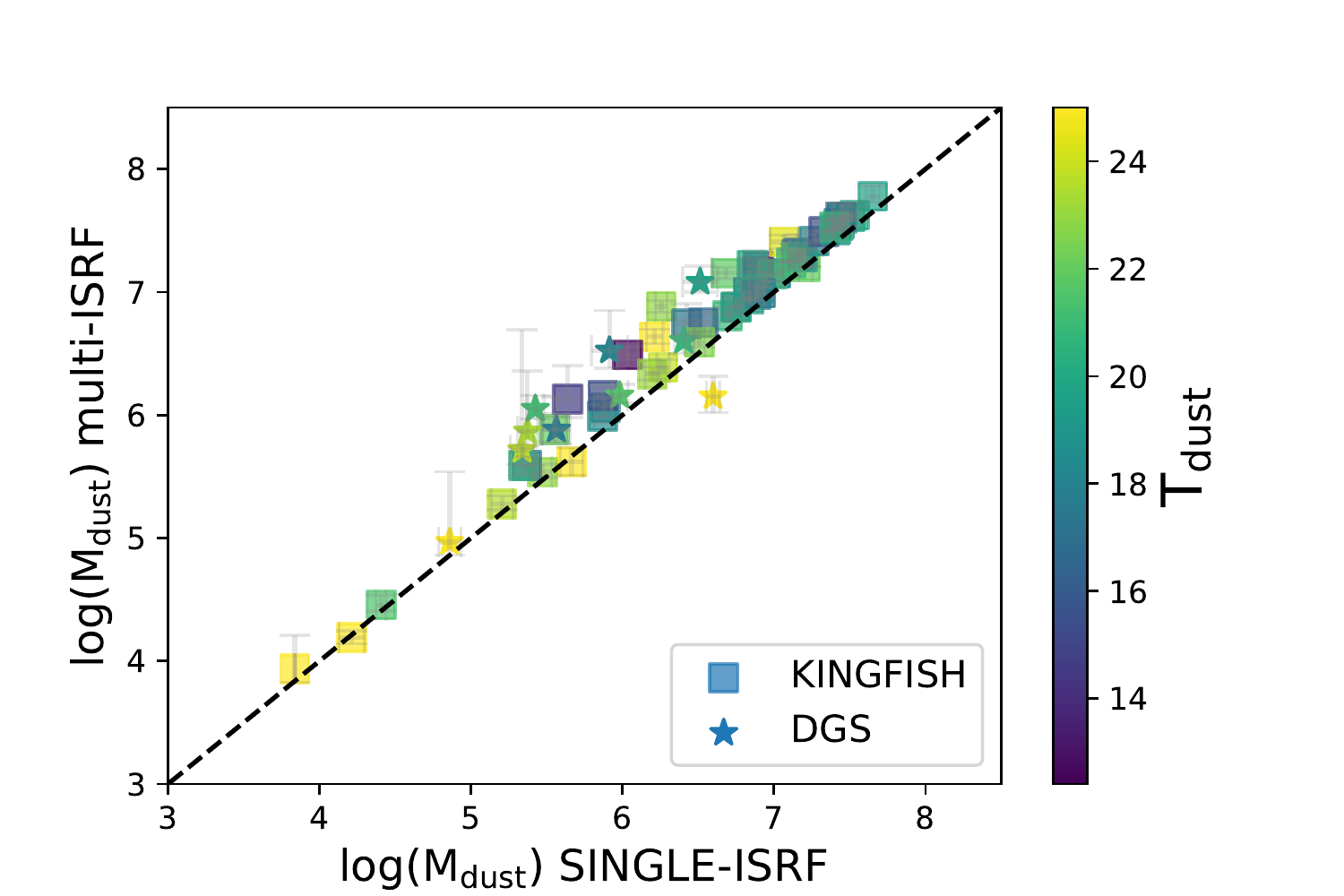}
\includegraphics[width=0.5\textwidth]{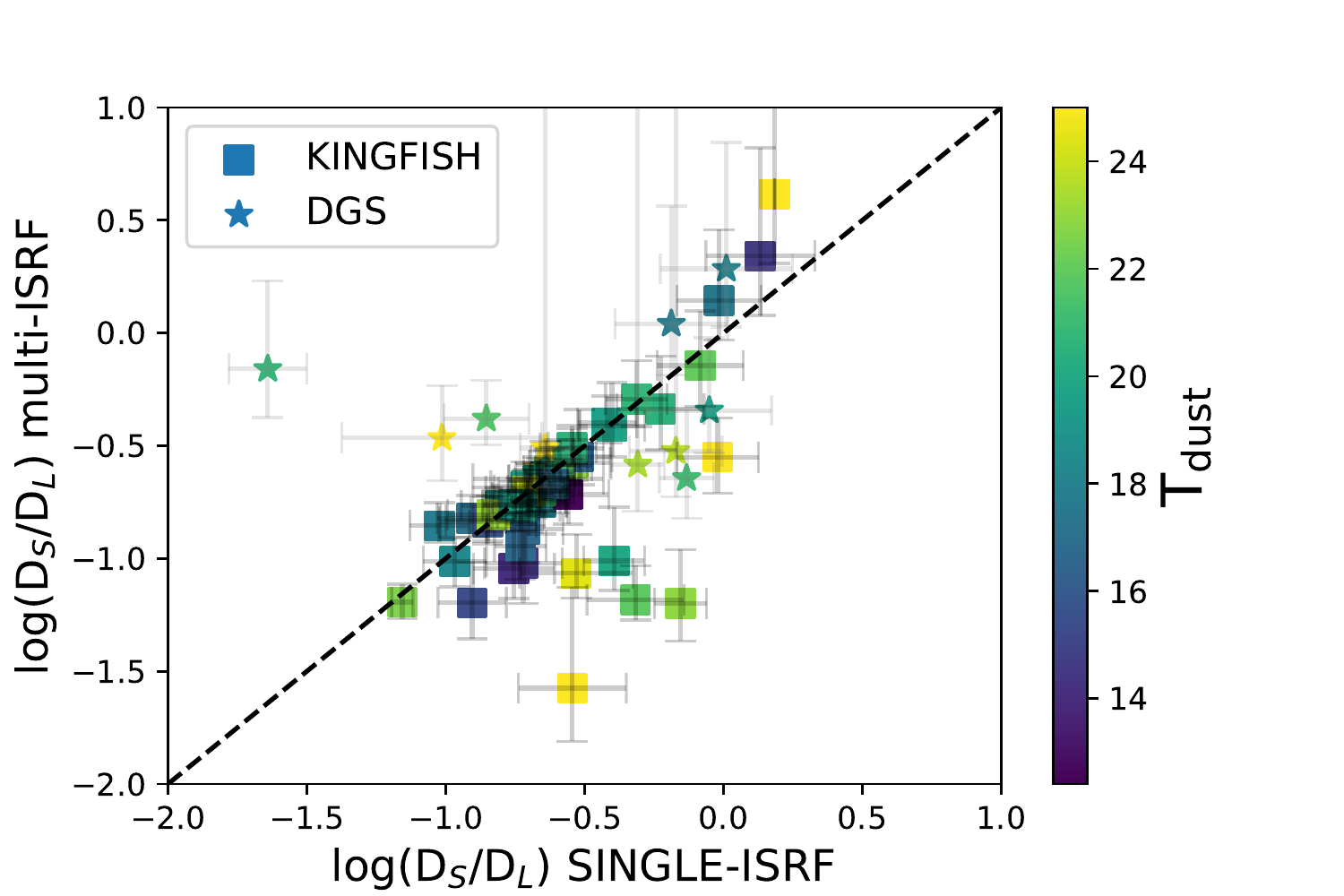}
\caption{Comparison of the dust masses (top panel) and the \stol\ grain mass ratio (\DstoDl, bottom panel) derived using a single ISRF (as derived in \citet{2020A&A...636A..18R}), and the multi-ISRF approach presented in this study. Colour bar corresponds to the dust temperature assuming the value of $U_{\rm min}$  derived from our best fits. }\label{app:compISRFRR15}
\end{figure}

We also compare our results, including the parametrisation introduced by \citet{2001ApJ...549..215D} where the dust mass elements heated by the ISRF are distributed in a power-law form, with previous results from \citet{2020A&A...636A..18R} using a single ISRF to fit the dust emission. In \citet{2020A&A...636A..18R} it was found that a significant number of galaxies were not properly fit with the strategy of a single ISRF, and therefore those galaxies having residuals in the 24\,\mi\ and 70\,\mi\ bands above 35\% were removed from the final sample. In Fig.\,\ref{app:compISRFRR15} we compare our results with those presented in \citet{2020A&A...636A..18R} derived using a single ISRF. The comparison is restricted to the galaxy sample common to both studies: 53 and 10 galaxies from KINGFISH and DGS samples. In the top panel of Fig.\,\ref{app:compISRFRR15} we see the comparison for the dust mass with a colour code indicating the dust temperature using the relation: $T_{\rm dust}=18\times\ U_{\rm min}^{1/(4+\beta)}$, with $\beta$\,=\, 2 and the normalisation of \citet{2014ApJ...780..172D} for $U$ = 1. The dust masses agree relatively well (the mean value of the difference in dust masses between the two methods is 0.18\,dex for KINGFISH galaxies and 0.39\,dex for galaxies in the DGS sample) with no relation between the deviations from the one-to-one relation and the dust temperature. 

The comparison of the \stol\ grain mass ratio when a single and multi-ISRF is assumed is especially interesting. \citet{2018ARA&A..56..673G} pointed to a degeneracy between the ISRF distribution and the mass fraction of small grains, in the sense that a single ISRF with a high mass fraction of small grains can fit the same SED as a multi-ISRF representing hotter environments (see figures 3b and 3c in \citealt{2018ARA&A..56..673G}). In the bottom panel of Fig.\,\ref{app:compISRFRR15} we show the \stol\ grain mass ratio (\DstoDl) derived with a single ISRF ($x$-axis) in \citet{2020A&A...636A..18R} and \DstoDl\ obtained using the approach applied in this paper, a multi-ISRF component ($y$-axis). The colour code represents the dust temperature estimated as explained above and the comparison is done for the galaxy sample common to both studies. There is agreement between the \DstoDl\ derived from the two approaches for a high fraction of galaxies. From 53 and 10 galaxies from KINGFISH and DGS samples, respectively, 37 and 7 show differences in the  \DstoDl\ obtained using a single and multi-ISRF that are less than the mean value of the uncertainties in the  \DstoDl\ obtained from the fit. However, there are outliers in the distribution mainly located in the lower side of the one-to-one correlation: 6 galaxies from KINGFISH sample and 3 from DGS, show differences in \DstoDl\ that are larger than 0.5\,dex. These galaxies have in general higher values of \DstoDl\ when a single ISRF is used than when a multi-ISRF approach is considered. This result goes in the same direction as the scenario claimed by \citet{2018ARA&A..56..673G}. We also see some hints that the galaxies outside the correlation tend to have slightly hotter dust than those on the one-to-one correlation. 

\section{Isolated disk galaxy simulation by GADGET4-Osaka code}\label{app:Rom}
Figs.~\ref{fig:D2G} -- \ref{fig:S2L} feature the evolutionary tracks from a new simulation of an isolated galaxy that has been performed with GADGET4-Osaka  \citep{2022MNRAS.tmp.1342R,2022MNRAS.tmp.1338R}, a modified version of the massively parallel TreeSPH/N-body cosmological hydrodynamic code GADGET-4 \citep{2021MNRAS.506.2871S}. In this Appendix we describe the setup and some of the main differences between this simulation and the ones performed by \citet{2017MNRAS.466..105A} and \citet{2017MNRAS.469..870H}. 

For the initial conditions (ICs), \citet{2022MNRAS.tmp.1342R,2022MNRAS.tmp.1338R} use the low-resolution isolated galaxy ICs from the AGORA code comparison project described by \citet{2016ApJ...833..202K}, but additionally they employ a hot gaseous halo component which was initialised by randomly sampling 40\,\% of the DM halo particles, mirroring them through the origin and assigning them as gas particles with particle mass set equal to the gas particle mass and temperature equal to $T_\text{halo} = 10^6$\,K

\citep{2021ApJ...917...12S}. The ICs feature a collisionless NFW halo with $R_{200} = 205.5\,\text{kpc}$, $M_{200} = 1.074 \times 10^{12} \text{M}_\odot$, $c = 10$ and $\lambda = 0.04$ \citep{1996ApJ...462..563N} enclosing a baryonic disk with stellar bulge. The disk is following an exponential density profile with scale radius $r_d = 3.432\, \text{kpc}$ and scale height $z_d = 0.1 r_d$ that is composed of a gas component making up 20 \% of its mass and a stellar component making up the rest. The bulge follows a Hernquist profile \citep{1990ApJ...356..359H} with bulge-to-disk ratio of $0.1$. The mass resolution and particle numbers are listed in Table\,\ref{tab:diskICs}. They employ a gravitational softening length of $\epsilon_\text{soft} = 80 \text{pc}$ and do not allow SPH smoothing lengths to drop below 10 \% of this value.
\begin{table}
\centering
\caption{Initial Conditions of Isolated Disk Galaxy in \citep{2022MNRAS.tmp.1342R,2022MNRAS.tmp.1338R}}
\label{tab:diskICs}
\begin{tabular}{lcc}\hline
Component & Mass resolution $\left[M_{\odot}\right]$ & N \\
\hline 
Gas (disk \& halo) & $8.593 \times 10^4$ & $1.4 \times 10^5$\\
DM halo & $1.254 \times 10^7$ & $10^5$\\
Stars (disk) & $3.4373 \times 10^5$ & $10^5$\\
Stars (bulge) & $3.4373 \times 10^5$ & $1.25 \times 10^4$ \\
\hline
\end{tabular}
\end{table}

They basically follow the stellar feedback and star formation prescriptions in GADGET3-Osaka code as described by \citet{2019MNRAS.484.2632S}. For the dust evolution, a modified version of the model described by \citet{2019MNRAS.482.2555H} was employed. Two minor modifications have been made addressing the underproduction of small grains and the too small amount of coagulation, which was reported in \citet{2019MNRAS.482.2555H}.
In order to address the latter, \citet{2022MNRAS.tmp.1342R,2022MNRAS.tmp.1338R} loosened the threshold for gas to host dense molecular clouds in order to reach higher global dense gas fractions, comparable to that of a Milky-Way-like galaxy ($\sim 20\,$\%, \citealt{2018MNRAS.476..875C}). 
Furthermore, they employed a model of diffusion following the Smagorinsky-Lilly model \citep{1963MWRv...91...99S}, which has an effect of smoothing the dust and metal distribution.
The implementation is inspired by the work of \citet{2010MNRAS.407.1581S}, who also modelled the diffusion of metals using the same subgrid turbulence model. 
In this model the large grains produced in the dense, star-forming regions are transported to the diffuse medium where they can efficiently shatter into smaller grains, globally boosting their production.  Further details of the treatment is described in \citet{2022MNRAS.tmp.1342R,2022MNRAS.tmp.1338R}.

As the resolution of simulation gets better, we expect that the simulations will better resolve the shear and turbulence, and the metal diffusion will be better captured naturally by the simulation. However, there will always be sub-resolution scales which we cannot resolve (e.g. sub-parsec scales) in galaxy simulations, and a subgrid diffusion model will probably remain necessary. In the future, we need to compare our results with high-resolution ISM simulations and seek for optimal resolution and parameters for the subgrid diffusion model.

\section{Molecular gas masses derived from scaling relations}\label{app:compCasaXCOLD}

We show in this section how we derived $\rm M_{\rm H_2}$ from the relation between $\rm M_{\rm HI}$/$\rm M_{\rm star}$ and $\rm M_{\rm H_2}/M_{\rm HI}$ presented in Eq.\,5 in \citet{2020A&A...633A.100C}. We estimate gas masses only for those galaxies that are within the range where this scaling relation was derived. Therefore, to be conservative, the estimations are done for our galaxies satisfying $-2.0<\log(\rm M_{\rm H_2}/M_{\rm HI})<0$. We use $\rm M_{\rm HI}$/$\rm M_{\rm star}$ derived from observations to estimate $\rm M_{\rm H_2}/M_{\rm HI}$ applying Eq.\,5 in \citet{2020A&A...633A.100C}. Then, $\rm M_{\rm H_2}$ is derived by multiplying with $\rm M_{\rm HI}$. We compare in this section the $\rm M_{\rm H_2}$ estimations with those derived using the scaling relations obtained from xCOLD GASS data \citep{2022arXiv220200690S}, and with gas mass estimations obtained from CO observations. 

\begin{figure}
\includegraphics[width=0.5\textwidth]{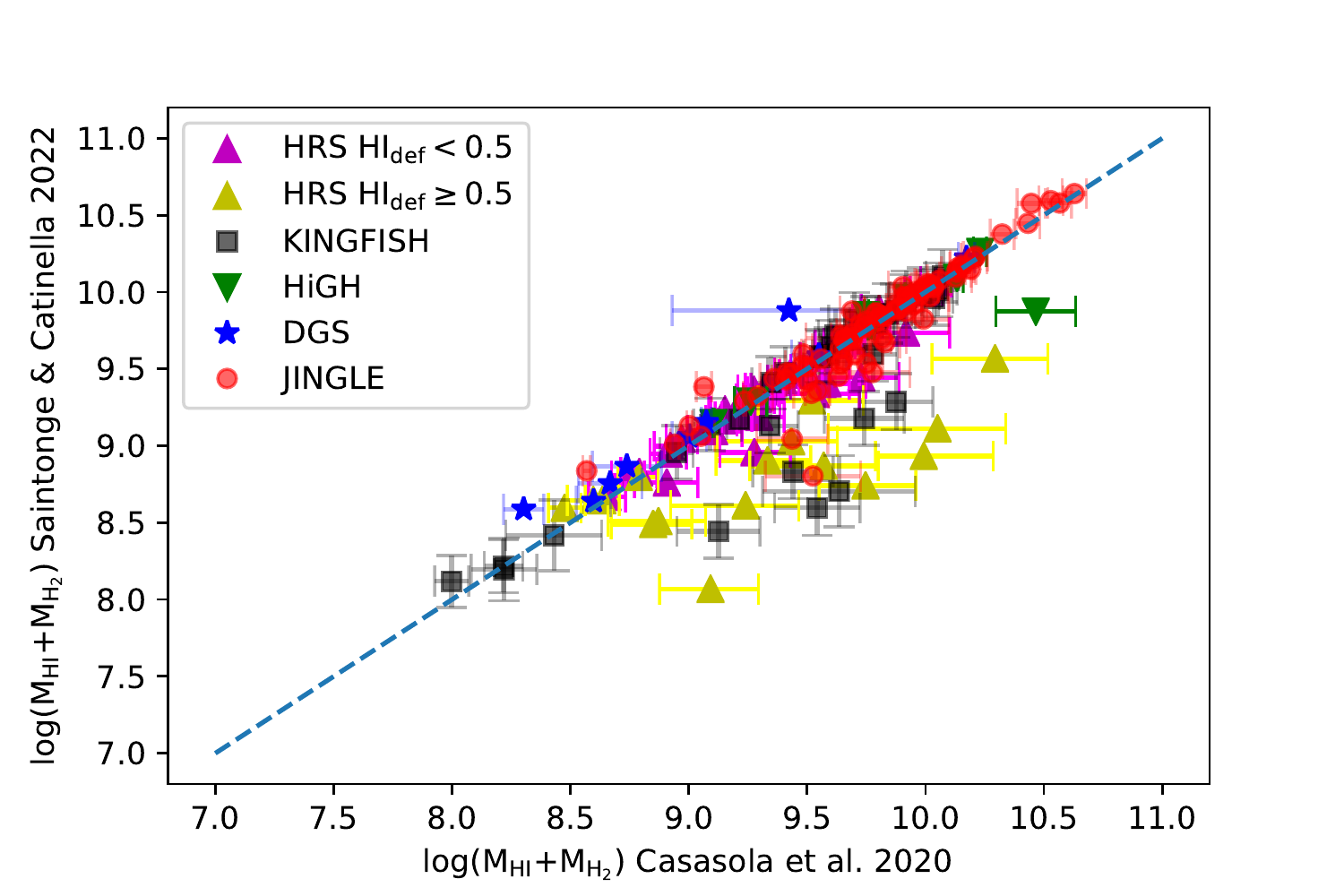}
\includegraphics[width=0.5\textwidth]{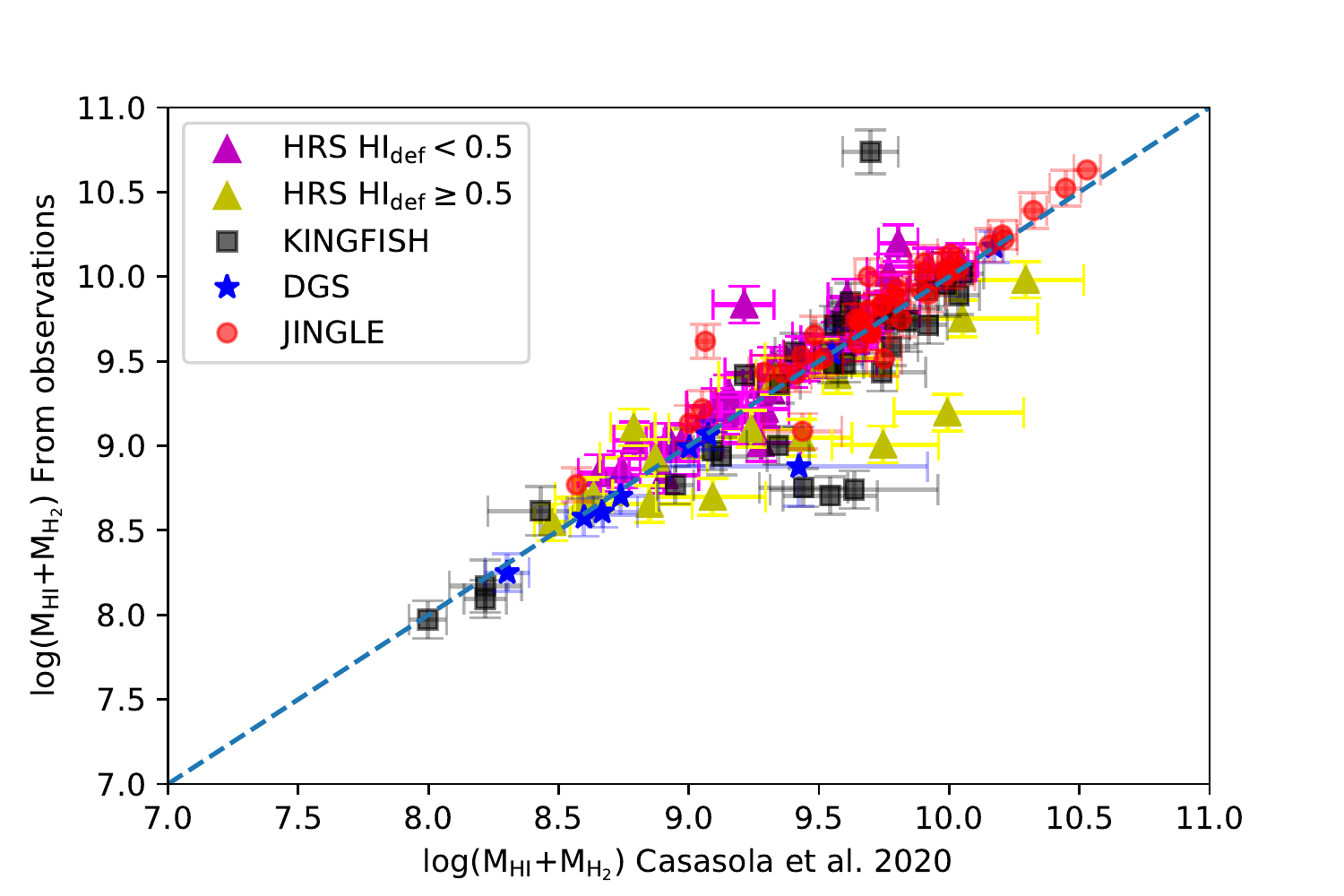}
\caption{Top: Comparison of $\log(\rm M_{\rm HI}+M_{\rm H_2})$ estimated from the prescription given in \citet{2020A&A...633A.100C} and those values derived using the relation presented in \citet{2022arXiv220200690S}. Bottom: Comparison of $\log(\rm M_{\rm HI}+M_{\rm H_2})$ estimated using \citet{2020A&A...633A.100C} with those galaxies that have CO observations.}\label{fig:compCasaXCOLD}
\end{figure}

To compare with the estimations from \citet{2022arXiv220200690S} we use the relation between $\log(\rm M_{\rm H_2}/M_{\rm star})$ and the sSFR presented in \citet[][]{2022arXiv220200690S}. Using the sSFR for our galaxies obtained from literature (see Section\,\ref{sec:samp} for details) we derive $\log(\rm M_{\rm H_2}/M_{\rm star})$.  Then, multiplying by $\rm M_{\rm star}$, we infer $\rm M_{\rm H_2}$. \citet{2022arXiv220200690S} suggest this relation should be applied for galaxies with $\log\rm(sSFR)> -11.5$, which, except for 4 galaxies, it is satisfied by all the objects in our sample (see bottom-left panel in Fig.\ref{fig:D2G}). In the top panel of Fig.\,\ref{fig:compCasaXCOLD} we compare $\rm M_{\rm HI}+M_{\rm H_2}$ estimated from the prescription given in \citet{2020A&A...633A.100C} with the values obtained from the relation in \citet{2022arXiv220200690S}. The mean values for the differences are 0.07, 0.08, 0.17, 0.13, and 0.1\,dex, for HRS H{\sc i} non-deficient galaxies, JINGLE, KINGFISH, DGS and HiGH samples, respectively. The differences are larger for HRS H{\sc i}-deficient galaxies (mean values of the differences is 0.5\,dex). 

In the bottom panel of Fig.\,\ref{fig:compCasaXCOLD} we compare $\rm M_{\rm HI}+M_{\rm H_2}$ obtained from \citet{2020A&A...633A.100C} relation with $\rm M_{\rm HI}+M_{\rm H_2}$ derived using H{\sc i} and CO observations\footnote{$\rm M_{\rm H_2}$ have been obtained using a Milky Way CO-to-H$_{2}$ conversion factor ($X_{\rm CO}=2.0\times10^{20}\rm cm^{-2} (K\,km\,s^{-1})^{-1}$, \citealt{2013ARA&A..51..207B})}. The mean differences are within 0.2\,dex for KINGFISH, DGS, JINGLE and HRS H{\sc i} non-deficient galaxies. H{\sc i}-deficient galaxies show larger deviations than the other galaxy samples. 
 
\section{Total and molecular gas mass fractions in observations and simulations}\label{app:compfmolSPH}

\begin{figure}
\includegraphics[width=0.5\textwidth]{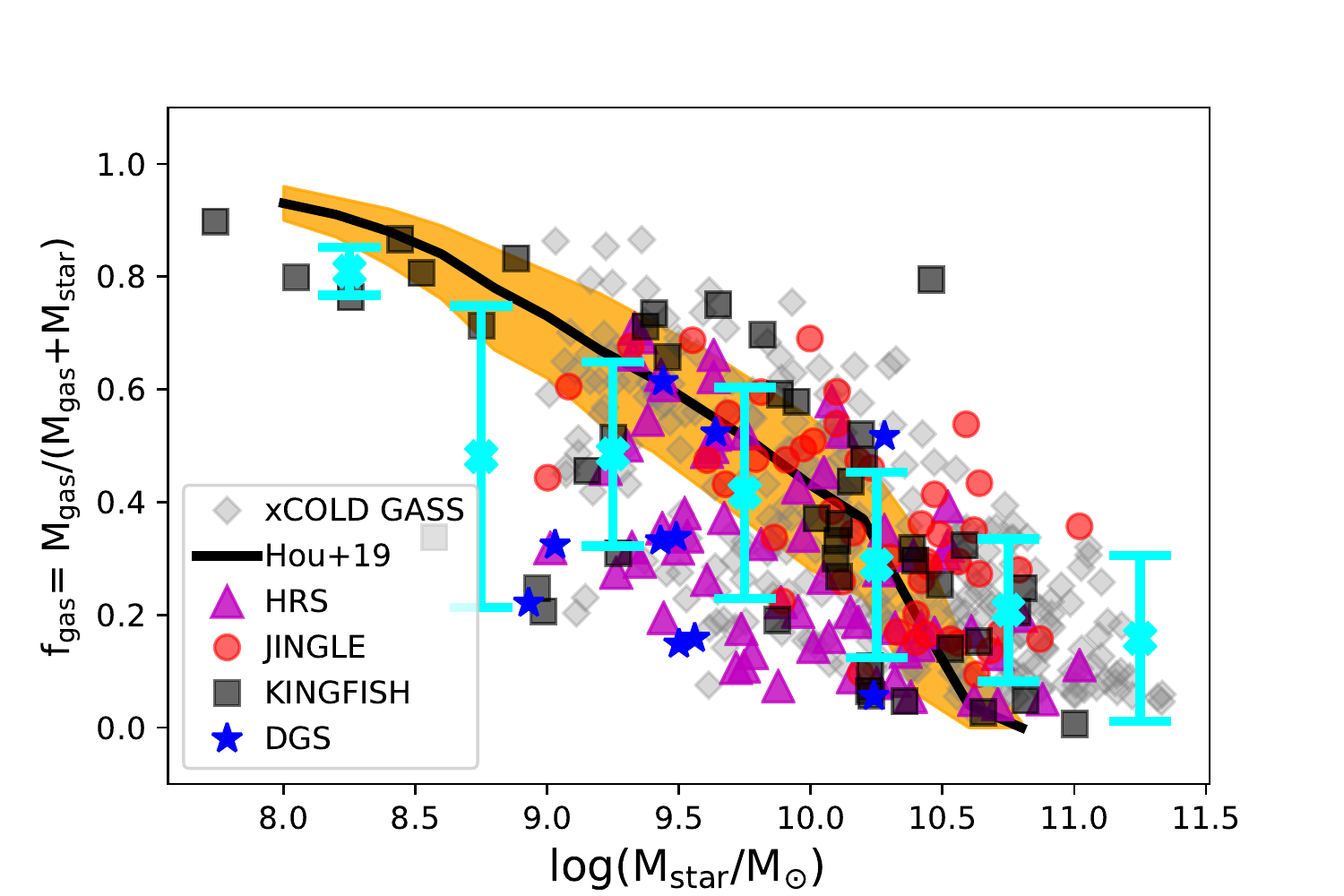}
\caption{Gas mass fraction (f$_{\rm gas}\rm = M_{\rm gas}/(M_{\rm gas}+M_{\rm star})$) versus log(M$_{\rm star}$) for the galaxies in our sample with reliable estimates of molecular gas masses via CO observations, $\rm M_{\rm gas}=1.36\,(M_{HI}+M_{H_{2}}$) to take into account He contribution. Colour code is the same as previous figures with magenta triangles including all (H{\sc i}-deficient and non-deficient) HRS galaxies. HiGH galaxies are not included in the sample as there are no CO observations for them. The black continuous line corresponds to the relation predicted in the simulations from \citet{2019MNRAS.485.1727H} and the yellow area delineates the uncertainty in the simulated results. Grey data points correspond to galaxies from the xCOLD GASS sample for which reliable estimates (no upper limits) of molecular gas masses have been done using CO observations \citep{2017ApJS..233...22S}. The cyan crosses and corresponding error bars represent mean values and standard deviations of magnitudes represented in the $y$-axis for bins in the $x$-axis.}
\label{fig:MgMs_Ms}
\end{figure}

In order to study how well the simulations describe the total gas mass content of our galaxy sample, we furthermore compare the gas mass fractions, f$_{\rm gas}\rm = M_{\rm gas}/(M_{\rm gas}+M_{\rm star})$, of our galaxy sample obtained using H{\sc i} and CO observations with the gas mass fractions predicted by simulations. In Fig.\,\ref{fig:MgMs_Ms} we show f$_{\rm gas}$ versus log(M$_{\rm star}$) for our galaxy sample. The gas masses are the combination of atomic and molecular gas masses with a factor of 1.36 to account for He contribution. The black continuous line represents the median of the distribution for the cosmological simulation in \citet{2019MNRAS.485.1727H} and the yellow colour represents the area within the 16th and 84th percentiles. We also add in this plot the xCOLD GASS sample \citep{2017ApJS..233...22S}, which covers a wider range of stellar mass and for which reliable estimates of molecular gas masses have been done using CO observations. The results of the simulations fall within the observed data and traces relatively well the observed trend in the data, showing that the simulations reproduce relatively well the gas mass content of our galaxy sample.

\begin{figure*}
\includegraphics[width=0.45\textwidth]{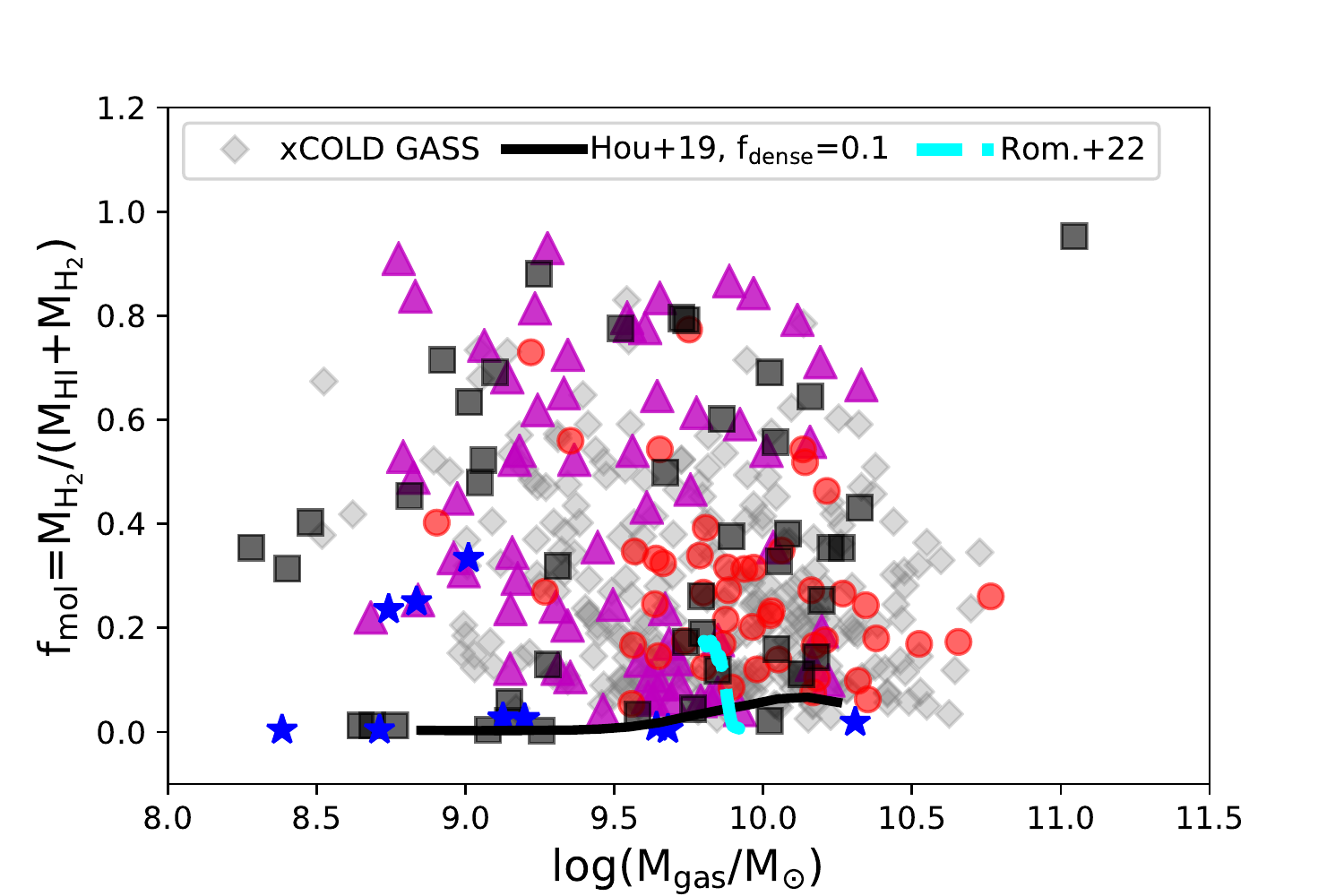}
\includegraphics[width=0.45\textwidth]{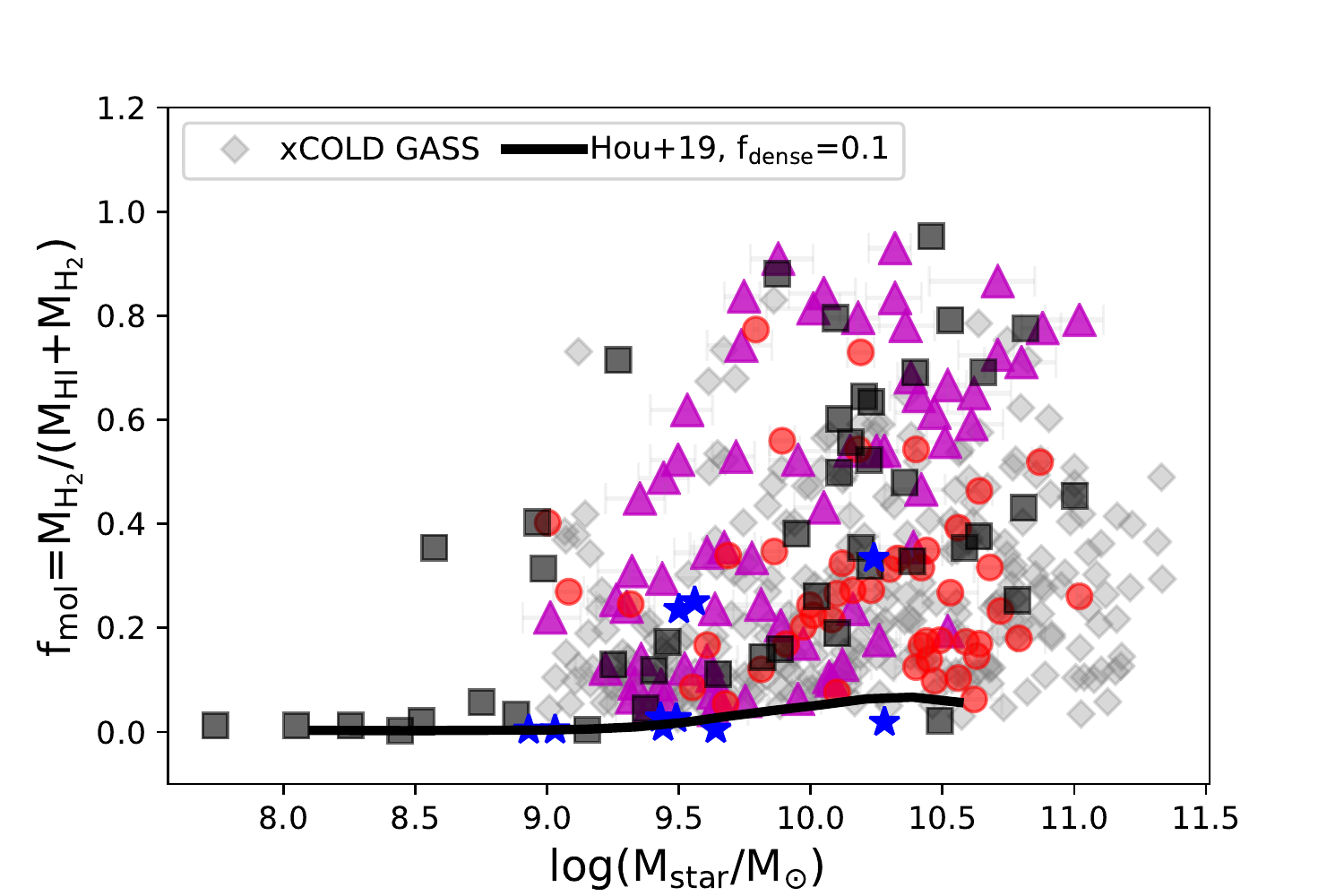}
\caption{Molecular gas mass fraction versus gas mass (left) and stellar mass (right) for our galaxy sample. Only galaxies with reliable estimates of molecular gas masses derived from CO observations have been included. We also add the xCOLD GASS sample (grey data points) which has reliable estimates (no upper limits included) of molecular gas masses using CO observations and cover a wider range of stellar masses. Colour code is the same as previous figures. The continuous black line corresponds to the global molecular gas mass estimates from the simulations in \citet{2019MNRAS.485.1727H} with a mass fraction of 10\% ($f_{\rm dense}$=0.1) in the form of dense clouds in cold and dense gas particles. The dashed black line corresponds to the simulation of  \citet{2022MNRAS.tmp.1338R} where a more suitable parametrisation of $f_{\rm dense}$ has been used in order to describe more reliable values of molecular gas mass fraction in galaxies.}\label{fig:fmol_MgMsXcold}
\end{figure*}

We also compare in this section the molecular gas mass fraction ($\rm f_{\rm mol}=M_{\rm H_2}/(M_{\rm HI}+M_{\rm H_2})$), with the global values of $\rm f_{\rm dense}$ obtained from the simulations of \citet{2019MNRAS.485.1727H}. In Fig.\,\ref{fig:fmol_MgMsXcold} we show the molecular gas mass fraction versus gas mass (left panel) and stellar mass (right panel) for our galaxy sample. We have also added the  xCOLD GASS sample to extend the range in stellar masses. We see that the continuous line representing the molecular gas mass fraction derived from simulations is significantly lower compared to the estimates of the molecular gas mass fractions for our galaxy sample and xCOLD GASS sample.  A proper characterisation of the molecular gas mass fraction requires observations and simulations at higher spatial resolution and therefore it is out of the scope of this study.

\section{\DstoDl\ and $\rm f_{\rm mol}$ for a metallicity dependent  $X_{\rm CO}$ factor.}\label{app:XCO}

\begin{figure*}
\includegraphics[width=0.45\textwidth]{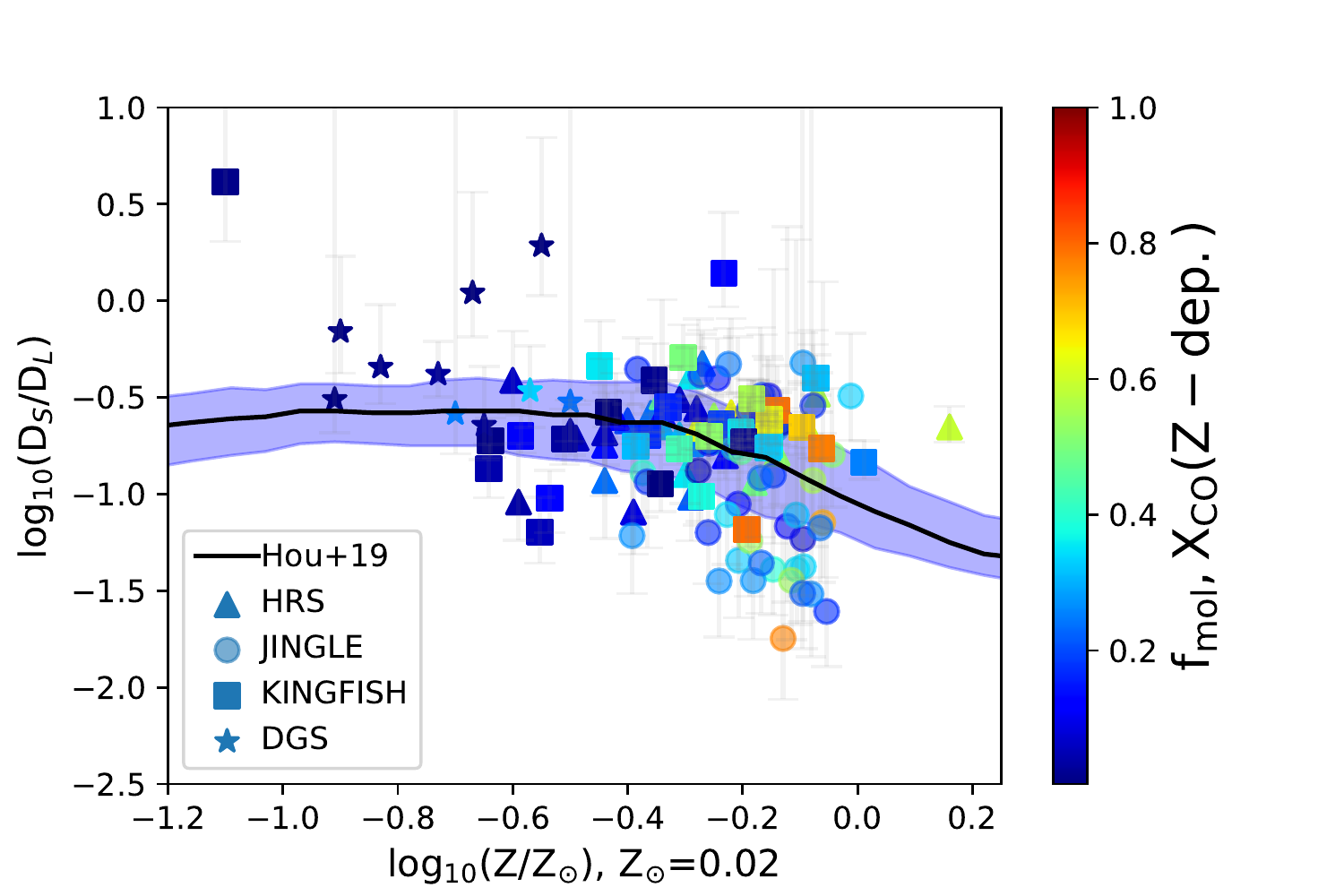}
\includegraphics[width=0.45\textwidth]{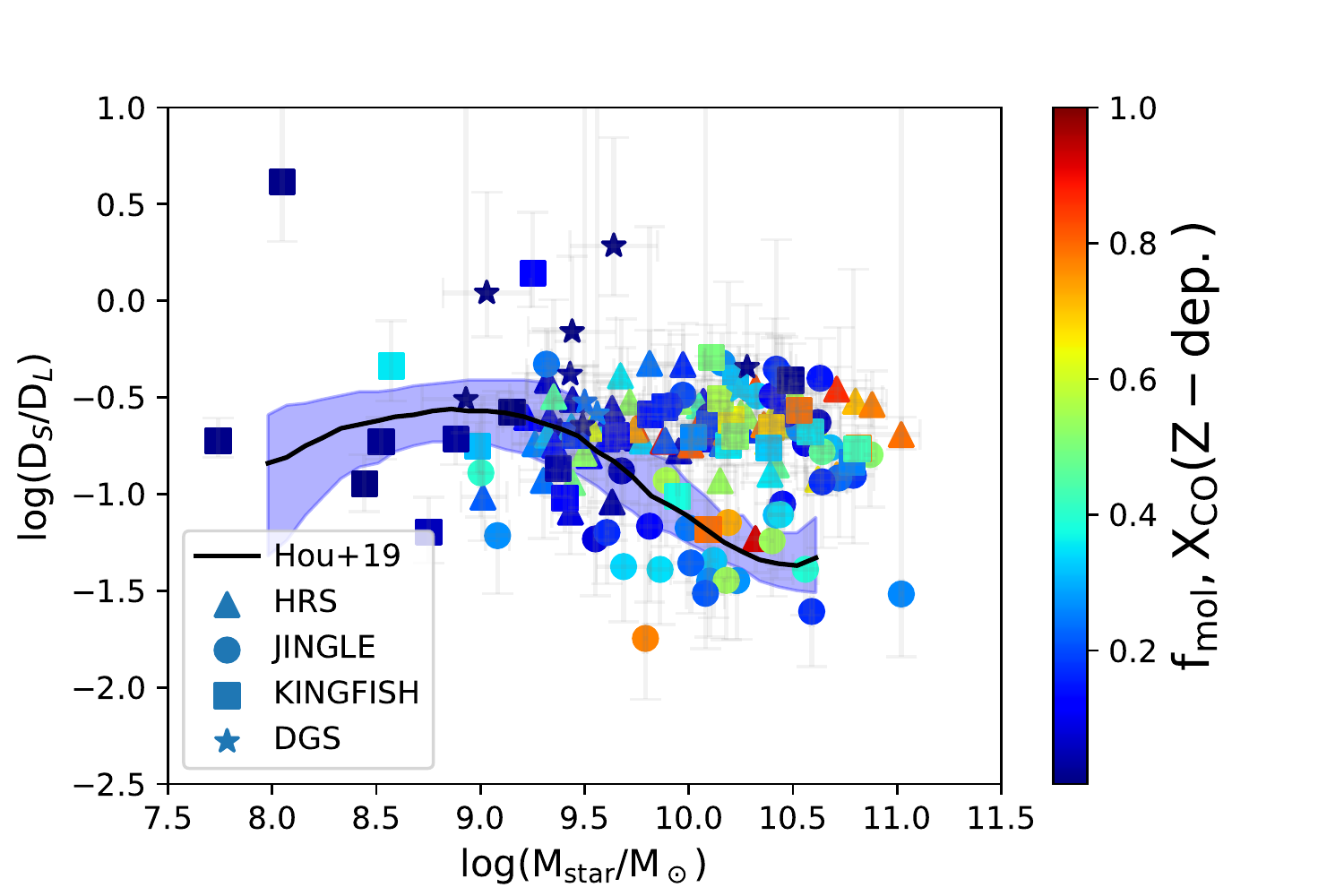}
\caption{\DstoDl\ versus metallicity (left) and stellar mass (right) colour coded with molecular gas mass fraction for our galaxy sample: JINGLE (circles), KINGFISH (squares), DGS (stars), and HRS (triangles). We only include galaxies with molecular gas mass estimates in the literature. These masses have been obtained using a metallicity dependent CO-to-H$_{2}$ conversion factor of $X_{\rm CO}\propto(Z/Z_{\odot})^{-1.55}$ parametrised by \citet{2020A&A...643A.180H}.}\label{fig:S2L_metfmol_app}
\end{figure*}

We show here in Fig.\,\ref{fig:S2L_metfmol_app} the comparison of the \DstoDl\ and molecular gas mass fraction assuming a metallicity dependent $X_{\rm CO}$ factor derived in \citet{2020A&A...643A.180H}. The results presented in Sextion\,\ref{sec:molgas} are not changed when a metallicity dependent $X_{\rm CO}$ factor is assumed. 

\section{SFMS for those galaxies presenting high \DstoDl.}\label{app:sfms}
In this section we identify those galaxies having high values of \DstoDl\ in the SFMS relation presented in left panel of Fig.\,\ref{fig:scarel}. We can see that albeit with some dispersion most of the galaxies with high values of \DstoDl\ are above the MS relation from \citet{2014ApJS..214...15S}. 

\begin{figure}
\includegraphics[width=0.45\textwidth]{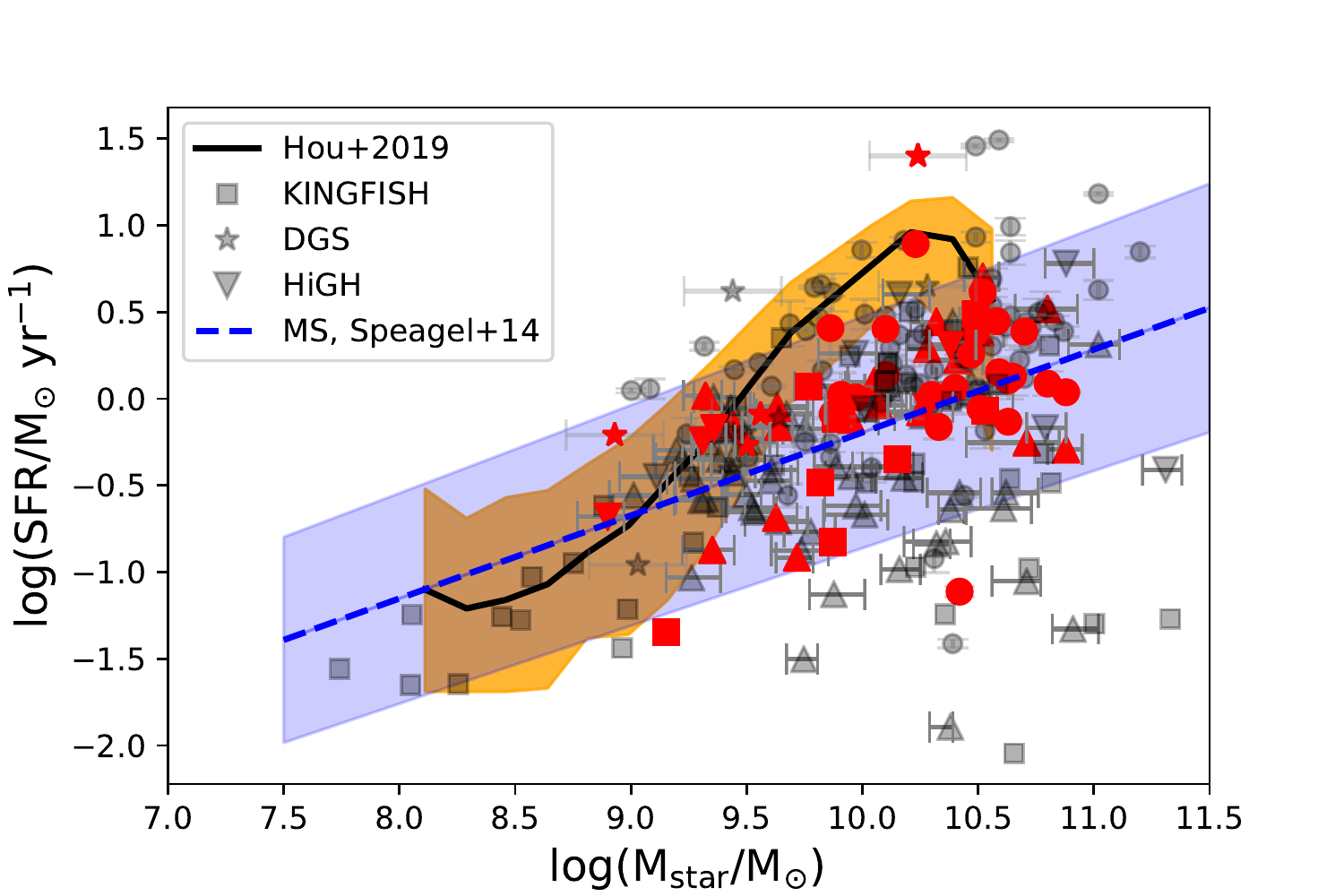}
\caption{SFR versus stellar mass predicted by the simulations (as in the left panel of Fig.\,\ref{fig:scarel}). Red points show the galaxies with \DstoDl\ values
 $-0.6<\log_{10}(D_{S}/D_{L})<-0.4$}\label{appfig:sfms}
\end{figure}

\section{Results of the SED fit}

\onecolumn
\begingroup
\renewcommand{\arraystretch}{1.5}
\LTcapwidth=\textwidth
\begin{longtable}{cccccc}
\caption{Dust masses, \stol\ grain mass ratios, $U_{\rm min}$ and $\alpha$ values derived using our fitting procedure (see details in Section\,\ref{sec:fit} in the main text). The top and bottom uncertainties correspond to the 16th and 84th percentiles of the posterior probability distribution for each free parameter.}\label{tab:results} \\
\hline
Galaxy & Sample & $\log_{10}$(M$_{\rm dust}/\rm M_{\odot})$ & $\log_{10}$(D$_{S}$/D$_{L}$)  & $U_{\rm min}$ & $\alpha$\\
\hline
\endfirsthead
Continuation of Table\,\ref{tab:results}.\\
\hline 
Galaxy & Sample & $\log_{10}$(M$_{\rm dust}/\rm M_{\odot})$ & $\log_{10}$(D$_{S}$/D$_{L}$)  & $U_{\rm min}$ & $\alpha$\\
\hline
\endhead
\hline 
\endfoot
\hline
\endlastfoot
JINGLE0&JINGLE &   6.59$^{+0.14}_{-0.07}$ & -0.23$^{+0.29}_{-0.15}$ &  2.03$^{+1.20}_{-0.93}$ &  3.63$^{+0.90}_{-0.95}$\\ 
JINGLE3&JINGLE &   6.51$^{+0.08}_{-0.07}$ & -0.89$^{+0.36}_{-0.23}$ &  1.46$^{+0.32}_{-0.28}$ &  2.13$^{+0.02}_{-0.02}$\\ 
JINGLE4&JINGLE &   7.28$^{+0.07}_{-0.10}$ & -1.39$^{+0.69}_{-0.28}$ &  0.22$^{+0.12}_{-0.05}$ &  2.06$^{+0.02}_{-0.01}$\\ 
JINGLE5&JINGLE &   7.26$^{+0.21}_{-0.08}$ & -0.49$^{+0.31}_{-0.16}$ &  1.72$^{+0.91}_{-1.16}$ &  3.46$^{+1.09}_{-1.19}$\\ 
JINGLE8&JINGLE &   7.14$^{+0.21}_{-0.13}$ & -0.88$^{+0.78}_{-0.28}$ &  0.10$^{+0.10}_{-0.05}$ &  2.08$^{+0.03}_{-0.02}$\\ 
JINGLE9&JINGLE &   7.10$^{+0.11}_{-0.07}$ & -0.55$^{+0.20}_{-0.13}$ &  0.77$^{+0.35}_{-0.32}$ &  3.50$^{+1.00}_{-0.79}$\\ 
JINGLE10&JINGLE &   7.54$^{+0.11}_{-0.22}$ & -1.05$^{+0.86}_{-0.34}$ &  0.24$^{+1.07}_{-0.08}$ &  2.13$^{+1.81}_{-0.03}$\\ 
JINGLE11&JINGLE &   7.58$^{+0.18}_{-0.15}$ & -0.75$^{+0.69}_{-0.33}$ &  0.15$^{+0.30}_{-0.07}$ &  2.17$^{+2.01}_{-0.05}$\\ 
JINGLE15&JINGLE &   7.32$^{+0.10}_{-0.11}$ & -1.23$^{+0.81}_{-0.30}$ &  0.26$^{+0.14}_{-0.09}$ &  2.08$^{+0.02}_{-0.01}$\\ 
JINGLE16&JINGLE &   7.29$^{+0.19}_{-0.11}$ & -0.53$^{+0.52}_{-0.27}$ &  0.19$^{+0.19}_{-0.09}$ &  2.10$^{+0.02}_{-0.02}$\\ 
JINGLE19&JINGLE &   7.46$^{+0.11}_{-0.07}$ & -0.40$^{+0.21}_{-0.13}$ &  0.85$^{+0.40}_{-0.34}$ &  3.76$^{+0.85}_{-0.84}$\\ 
JINGLE20&JINGLE &   7.10$^{+0.14}_{-0.09}$ & -0.64$^{+0.25}_{-0.16}$ &  0.71$^{+0.53}_{-0.35}$ &  3.14$^{+1.17}_{-0.63}$\\ 
JINGLE22&JINGLE &   7.47$^{+0.19}_{-0.08}$ & -0.74$^{+0.31}_{-0.15}$ &  1.22$^{+0.91}_{-0.75}$ &  2.95$^{+0.93}_{-0.53}$\\ 
JINGLE23&JINGLE &   7.31$^{+0.21}_{-0.09}$ & -0.49$^{+0.32}_{-0.23}$ &  1.19$^{+0.61}_{-0.89}$ &  3.46$^{+1.10}_{-1.31}$\\ 
JINGLE25&JINGLE &   7.36$^{+0.09}_{-0.08}$ & -1.34$^{+0.62}_{-0.30}$ &  0.53$^{+0.16}_{-0.14}$ &  2.08$^{+0.02}_{-0.01}$\\ 
JINGLE26&JINGLE &   7.03$^{+0.10}_{-0.07}$ & -1.20$^{+0.58}_{-0.27}$ &  0.81$^{+0.28}_{-0.23}$ &  2.11$^{+0.02}_{-0.01}$\\ 
JINGLE28&JINGLE &   7.46$^{+0.26}_{-0.11}$ & -0.54$^{+0.39}_{-0.27}$ &  0.62$^{+0.51}_{-0.48}$ &  3.36$^{+1.12}_{-1.23}$\\ 
JINGLE36&JINGLE &   7.16$^{+0.09}_{-0.14}$ & -1.27$^{+0.61}_{-0.28}$ &  0.14$^{+0.10}_{-0.05}$ &  2.10$^{+0.02}_{-0.02}$\\ 
JINGLE37&JINGLE &   7.19$^{+0.09}_{-0.08}$ & -1.01$^{+0.55}_{-0.25}$ &  1.55$^{+0.51}_{-0.36}$ &  2.15$^{+0.02}_{-0.02}$\\ 
JINGLE40&JINGLE &   7.72$^{+0.09}_{-0.06}$ & -0.51$^{+0.17}_{-0.11}$ &  0.99$^{+0.34}_{-0.30}$ &  3.87$^{+0.76}_{-0.78}$\\ 
JINGLE41&JINGLE &   7.77$^{+0.07}_{-0.07}$ & -0.85$^{+0.13}_{-0.12}$ &  0.98$^{+0.48}_{-0.29}$ &  2.59$^{+0.33}_{-0.15}$\\ 
JINGLE42&JINGLE &   7.22$^{+0.12}_{-0.23}$ & -1.07$^{+0.39}_{-0.33}$ &  0.14$^{+0.94}_{-0.06}$ &  2.08$^{+1.56}_{-0.02}$\\ 
JINGLE43&JINGLE &   8.07$^{+0.05}_{-0.05}$ & -1.61$^{+0.63}_{-0.28}$ &  0.56$^{+0.11}_{-0.10}$ &  2.09$^{+0.01}_{-0.01}$\\ 
JINGLE44&JINGLE &   7.82$^{+0.09}_{-0.07}$ & -0.46$^{+0.17}_{-0.13}$ &  0.55$^{+0.27}_{-0.19}$ &  3.58$^{+1.00}_{-0.82}$\\ 
JINGLE45&JINGLE &   7.95$^{+0.06}_{-0.06}$ & -0.59$^{+0.12}_{-0.10}$ &  0.58$^{+0.15}_{-0.14}$ &  4.11$^{+0.62}_{-0.75}$\\ 
JINGLE47&JINGLE &   7.57$^{+0.07}_{-0.06}$ & -1.39$^{+0.53}_{-0.24}$ &  0.86$^{+0.22}_{-0.14}$ &  2.13$^{+0.02}_{-0.01}$\\ 
JINGLE48&JINGLE &   7.48$^{+0.06}_{-0.06}$ & -1.45$^{+0.60}_{-0.30}$ &  1.30$^{+0.26}_{-0.22}$ &  2.16$^{+0.02}_{-0.01}$\\ 
JINGLE49&JINGLE &   7.54$^{+0.10}_{-0.10}$ & -1.24$^{+0.75}_{-0.29}$ &  0.53$^{+0.24}_{-0.15}$ &  2.11$^{+0.02}_{-0.01}$\\ 
JINGLE51&JINGLE &   7.44$^{+0.07}_{-0.09}$ & -1.36$^{+0.73}_{-0.26}$ &  0.72$^{+0.29}_{-0.16}$ &  2.09$^{+0.02}_{-0.01}$\\ 
JINGLE55&JINGLE &   7.61$^{+0.09}_{-0.12}$ & -0.80$^{+0.34}_{-0.27}$ &  1.65$^{+1.55}_{-0.39}$ &  2.28$^{+0.71}_{-0.07}$\\ 
JINGLE57&JINGLE &   7.67$^{+0.26}_{-0.21}$ & -0.93$^{+0.70}_{-0.34}$ &  0.08$^{+0.46}_{-0.04}$ &  2.08$^{+2.26}_{-0.03}$\\ 
JINGLE58&JINGLE &   7.64$^{+0.19}_{-0.19}$ & -0.99$^{+0.64}_{-0.35}$ &  0.04$^{+0.05}_{-0.02}$ &  2.04$^{+0.03}_{-0.02}$\\ 
JINGLE60&JINGLE &   7.75$^{+0.22}_{-0.17}$ & -0.83$^{+0.75}_{-0.35}$ &  0.04$^{+0.08}_{-0.02}$ &  2.05$^{+0.03}_{-0.02}$\\ 
JINGLE61&JINGLE &   7.11$^{+0.12}_{-0.08}$ & -0.54$^{+0.20}_{-0.15}$ &  1.06$^{+0.49}_{-0.46}$ &  3.62$^{+0.92}_{-0.87}$\\ 
JINGLE64&JINGLE &   7.31$^{+0.22}_{-0.11}$ & -1.22$^{+0.75}_{-0.30}$ &  0.21$^{+0.14}_{-0.09}$ &  2.06$^{+0.02}_{-0.02}$\\ 
JINGLE66&JINGLE &   7.80$^{+0.08}_{-0.06}$ & -0.63$^{+0.15}_{-0.11}$ &  1.04$^{+0.39}_{-0.34}$ &  3.89$^{+0.75}_{-0.82}$\\ 
JINGLE68&JINGLE &   7.54$^{+0.10}_{-0.22}$ & -1.08$^{+0.52}_{-0.34}$ &  0.24$^{+1.14}_{-0.08}$ &  2.12$^{+1.52}_{-0.02}$\\ 
JINGLE70&JINGLE &   7.45$^{+0.10}_{-0.07}$ & -0.46$^{+0.20}_{-0.13}$ &  1.35$^{+0.50}_{-0.54}$ &  3.96$^{+0.72}_{-0.91}$\\ 
JINGLE71&JINGLE &   7.12$^{+0.10}_{-0.09}$ & -0.99$^{+0.55}_{-0.26}$ &  0.93$^{+0.48}_{-0.23}$ &  2.14$^{+0.03}_{-0.02}$\\ 
JINGLE72&JINGLE &   7.55$^{+0.05}_{-0.05}$ & -1.75$^{+0.69}_{-0.31}$ &  0.80$^{+0.14}_{-0.12}$ &  2.10$^{+0.01}_{-0.01}$\\ 
JINGLE74&JINGLE &   7.23$^{+0.10}_{-0.07}$ & -0.33$^{+0.19}_{-0.14}$ &  1.91$^{+0.88}_{-0.74}$ &  3.70$^{+0.88}_{-0.92}$\\ 
JINGLE76&JINGLE &   7.14$^{+0.06}_{-0.04}$ & -0.66$^{+0.10}_{-0.08}$ &  5.11$^{+1.04}_{-1.17}$ &  3.85$^{+0.81}_{-0.66}$\\ 
JINGLE77&JINGLE &   7.66$^{+0.08}_{-0.10}$ & -1.45$^{+0.71}_{-0.29}$ &  0.72$^{+0.32}_{-0.16}$ &  2.09$^{+0.02}_{-0.01}$\\ 
JINGLE81&JINGLE &   7.37$^{+0.14}_{-0.11}$ & -1.15$^{+1.25}_{-0.28}$ &  0.44$^{+0.28}_{-0.14}$ &  2.11$^{+0.04}_{-0.01}$\\ 
JINGLE82&JINGLE &   7.47$^{+0.25}_{-0.12}$ & -1.14$^{+0.89}_{-0.27}$ &  0.08$^{+0.07}_{-0.03}$ &  2.09$^{+0.03}_{-0.02}$\\ 
JINGLE83&JINGLE &   7.75$^{+0.08}_{-0.06}$ & -0.35$^{+0.16}_{-0.12}$ &  1.05$^{+0.35}_{-0.35}$ &  3.96$^{+0.80}_{-0.77}$\\ 
JINGLE84&JINGLE &   7.53$^{+0.10}_{-0.07}$ & -0.49$^{+0.19}_{-0.13}$ &  0.84$^{+0.39}_{-0.33}$ &  3.53$^{+1.03}_{-0.73}$\\ 
JINGLE86&JINGLE &   7.56$^{+0.09}_{-0.06}$ & -0.57$^{+0.17}_{-0.12}$ &  1.52$^{+0.57}_{-0.60}$ &  3.58$^{+0.98}_{-0.82}$\\ 
JINGLE87&JINGLE &   7.46$^{+0.11}_{-0.07}$ & -0.76$^{+0.21}_{-0.13}$ &  1.60$^{+0.94}_{-0.71}$ &  3.14$^{+1.06}_{-0.51}$\\ 
JINGLE89&JINGLE &   7.54$^{+0.08}_{-0.06}$ & -0.32$^{+0.16}_{-0.11}$ &  1.49$^{+0.41}_{-0.43}$ &  3.96$^{+0.74}_{-0.72}$\\ 
JINGLE90&JINGLE &   7.88$^{+0.18}_{-0.26}$ & -1.17$^{+1.55}_{-0.39}$ &  0.11$^{+0.17}_{-0.04}$ &  2.05$^{+1.49}_{-0.01}$\\ 
JINGLE92&JINGLE &   7.87$^{+0.07}_{-0.06}$ & -0.59$^{+0.15}_{-0.13}$ &  0.54$^{+0.22}_{-0.18}$ &  3.50$^{+1.02}_{-0.79}$\\ 
JINGLE98&JINGLE &   7.41$^{+0.15}_{-0.09}$ & -0.84$^{+0.25}_{-0.15}$ &  0.79$^{+0.54}_{-0.38}$ &  2.95$^{+1.19}_{-0.47}$\\ 
JINGLE99&JINGLE &   8.14$^{+0.09}_{-0.06}$ & -0.48$^{+0.17}_{-0.13}$ &  0.65$^{+0.26}_{-0.22}$ &  3.90$^{+0.76}_{-0.86}$\\ 
JINGLE100&JINGLE &   7.49$^{+0.08}_{-0.08}$ & -1.38$^{+0.62}_{-0.29}$ &  0.33$^{+0.11}_{-0.07}$ &  2.06$^{+0.01}_{-0.01}$\\ 
JINGLE101&JINGLE &   7.84$^{+0.07}_{-0.09}$ & -1.45$^{+0.65}_{-0.31}$ &  1.05$^{+0.37}_{-0.18}$ &  2.11$^{+0.02}_{-0.01}$\\ 
JINGLE102&JINGLE &   7.79$^{+0.11}_{-0.12}$ & -0.85$^{+0.49}_{-0.28}$ &  0.34$^{+0.29}_{-0.12}$ &  2.09$^{+0.03}_{-0.02}$\\ 
JINGLE108&JINGLE &   7.80$^{+0.08}_{-0.08}$ & -1.18$^{+0.36}_{-0.28}$ &  0.54$^{+0.23}_{-0.14}$ &  2.08$^{+0.02}_{-0.01}$\\ 
JINGLE111&JINGLE &   7.64$^{+0.16}_{-0.22}$ & -1.11$^{+1.42}_{-0.33}$ &  0.30$^{+0.94}_{-0.12}$ &  2.11$^{+1.90}_{-0.02}$\\ 
JINGLE118&JINGLE &   8.25$^{+0.11}_{-0.19}$ & -1.52$^{+2.86}_{-0.33}$ &  0.67$^{+1.21}_{-0.13}$ &  2.14$^{+1.11}_{-0.01}$\\ 
JINGLE121&JINGLE &   7.97$^{+0.11}_{-0.08}$ & -0.44$^{+0.21}_{-0.15}$ &  0.62$^{+0.34}_{-0.27}$ &  3.46$^{+1.03}_{-0.77}$\\ 
JINGLE122&JINGLE &   8.07$^{+0.16}_{-0.15}$ & -0.90$^{+1.07}_{-0.35}$ &  0.29$^{+0.73}_{-0.12}$ &  2.13$^{+2.47}_{-0.03}$\\ 
JINGLE123&JINGLE &   7.74$^{+0.12}_{-0.11}$ & -1.33$^{+0.62}_{-0.30}$ &  0.29$^{+0.19}_{-0.09}$ &  2.07$^{+0.02}_{-0.01}$\\ 
JINGLE125&JINGLE &   7.82$^{+0.13}_{-0.15}$ & -0.92$^{+0.78}_{-0.31}$ &  0.39$^{+0.59}_{-0.13}$ &  2.14$^{+1.28}_{-0.02}$\\ 
JINGLE127&JINGLE &   7.62$^{+0.10}_{-0.07}$ & -0.52$^{+0.20}_{-0.13}$ &  2.30$^{+1.15}_{-1.02}$ &  3.31$^{+1.05}_{-0.59}$\\ 
JINGLE128&JINGLE &   7.76$^{+0.09}_{-0.07}$ & -0.63$^{+0.17}_{-0.12}$ &  0.95$^{+0.42}_{-0.33}$ &  3.68$^{+0.89}_{-0.71}$\\ 
JINGLE131&JINGLE &   7.62$^{+0.09}_{-0.12}$ & -1.24$^{+0.98}_{-0.28}$ &  0.23$^{+0.15}_{-0.06}$ &  2.07$^{+0.03}_{-0.01}$\\ 
JINGLE135&JINGLE &   7.65$^{+0.15}_{-0.20}$ & -1.30$^{+1.43}_{-0.36}$ &  0.32$^{+0.20}_{-0.12}$ &  2.09$^{+0.89}_{-0.01}$\\ 
JINGLE136&JINGLE &   7.70$^{+0.08}_{-0.06}$ & -0.56$^{+0.15}_{-0.12}$ &  0.99$^{+0.48}_{-0.36}$ &  2.97$^{+1.09}_{-0.40}$\\ 
JINGLE139&JINGLE &   7.89$^{+0.21}_{-0.22}$ & -1.19$^{+2.08}_{-0.36}$ &  0.12$^{+0.30}_{-0.05}$ &  2.06$^{+1.82}_{-0.02}$\\ 
JINGLE143&JINGLE &   7.39$^{+0.12}_{-0.07}$ & -0.54$^{+0.21}_{-0.13}$ &  2.08$^{+1.17}_{-0.82}$ &  3.40$^{+0.95}_{-0.61}$\\ 
JINGLE144&JINGLE &   7.64$^{+0.07}_{-0.07}$ & -1.39$^{+0.72}_{-0.31}$ &  0.57$^{+0.18}_{-0.11}$ &  2.10$^{+0.02}_{-0.01}$\\ 
JINGLE146&JINGLE &   7.54$^{+0.08}_{-0.07}$ & -0.92$^{+0.30}_{-0.24}$ &  0.73$^{+0.24}_{-0.16}$ &  2.11$^{+0.02}_{-0.01}$\\ 
JINGLE147&JINGLE &   7.73$^{+0.11}_{-0.18}$ & -1.11$^{+1.00}_{-0.35}$ &  1.76$^{+2.14}_{-0.38}$ &  2.23$^{+1.17}_{-0.03}$\\ 
JINGLE148&JINGLE &   7.46$^{+0.09}_{-0.10}$ & -1.05$^{+0.35}_{-0.23}$ &  0.58$^{+0.37}_{-0.13}$ &  2.08$^{+0.02}_{-0.02}$\\ 
JINGLE149&JINGLE &   7.42$^{+0.16}_{-0.17}$ & -1.51$^{+4.13}_{-0.29}$ &  0.32$^{+0.55}_{-0.09}$ &  2.08$^{+1.43}_{-0.01}$\\ 
JINGLE150&JINGLE &   8.14$^{+0.10}_{-0.14}$ & -1.19$^{+0.72}_{-0.33}$ &  0.51$^{+0.30}_{-0.14}$ &  2.11$^{+0.04}_{-0.01}$\\ 
JINGLE151&JINGLE &   7.94$^{+0.12}_{-0.07}$ & -0.78$^{+0.20}_{-0.13}$ &  1.32$^{+0.79}_{-0.58}$ &  2.98$^{+0.96}_{-0.40}$\\ 
JINGLE152&JINGLE &   7.70$^{+0.07}_{-0.05}$ & -0.45$^{+0.15}_{-0.11}$ &  0.91$^{+0.23}_{-0.24}$ &  4.14$^{+0.52}_{-0.74}$\\ 
JINGLE155&JINGLE &   7.79$^{+0.11}_{-0.07}$ & -0.63$^{+0.21}_{-0.14}$ &  1.08$^{+0.48}_{-0.39}$ &  3.94$^{+0.74}_{-0.86}$\\ 
JINGLE156&JINGLE &   7.75$^{+0.09}_{-0.08}$ & -0.78$^{+0.30}_{-0.22}$ &  1.94$^{+0.61}_{-0.45}$ &  2.17$^{+0.02}_{-0.02}$\\ 
JINGLE159&JINGLE &   7.52$^{+0.08}_{-0.07}$ & -0.99$^{+0.45}_{-0.27}$ &  0.63$^{+0.21}_{-0.16}$ &  2.10$^{+0.02}_{-0.02}$\\ 
JINGLE165&JINGLE &   7.52$^{+0.06}_{-0.05}$ & -0.49$^{+0.13}_{-0.10}$ &  6.61$^{+1.48}_{-1.30}$ &  4.16$^{+0.61}_{-0.67}$\\ 
JINGLE166&JINGLE &   7.62$^{+0.40}_{-0.10}$ & -0.63$^{+0.55}_{-0.26}$ &  1.26$^{+0.90}_{-1.07}$ &  3.24$^{+1.18}_{-1.14}$\\ 
JINGLE167&JINGLE &   8.28$^{+0.06}_{-0.24}$ & -0.94$^{+0.25}_{-0.34}$ &  0.22$^{+1.14}_{-0.06}$ &  2.08$^{+0.08}_{-0.01}$\\ 
JINGLE168&JINGLE &   8.07$^{+0.17}_{-0.09}$ & -0.62$^{+0.29}_{-0.16}$ &  0.53$^{+0.38}_{-0.30}$ &  3.17$^{+1.10}_{-0.62}$\\ 
JINGLE170&JINGLE &   7.69$^{+0.12}_{-0.09}$ & -0.69$^{+0.23}_{-0.16}$ &  1.37$^{+0.67}_{-0.58}$ &  3.57$^{+0.96}_{-0.74}$\\ 
JINGLE173&JINGLE &   7.20$^{+0.20}_{-0.07}$ & -0.60$^{+0.31}_{-0.14}$ &  1.98$^{+1.03}_{-1.26}$ &  3.62$^{+0.87}_{-1.00}$\\ 
JINGLE175&JINGLE &   8.09$^{+0.14}_{-0.31}$ & -1.20$^{+1.17}_{-0.42}$ &  0.14$^{+0.37}_{-0.06}$ &  2.06$^{+1.81}_{-0.01}$\\ 
JINGLE176&JINGLE &   7.89$^{+0.12}_{-0.14}$ & -0.29$^{+0.29}_{-0.36}$ &  0.50$^{+0.33}_{-0.32}$ &  3.51$^{+1.12}_{-1.45}$\\ 
JINGLE177&JINGLE &   7.57$^{+0.09}_{-0.07}$ & -0.41$^{+0.25}_{-0.14}$ &  1.55$^{+0.46}_{-0.40}$ &  2.10$^{+0.01}_{-0.01}$\\ 
JINGLE178&JINGLE &   7.91$^{+0.33}_{-0.25}$ & -1.36$^{+6.33}_{-0.38}$ &  0.15$^{+0.22}_{-0.06}$ &  2.08$^{+1.94}_{-0.01}$\\ 
JINGLE181&JINGLE &   7.89$^{+0.18}_{-0.23}$ & -1.35$^{+3.09}_{-0.36}$ &  0.12$^{+0.27}_{-0.04}$ &  2.07$^{+1.78}_{-0.01}$\\ 
JINGLE183&JINGLE &   7.20$^{+0.11}_{-0.07}$ & -0.38$^{+0.21}_{-0.13}$ &  1.69$^{+0.62}_{-0.70}$ &  3.94$^{+0.71}_{-0.98}$\\ 
JINGLE184&JINGLE &   8.11$^{+0.21}_{-0.37}$ & -1.30$^{+1.10}_{-0.50}$ &  0.18$^{+0.75}_{-0.08}$ &  2.07$^{+1.32}_{-0.01}$\\ 
JINGLE186&JINGLE &   8.02$^{+0.09}_{-0.06}$ & -1.59$^{+3.11}_{-0.30}$ &  3.85$^{+3.98}_{-0.45}$ &  2.23$^{+1.13}_{-0.01}$\\ 
JINGLE191&JINGLE &   7.62$^{+0.08}_{-0.07}$ & -1.11$^{+0.40}_{-0.26}$ &  1.42$^{+0.40}_{-0.34}$ &  2.17$^{+0.02}_{-0.02}$\\ 
JINGLE192&JINGLE &   8.35$^{+0.08}_{-0.07}$ & -1.55$^{+0.73}_{-0.31}$ &  0.64$^{+0.19}_{-0.14}$ &  2.14$^{+0.02}_{-0.01}$\\ 
NGC0337&KINGFISH &   6.88$^{+0.08}_{-0.06}$ & -0.70$^{+0.11}_{-0.10}$ &  2.44$^{+1.18}_{-0.82}$ &  2.88$^{+0.82}_{-0.36}$\\ 
NGC0628&KINGFISH &   7.15$^{+0.11}_{-0.06}$ & -0.64$^{+0.16}_{-0.10}$ &  0.87$^{+0.43}_{-0.41}$ &  3.02$^{+1.01}_{-0.51}$\\ 
NGC0855&KINGFISH &   5.28$^{+0.06}_{-0.05}$ & -0.80$^{+0.10}_{-0.08}$ &  4.61$^{+1.56}_{-1.33}$ &  3.23$^{+1.00}_{-0.48}$\\ 
NGC0925&KINGFISH &   7.21$^{+0.05}_{-0.09}$ & -1.02$^{+0.14}_{-0.18}$ &  0.24$^{+0.11}_{-0.07}$ &  2.34$^{+0.12}_{-0.06}$\\ 
NGC1097&KINGFISH &   7.60$^{+0.12}_{-0.06}$ & -0.41$^{+0.19}_{-0.14}$ &  1.63$^{+0.80}_{-0.87}$ &  3.30$^{+1.17}_{-0.97}$\\ 
NGC1266&KINGFISH &   6.64$^{+0.06}_{-0.06}$ & -1.57$^{+0.45}_{-0.24}$ &  7.00$^{+1.63}_{-1.21}$ &  2.33$^{+0.04}_{-0.03}$\\ 
NGC1291&KINGFISH &   6.98$^{+0.05}_{-0.05}$ & -0.82$^{+0.10}_{-0.08}$ &  0.66$^{+0.15}_{-0.16}$ &  3.96$^{+0.72}_{-0.66}$\\ 
NGC1316&KINGFISH &   6.81$^{+0.07}_{-0.05}$ & -0.65$^{+0.13}_{-0.10}$ &  2.19$^{+0.57}_{-0.72}$ &  3.78$^{+0.88}_{-0.86}$\\ 
NGC1377&KINGFISH &   5.62$^{+0.12}_{-0.11}$ & -0.55$^{+0.17}_{-0.16}$ & 17.37$^{+7.04}_{-4.86}$ &  2.14$^{+0.05}_{-0.03}$\\ 
IC0342&KINGFISH &   7.21$^{+0.00}_{-0.00}$ & -0.59$^{+0.00}_{-0.00}$ &  3.04$^{+1.02}_{-0.85}$ &  1.00$^{+0.11}_{-0.09}$\\ 
NGC1482&KINGFISH &   7.15$^{+0.05}_{-0.04}$ & -1.18$^{+0.15}_{-0.09}$ &  2.93$^{+0.51}_{-0.47}$ &  2.22$^{+0.02}_{-0.02}$\\ 
NGC1512&KINGFISH &   7.16$^{+0.17}_{-0.08}$ & -1.05$^{+0.25}_{-0.13}$ &  0.21$^{+0.08}_{-0.07}$ &  2.31$^{+0.07}_{-0.04}$\\ 
NGC2146&KINGFISH &   7.41$^{+0.06}_{-0.05}$ & -1.07$^{+0.17}_{-0.11}$ &  5.67$^{+1.22}_{-0.87}$ &  2.37$^{+0.07}_{-0.04}$\\ 
HoII&KINGFISH &   4.19$^{+0.06}_{-0.05}$ & -0.65$^{+0.10}_{-0.09}$ & 10.18$^{+1.80}_{-2.47}$ &  4.02$^{+0.71}_{-0.82}$\\ 
NGC2798&KINGFISH &   6.88$^{+0.05}_{-0.05}$ & -1.20$^{+0.24}_{-0.17}$ &  3.86$^{+0.57}_{-0.57}$ &  2.20$^{+0.03}_{-0.02}$\\ 
NGC2841&KINGFISH &   7.55$^{+0.07}_{-0.05}$ & -0.84$^{+0.11}_{-0.09}$ &  0.57$^{+0.25}_{-0.19}$ &  3.12$^{+0.83}_{-0.35}$\\ 
NGC2915&KINGFISH &   4.46$^{+0.08}_{-0.05}$ & -0.72$^{+0.12}_{-0.09}$ &  2.61$^{+1.10}_{-0.92}$ &  3.13$^{+1.08}_{-0.47}$\\ 
NGC2976&KINGFISH &   6.06$^{+0.08}_{-0.08}$ & -0.75$^{+0.12}_{-0.14}$ &  0.85$^{+0.52}_{-0.25}$ &  2.55$^{+0.37}_{-0.19}$\\ 
NGC3049&KINGFISH &   6.74$^{+0.16}_{-0.09}$ &  0.14$^{+0.31}_{-0.18}$ &  0.77$^{+0.57}_{-0.37}$ &  3.66$^{+0.95}_{-0.91}$\\ 
NGC3077&KINGFISH &   5.54$^{+0.07}_{-0.05}$ & -0.58$^{+0.10}_{-0.08}$ &  3.89$^{+1.44}_{-1.15}$ &  3.21$^{+1.05}_{-0.53}$\\ 
NGC3190&KINGFISH &   7.00$^{+0.07}_{-0.06}$ & -1.01$^{+0.10}_{-0.11}$ &  1.01$^{+0.50}_{-0.23}$ &  2.63$^{+0.37}_{-0.16}$\\ 
NGC3184&KINGFISH &   7.30$^{+0.09}_{-0.05}$ & -0.70$^{+0.13}_{-0.09}$ &  0.80$^{+0.38}_{-0.33}$ &  2.99$^{+0.76}_{-0.40}$\\ 
NGC3198&KINGFISH &   7.32$^{+0.15}_{-0.07}$ & -0.55$^{+0.21}_{-0.12}$ &  0.55$^{+0.40}_{-0.27}$ &  2.86$^{+0.97}_{-0.41}$\\ 
IC2574&KINGFISH &   5.59$^{+0.07}_{-0.06}$ & -0.95$^{+0.15}_{-0.15}$ &  0.59$^{+0.19}_{-0.14}$ &  2.43$^{+0.12}_{-0.07}$\\ 
NGC3265&KINGFISH &   5.88$^{+0.13}_{-0.10}$ & -0.14$^{+0.24}_{-0.18}$ &  3.04$^{+3.17}_{-1.00}$ &  2.64$^{+1.30}_{-0.30}$\\ 
NGC3351&KINGFISH &   6.94$^{+0.06}_{-0.05}$ & -0.40$^{+0.12}_{-0.09}$ &  1.48$^{+0.39}_{-0.52}$ &  3.89$^{+0.82}_{-1.00}$\\ 
NGC3521&KINGFISH &   7.63$^{+0.07}_{-0.04}$ & -0.68$^{+0.11}_{-0.07}$ &  1.63$^{+0.45}_{-0.62}$ &  3.54$^{+1.04}_{-0.72}$\\ 
NGC3621&KINGFISH &   7.03$^{+0.09}_{-0.06}$ & -0.69$^{+0.13}_{-0.10}$ &  1.12$^{+0.66}_{-0.45}$ &  2.93$^{+1.03}_{-0.42}$\\ 
NGC3627&KINGFISH &   7.28$^{+0.06}_{-0.04}$ & -0.57$^{+0.09}_{-0.07}$ &  2.76$^{+0.57}_{-0.96}$ &  3.89$^{+0.76}_{-1.06}$\\ 
NGC3773&KINGFISH &   5.59$^{+0.15}_{-0.08}$ & -0.34$^{+0.23}_{-0.18}$ &  1.98$^{+1.11}_{-0.98}$ &  3.04$^{+1.35}_{-0.77}$\\ 
NGC3938&KINGFISH &   7.41$^{+0.09}_{-0.07}$ & -0.75$^{+0.12}_{-0.11}$ &  0.97$^{+0.72}_{-0.34}$ &  2.78$^{+0.95}_{-0.34}$\\ 
NGC4236&KINGFISH &   6.49$^{+0.07}_{-0.04}$ & -0.72$^{+0.16}_{-0.13}$ &  0.11$^{+0.03}_{-0.03}$ &  2.35$^{+0.07}_{-0.06}$\\ 
NGC4254&KINGFISH &   7.54$^{+0.06}_{-0.04}$ & -0.62$^{+0.10}_{-0.07}$ &  2.23$^{+0.51}_{-0.74}$ &  3.62$^{+0.86}_{-0.74}$\\ 
NGC4321&KINGFISH &   7.61$^{+0.05}_{-0.03}$ & -0.65$^{+0.08}_{-0.06}$ &  1.67$^{+0.32}_{-0.49}$ &  3.98$^{+0.69}_{-0.86}$\\ 
NGC4536&KINGFISH &   7.14$^{+0.10}_{-0.06}$ & -0.29$^{+0.17}_{-0.12}$ &  2.14$^{+0.93}_{-0.95}$ &  3.14$^{+1.08}_{-0.65}$\\ 
NGC4559&KINGFISH &   6.75$^{+0.08}_{-0.07}$ & -0.87$^{+0.13}_{-0.15}$ &  0.56$^{+0.26}_{-0.16}$ &  2.46$^{+0.22}_{-0.14}$\\ 
NGC4569&KINGFISH &   6.88$^{+0.05}_{-0.04}$ & -0.51$^{+0.09}_{-0.07}$ &  1.44$^{+0.23}_{-0.27}$ &  4.29$^{+0.50}_{-0.69}$\\ 
NGC4579&KINGFISH &   7.29$^{+0.05}_{-0.03}$ & -0.76$^{+0.08}_{-0.06}$ &  1.31$^{+0.20}_{-0.25}$ &  4.19$^{+0.52}_{-0.70}$\\ 
NGC4594&KINGFISH &   6.99$^{+0.06}_{-0.04}$ & -0.85$^{+0.10}_{-0.07}$ &  0.83$^{+0.17}_{-0.21}$ &  3.98$^{+0.71}_{-0.72}$\\ 
NGC4625&KINGFISH &   5.99$^{+0.09}_{-0.06}$ & -0.73$^{+0.13}_{-0.09}$ &  1.06$^{+0.50}_{-0.40}$ &  3.18$^{+1.19}_{-0.47}$\\ 
NGC4631&KINGFISH &   7.24$^{+0.07}_{-0.05}$ & -0.70$^{+0.11}_{-0.09}$ &  2.33$^{+1.03}_{-0.91}$ &  3.14$^{+1.13}_{-0.58}$\\ 
NGC4725&KINGFISH &   7.49$^{+0.10}_{-0.07}$ & -0.83$^{+0.14}_{-0.11}$ &  0.35$^{+0.17}_{-0.12}$ &  2.77$^{+0.36}_{-0.19}$\\ 
NGC4736&KINGFISH &   6.39$^{+0.06}_{-0.04}$ & -0.70$^{+0.09}_{-0.07}$ &  4.96$^{+1.05}_{-1.58}$ &  3.66$^{+0.83}_{-0.79}$\\ 
NGC4826&KINGFISH &   6.33$^{+0.06}_{-0.05}$ & -0.81$^{+0.08}_{-0.07}$ &  4.04$^{+0.78}_{-1.28}$ &  3.70$^{+0.86}_{-0.85}$\\ 
NGC5055&KINGFISH &   7.51$^{+0.07}_{-0.05}$ & -0.76$^{+0.10}_{-0.08}$ &  1.28$^{+0.33}_{-0.39}$ &  3.60$^{+0.91}_{-0.70}$\\ 
NGC5398&KINGFISH &   6.13$^{+0.27}_{-0.15}$ &  0.34$^{+0.48}_{-0.26}$ &  0.26$^{+0.35}_{-0.13}$ &  3.31$^{+1.07}_{-0.60}$\\ 
NGC5457&KINGFISH &   7.61$^{+0.11}_{-0.08}$ & -0.67$^{+0.16}_{-0.13}$ &  0.64$^{+0.60}_{-0.26}$ &  2.68$^{+0.89}_{-0.29}$\\ 
NGC5408&KINGFISH &   3.93$^{+0.27}_{-0.11}$ &  0.62$^{+1.01}_{-0.31}$ & 16.05$^{+14.06}_{-9.70}$ &  2.89$^{+1.14}_{-0.47}$\\ 
NGC5474&KINGFISH &   6.16$^{+0.08}_{-0.07}$ & -1.20$^{+0.18}_{-0.16}$ &  0.36$^{+0.12}_{-0.08}$ &  2.37$^{+0.09}_{-0.05}$\\ 
NGC5713&KINGFISH &   7.22$^{+0.06}_{-0.06}$ & -1.01$^{+0.24}_{-0.13}$ &  1.68$^{+0.41}_{-0.30}$ &  2.25$^{+0.05}_{-0.03}$\\ 
NGC5866&KINGFISH &   6.59$^{+0.05}_{-0.04}$ & -1.19$^{+0.08}_{-0.07}$ &  3.53$^{+0.63}_{-0.75}$ &  3.99$^{+0.67}_{-0.78}$\\ 
NGC6946&KINGFISH &   7.53$^{+0.06}_{-0.04}$ & -0.51$^{+0.10}_{-0.08}$ &  2.02$^{+0.52}_{-0.54}$ &  3.89$^{+0.68}_{-0.85}$\\ 
NGC7331&KINGFISH &   7.78$^{+0.09}_{-0.05}$ & -0.77$^{+0.12}_{-0.08}$ &  1.53$^{+0.60}_{-0.60}$ &  3.20$^{+1.06}_{-0.57}$\\ 
Haro11&DGS &   6.15$^{+0.16}_{-0.13}$ & -0.47$^{+0.23}_{-0.19}$ & 62.31$^{+27.03}_{-19.03}$ &  2.14$^{+0.04}_{-0.03}$\\ 
Haro2&DGS &   5.87$^{+0.49}_{-0.11}$ & -0.58$^{+2.35}_{-0.21}$ &  4.31$^{+1.52}_{-3.46}$ &  2.16$^{+0.03}_{-0.03}$\\ 
Haro3&DGS &   6.05$^{+0.11}_{-0.08}$ & -0.64$^{+0.30}_{-0.18}$ &  2.38$^{+0.76}_{-0.85}$ &  2.15$^{+0.03}_{-0.03}$\\ 
He2-10&DGS &   5.72$^{+0.98}_{-0.12}$ & -0.52$^{+4.21}_{-0.20}$ &  4.69$^{+1.94}_{-4.04}$ &  2.13$^{+0.03}_{-0.03}$\\ 
Mrk 1089&DGS &   7.08$^{+0.13}_{-0.08}$ & -0.34$^{+0.32}_{-0.19}$ &  1.55$^{+0.63}_{-0.52}$ &  2.22$^{+0.05}_{-0.03}$\\ 
Mrk 930&DGS &   6.60$^{+0.17}_{-0.10}$ & -0.16$^{+0.39}_{-0.22}$ &  2.15$^{+1.01}_{-0.85}$ &  2.20$^{+0.03}_{-0.02}$\\ 
NGC 1140&DGS &   6.52$^{+0.33}_{-0.14}$ &  0.28$^{+0.56}_{-0.26}$ &  0.92$^{+1.00}_{-0.48}$ &  2.98$^{+1.34}_{-0.63}$\\ 
NGC 1569&DGS &   4.97$^{+0.57}_{-0.11}$ & -0.51$^{+2.37}_{-0.17}$ &  9.47$^{+4.33}_{-8.00}$ &  2.26$^{+0.07}_{-0.05}$\\ 
NGC 4214&DGS &   5.88$^{+0.27}_{-0.12}$ &  0.04$^{+0.52}_{-0.23}$ &  0.81$^{+0.71}_{-0.49}$ &  3.23$^{+1.04}_{-0.62}$\\ 
NGC 4449&DGS &   6.16$^{+0.09}_{-0.06}$ & -0.38$^{+0.17}_{-0.12}$ &  2.68$^{+0.94}_{-0.99}$ &  3.81$^{+0.84}_{-0.97}$\\ 
NGC4030&HIGH &   8.07$^{+0.05}_{-0.04}$ & -0.75$^{+0.08}_{-0.07}$ &  2.20$^{+0.32}_{-0.32}$ &  4.51$^{+0.36}_{-0.59}$\\ 
NGC5496&HIGH &   7.32$^{+0.07}_{-0.11}$ & -0.88$^{+0.20}_{-0.22}$ &  0.14$^{+0.10}_{-0.04}$ &  2.40$^{+0.27}_{-0.12}$\\ 
NGC5584&HIGH &   7.54$^{+0.06}_{-0.05}$ & -1.12$^{+0.53}_{-0.18}$ &  0.20$^{+0.05}_{-0.04}$ &  2.17$^{+0.07}_{-0.03}$\\ 
UGC09215&HIGH &   7.14$^{+0.14}_{-0.08}$ & -0.41$^{+0.24}_{-0.16}$ &  0.25$^{+0.18}_{-0.11}$ &  2.82$^{+1.21}_{-0.48}$\\ 
NGC5690&HIGH &   7.63$^{+0.04}_{-0.04}$ & -0.57$^{+0.09}_{-0.07}$ &  1.82$^{+0.28}_{-0.28}$ &  4.31$^{+0.50}_{-0.64}$\\ 
NGC5691&HIGH &   7.00$^{+0.06}_{-0.07}$ & -1.52$^{+0.63}_{-0.22}$ &  0.82$^{+0.27}_{-0.15}$ &  2.14$^{+0.03}_{-0.01}$\\ 
NGC5719&HIGH &   7.50$^{+0.09}_{-0.11}$ & -1.06$^{+0.56}_{-0.20}$ &  0.84$^{+1.69}_{-0.19}$ &  2.28$^{+2.01}_{-0.07}$\\ 
NGC5740&HIGH &   7.13$^{+0.05}_{-0.04}$ & -0.44$^{+0.10}_{-0.08}$ &  1.54$^{+0.35}_{-0.35}$ &  3.92$^{+0.74}_{-0.83}$\\ 
NGC5746&HIGH &   7.96$^{+0.04}_{-0.03}$ & -0.86$^{+0.07}_{-0.06}$ &  0.70$^{+0.11}_{-0.10}$ &  4.35$^{+0.48}_{-0.63}$\\ 
UGC07000&HIGH &   6.43$^{+0.12}_{-0.09}$ & -0.64$^{+0.25}_{-0.16}$ &  1.00$^{+0.59}_{-0.43}$ &  3.75$^{+0.83}_{-0.98}$\\ 
UGC09470&HIGH &   6.46$^{+0.49}_{-0.15}$ & -0.43$^{+1.15}_{-0.32}$ &  0.28$^{+0.50}_{-0.18}$ &  3.46$^{+1.05}_{-0.82}$\\ 
UGC04996&HIGH &   7.25$^{+0.20}_{-0.12}$ & -0.54$^{+0.42}_{-0.25}$ &  0.24$^{+0.28}_{-0.13}$ &  3.09$^{+1.21}_{-0.63}$\\ 
IC1011&HIGH &   7.64$^{+0.16}_{-0.24}$ & -1.25$^{+1.30}_{-0.34}$ &  0.38$^{+1.69}_{-0.14}$ &  2.10$^{+1.86}_{-0.02}$\\ 
NGC3254&HRS &   7.76$^{+0.09}_{-0.06}$ & -0.78$^{+0.14}_{-0.10}$ &  0.28$^{+0.07}_{-0.09}$ &  4.21$^{+0.55}_{-0.69}$\\ 
NGC3338&HRS &   7.62$^{+0.09}_{-0.06}$ & -0.71$^{+0.15}_{-0.11}$ &  0.64$^{+0.25}_{-0.22}$ &  3.83$^{+0.80}_{-0.82}$\\ 
NGC3370&HRS &   7.27$^{+0.08}_{-0.05}$ & -0.58$^{+0.13}_{-0.10}$ &  1.76$^{+0.54}_{-0.58}$ &  3.87$^{+0.78}_{-0.82}$\\ 
NGC3381&HRS &   7.30$^{+0.19}_{-0.17}$ & -0.76$^{+0.46}_{-0.28}$ &  0.11$^{+0.37}_{-0.06}$ &  2.12$^{+1.65}_{-0.03}$\\ 
NGC3424&HRS &   7.27$^{+0.08}_{-0.06}$ & -0.52$^{+0.13}_{-0.10}$ &  2.86$^{+0.73}_{-0.88}$ &  4.01$^{+0.67}_{-0.91}$\\ 
NGC3430&HRS &   7.45$^{+0.06}_{-0.06}$ & -0.63$^{+0.11}_{-0.09}$ &  1.38$^{+0.34}_{-0.33}$ &  4.03$^{+0.65}_{-0.70}$\\ 
NGC3437&HRS &   7.57$^{+0.08}_{-0.09}$ & -0.32$^{+0.17}_{-0.16}$ &  0.21$^{+0.13}_{-0.06}$ &  2.04$^{+0.01}_{-0.01}$\\ 
NGC3448&HRS &   7.18$^{+0.10}_{-0.09}$ & -1.04$^{+0.30}_{-0.20}$ &  0.38$^{+0.19}_{-0.11}$ &  2.15$^{+0.03}_{-0.02}$\\ 
NGC3504&HRS &   7.72$^{+0.08}_{-0.07}$ & -0.45$^{+0.17}_{-0.15}$ &  0.29$^{+0.13}_{-0.09}$ &  2.04$^{+0.01}_{-0.01}$\\ 
NGC3512&HRS &   6.81$^{+0.10}_{-0.07}$ & -0.65$^{+0.15}_{-0.11}$ &  1.29$^{+0.58}_{-0.53}$ &  3.57$^{+0.99}_{-0.87}$\\ 
NGC3655&HRS &   7.00$^{+0.08}_{-0.05}$ & -0.54$^{+0.14}_{-0.10}$ &  3.82$^{+1.16}_{-1.01}$ &  4.06$^{+0.63}_{-0.68}$\\ 
NGC3659&HRS &   6.75$^{+0.09}_{-0.08}$ & -0.64$^{+0.17}_{-0.13}$ &  1.06$^{+0.53}_{-0.42}$ &  3.55$^{+0.97}_{-0.78}$\\ 
NGC3666&HRS &   6.83$^{+0.08}_{-0.06}$ & -0.71$^{+0.13}_{-0.10}$ &  1.17$^{+0.44}_{-0.42}$ &  3.68$^{+0.94}_{-0.87}$\\ 
NGC3683&HRS &   7.36$^{+0.37}_{-0.08}$ & -0.58$^{+0.49}_{-0.22}$ &  4.33$^{+1.91}_{-3.61}$ &  3.71$^{+0.89}_{-1.53}$\\ 
NGC3686&HRS &   6.98$^{+0.07}_{-0.05}$ & -0.52$^{+0.13}_{-0.09}$ &  1.20$^{+0.36}_{-0.28}$ &  4.10$^{+0.59}_{-0.66}$\\ 
NGC3729&HRS &   6.96$^{+0.11}_{-0.07}$ & -0.33$^{+0.21}_{-0.14}$ &  1.25$^{+0.52}_{-0.50}$ &  4.12$^{+0.63}_{-1.06}$\\ 
NGC3953&HRS &   7.51$^{+0.06}_{-0.04}$ & -0.65$^{+0.10}_{-0.08}$ &  0.61$^{+0.14}_{-0.13}$ &  4.33$^{+0.50}_{-0.64}$\\ 
NGC3982&HRS &   6.98$^{+0.08}_{-0.06}$ & -0.39$^{+0.15}_{-0.11}$ &  2.85$^{+0.92}_{-0.91}$ &  3.96$^{+0.75}_{-0.87}$\\ 
NGC4116&HRS &   7.19$^{+0.14}_{-0.09}$ & -0.61$^{+0.26}_{-0.17}$ &  0.25$^{+0.20}_{-0.12}$ &  3.10$^{+1.20}_{-0.75}$\\ 
NGC4178&HRS &   7.54$^{+0.12}_{-0.07}$ & -0.71$^{+0.17}_{-0.12}$ &  0.63$^{+0.31}_{-0.33}$ &  3.21$^{+1.14}_{-0.64}$\\ 
NGC4206&HRS &   7.25$^{+0.09}_{-0.07}$ & -0.58$^{+0.17}_{-0.14}$ &  0.23$^{+0.10}_{-0.07}$ &  4.03$^{+0.67}_{-0.73}$\\ 
NGC4207&HRS &   6.32$^{+0.07}_{-0.06}$ & -0.73$^{+0.12}_{-0.10}$ &  4.58$^{+1.32}_{-1.18}$ &  4.13$^{+0.63}_{-0.70}$\\ 
NGC4237&HRS &   6.95$^{+0.06}_{-0.05}$ & -0.75$^{+0.10}_{-0.08}$ &  1.73$^{+0.37}_{-0.38}$ &  4.17$^{+0.59}_{-0.64}$\\ 
NGC4254&HRS &   7.51$^{+0.06}_{-0.05}$ & -0.52$^{+0.10}_{-0.08}$ &  1.83$^{+0.42}_{-0.36}$ &  4.29$^{+0.49}_{-0.77}$\\ 
NGC4294&HRS &   6.58$^{+0.11}_{-0.07}$ & -0.68$^{+0.16}_{-0.12}$ &  1.31$^{+0.57}_{-0.60}$ &  3.29$^{+1.03}_{-0.70}$\\ 
NGC4298&HRS &   7.35$^{+0.05}_{-0.04}$ & -0.93$^{+0.08}_{-0.06}$ &  1.37$^{+0.23}_{-0.24}$ &  4.27$^{+0.54}_{-0.70}$\\ 
NGC4302&HRS &   7.22$^{+0.06}_{-0.06}$ & -0.85$^{+0.09}_{-0.09}$ &  0.92$^{+0.21}_{-0.19}$ &  4.06$^{+0.64}_{-0.73}$\\ 
NGC4321&HRS &   7.83$^{+0.06}_{-0.05}$ & -0.52$^{+0.11}_{-0.09}$ &  1.15$^{+0.29}_{-0.22}$ &  4.21$^{+0.55}_{-0.67}$\\ 
NGC4351&HRS &   6.04$^{+0.11}_{-0.08}$ & -0.74$^{+0.21}_{-0.14}$ &  0.90$^{+0.37}_{-0.42}$ &  3.55$^{+1.00}_{-0.83}$\\ 
NGC4378&HRS &   7.71$^{+0.10}_{-0.07}$ & -0.74$^{+0.22}_{-0.15}$ &  0.41$^{+0.18}_{-0.14}$ &  4.21$^{+0.56}_{-0.58}$\\ 
NGC4380&HRS &   7.04$^{+0.08}_{-0.06}$ & -0.73$^{+0.15}_{-0.11}$ &  0.59$^{+0.16}_{-0.14}$ &  4.18$^{+0.58}_{-0.70}$\\ 
NGC4383&HRS &   6.75$^{+0.09}_{-0.09}$ & -1.09$^{+0.43}_{-0.22}$ &  0.66$^{+0.30}_{-0.21}$ &  2.12$^{+0.02}_{-0.02}$\\ 
NGC4388&HRS &   7.14$^{+0.08}_{-0.07}$ & -0.58$^{+0.19}_{-0.15}$ &  0.22$^{+0.11}_{-0.06}$ &  2.04$^{+0.01}_{-0.01}$\\ 
NGC4396&HRS &   7.04$^{+0.17}_{-0.15}$ & -0.93$^{+0.32}_{-0.30}$ &  0.12$^{+0.21}_{-0.05}$ &  2.33$^{+0.71}_{-0.13}$\\ 
NGC4402&HRS &   6.87$^{+0.05}_{-0.05}$ & -0.68$^{+0.09}_{-0.08}$ &  1.53$^{+0.33}_{-0.32}$ &  4.16$^{+0.59}_{-0.82}$\\ 
NGC4413&HRS &   6.47$^{+0.10}_{-0.06}$ & -0.52$^{+0.17}_{-0.11}$ &  0.85$^{+0.31}_{-0.30}$ &  3.82$^{+0.79}_{-0.86}$\\ 
NGC4412&HRS &   7.15$^{+0.11}_{-0.07}$ & -0.94$^{+0.32}_{-0.17}$ &  0.44$^{+0.22}_{-0.14}$ &  2.11$^{+0.03}_{-0.02}$\\ 
NGC4419&HRS &   6.97$^{+0.09}_{-0.07}$ & -1.23$^{+0.29}_{-0.18}$ &  0.47$^{+0.17}_{-0.11}$ &  2.11$^{+0.02}_{-0.02}$\\ 
NGC4409&HRS &   6.49$^{+0.08}_{-0.05}$ & -0.65$^{+0.13}_{-0.09}$ &  1.77$^{+0.56}_{-0.52}$ &  3.75$^{+0.88}_{-0.72}$\\ 
NGC4435&HRS &   6.01$^{+0.09}_{-0.06}$ & -0.74$^{+0.13}_{-0.10}$ &  3.62$^{+0.95}_{-0.87}$ &  4.12$^{+0.61}_{-0.80}$\\ 
NGC4438&HRS &   6.47$^{+0.07}_{-0.05}$ & -0.92$^{+0.10}_{-0.08}$ &  1.80$^{+0.43}_{-0.53}$ &  3.80$^{+0.87}_{-0.73}$\\ 
NGC4450&HRS &   7.02$^{+0.07}_{-0.05}$ & -0.90$^{+0.11}_{-0.09}$ &  0.89$^{+0.20}_{-0.21}$ &  4.14$^{+0.60}_{-0.68}$\\ 
IC3392&HRS &   6.07$^{+0.08}_{-0.06}$ & -0.66$^{+0.13}_{-0.10}$ &  1.30$^{+0.37}_{-0.41}$ &  3.98$^{+0.65}_{-0.84}$\\ 
NGC4470&HRS &   6.35$^{+0.08}_{-0.06}$ & -0.69$^{+0.12}_{-0.10}$ &  2.09$^{+0.66}_{-0.74}$ &  3.59$^{+0.92}_{-0.83}$\\ 
NGC4496A&HRS &   6.94$^{+0.10}_{-0.06}$ & -0.51$^{+0.17}_{-0.11}$ &  0.87$^{+0.37}_{-0.30}$ &  3.73$^{+0.84}_{-0.85}$\\ 
NGC4498&HRS &   6.78$^{+0.13}_{-0.08}$ & -0.80$^{+0.19}_{-0.12}$ &  0.57$^{+0.46}_{-0.29}$ &  2.92$^{+1.00}_{-0.48}$\\ 
NGC4501&HRS &   7.42$^{+0.06}_{-0.05}$ & -0.69$^{+0.09}_{-0.08}$ &  1.28$^{+0.22}_{-0.23}$ &  4.30$^{+0.52}_{-0.59}$\\ 
IC3476&HRS &   6.32$^{+0.10}_{-0.14}$ & -1.02$^{+0.52}_{-0.23}$ &  0.25$^{+0.48}_{-0.08}$ &  2.20$^{+0.89}_{-0.03}$\\ 
NGC4437&HRS &   7.19$^{+0.05}_{-0.05}$ & -0.62$^{+0.09}_{-0.09}$ &  0.52$^{+0.13}_{-0.12}$ &  4.09$^{+0.64}_{-0.74}$\\ 
NGC4522&HRS &   6.47$^{+0.10}_{-0.06}$ & -0.66$^{+0.16}_{-0.11}$ &  1.14$^{+0.40}_{-0.41}$ &  3.79$^{+0.80}_{-0.86}$\\ 
NGC4526&HRS &   6.60$^{+0.08}_{-0.06}$ & -0.99$^{+0.12}_{-0.10}$ &  3.02$^{+0.88}_{-1.09}$ &  3.53$^{+1.00}_{-0.79}$\\ 
NGC4532&HRS &   6.32$^{+0.16}_{-0.08}$ & -0.41$^{+0.25}_{-0.17}$ &  2.77$^{+1.19}_{-1.56}$ &  3.79$^{+0.86}_{-1.41}$\\ 
NGC4536&HRS &   7.46$^{+0.09}_{-0.06}$ & -0.90$^{+0.23}_{-0.15}$ &  0.29$^{+0.12}_{-0.09}$ &  2.11$^{+0.02}_{-0.01}$\\ 
NGC4567&HRS &   8.00$^{+0.06}_{-0.05}$ & -0.64$^{+0.09}_{-0.09}$ &  2.08$^{+0.46}_{-0.44}$ &  4.20$^{+0.55}_{-0.65}$\\ 
NGC4568&HRS &   7.64$^{+0.05}_{-0.04}$ & -0.64$^{+0.09}_{-0.07}$ &  1.89$^{+0.39}_{-0.39}$ &  4.22$^{+0.56}_{-0.72}$\\ 
NGC4569&HRS &   7.06$^{+0.06}_{-0.05}$ & -0.46$^{+0.12}_{-0.09}$ &  1.21$^{+0.32}_{-0.24}$ &  4.25$^{+0.54}_{-0.67}$\\ 
NGC4579&HRS &   7.63$^{+0.10}_{-0.07}$ & -0.54$^{+0.17}_{-0.12}$ &  0.68$^{+0.24}_{-0.19}$ &  4.24$^{+0.52}_{-0.75}$\\ 
NGC4580&HRS &   6.81$^{+0.07}_{-0.05}$ & -0.73$^{+0.11}_{-0.09}$ &  0.98$^{+0.27}_{-0.29}$ &  3.84$^{+0.82}_{-0.74}$\\ 
NGC4592&HRS &   7.06$^{+0.13}_{-0.08}$ & -0.62$^{+0.31}_{-0.17}$ &  0.11$^{+0.09}_{-0.04}$ &  2.70$^{+1.20}_{-0.34}$\\ 
NGC4607&HRS &   6.84$^{+0.08}_{-0.06}$ & -0.67$^{+0.12}_{-0.09}$ &  1.66$^{+0.45}_{-0.48}$ &  3.88$^{+0.81}_{-0.79}$\\ 
NGC4630&HRS &   6.60$^{+0.17}_{-0.16}$ & -0.79$^{+0.45}_{-0.28}$ &  0.49$^{+1.06}_{-0.24}$ &  2.22$^{+1.89}_{-0.06}$\\ 
NGC4639&HRS &   7.08$^{+0.11}_{-0.08}$ & -0.72$^{+0.18}_{-0.13}$ &  0.89$^{+0.43}_{-0.38}$ &  3.51$^{+1.07}_{-0.73}$\\ 
NGC4651&HRS &   7.42$^{+0.07}_{-0.06}$ & -0.67$^{+0.12}_{-0.10}$ &  1.28$^{+0.37}_{-0.44}$ &  3.79$^{+0.82}_{-0.86}$\\ 
NGC4654&HRS &   7.20$^{+0.06}_{-0.05}$ & -0.60$^{+0.10}_{-0.09}$ &  1.90$^{+0.47}_{-0.53}$ &  3.94$^{+0.75}_{-0.87}$\\ 
NGC4689&HRS &   6.98$^{+0.06}_{-0.06}$ & -0.64$^{+0.10}_{-0.09}$ &  1.07$^{+0.26}_{-0.23}$ &  4.17$^{+0.56}_{-0.68}$\\ 
NGC4713&HRS &   6.92$^{+0.12}_{-0.14}$ & -0.75$^{+0.45}_{-0.26}$ &  0.28$^{+0.52}_{-0.11}$ &  2.17$^{+1.30}_{-0.04}$\\ 
NGC4747&HRS &   6.70$^{+0.19}_{-0.13}$ & -0.37$^{+0.38}_{-0.27}$ &  0.27$^{+0.30}_{-0.16}$ &  3.03$^{+1.29}_{-0.85}$\\ 
NGC4808&HRS &   6.91$^{+0.08}_{-0.05}$ & -0.56$^{+0.13}_{-0.09}$ &  2.05$^{+0.57}_{-0.62}$ &  3.82$^{+0.79}_{-0.83}$\\ 
NGC5014&HRS &   5.77$^{+0.20}_{-0.11}$ & -0.50$^{+0.50}_{-0.22}$ &  0.12$^{+0.12}_{-0.07}$ &  2.08$^{+0.03}_{-0.02}$\\ 
NGC5147&HRS &   6.88$^{+0.15}_{-0.08}$ & -0.60$^{+0.23}_{-0.14}$ &  1.31$^{+0.63}_{-0.75}$ &  3.48$^{+0.93}_{-0.96}$\\ 
NGC5248&HRS &   7.24$^{+0.06}_{-0.05}$ & -0.48$^{+0.11}_{-0.09}$ &  1.76$^{+0.39}_{-0.43}$ &  4.20$^{+0.57}_{-0.69}$\\ 
NGC5669&HRS &   6.80$^{+0.14}_{-0.09}$ & -0.68$^{+0.25}_{-0.17}$ &  0.53$^{+0.44}_{-0.26}$ &  3.33$^{+1.09}_{-0.79}$\\ 
 \\
\end{longtable}
\endgroup
\clearpage
\twocolumn

%%%%%%%%%%%%%%%%%%%%%%%%%%%%%%%%%%%%%%%%%%%%%%%%%%

% Don't change these lines
\bsp	% typesetting comment
\label{lastpage}
\end{document}

% End of mnras_template.tex